\documentclass[prb,twocolumn,superscriptaddress,floatfix,longbibliography,a4paper]{revtex4-1}
\usepackage[english]{babel}
\usepackage[utf8]{inputenc}
\usepackage{graphicx}
\usepackage{amsmath}
\usepackage{amssymb}
\usepackage{color}
\usepackage[normalem]{ulem}
\usepackage{comment}
\usepackage{hyperref}
\usepackage{amsthm}
\usepackage{mathtools}
\usepackage{physics}
\usepackage{xcolor}
\usepackage{graphicx}
\usepackage[left=23mm,right=13mm,top=35mm,columnsep=15pt]{geometry} 
\usepackage{adjustbox}
\usepackage{placeins}
\usepackage[T1]{fontenc}
\usepackage{lipsum}
\usepackage{csquotes}

\begin{document}
\title{Spectral and thermodynamic properties of the Holstein polaron: Hierarchical equations of motion approach}

\author{Veljko Jankovi\'c}
    \email{veljko.jankovic@ipb.ac.rs}
    \affiliation{Institute of Physics Belgrade, University of Belgrade, Pregrevica 118, 11080 Belgrade, Serbia}
\author{Nenad Vukmirovi\'c}
    \email{nenad.vukmirovic@ipb.ac.rs}
    \affiliation{Institute of Physics Belgrade, University of Belgrade, Pregrevica 118, 11080 Belgrade, Serbia}

\begin{abstract}
We develop a hierarchical equations of motion (HEOM) approach to compute real-time single-particle correlation functions and thermodynamic properties of the Holstein model at finite temperature. We exploit the conservation of the total momentum of the system to formulate the momentum-space HEOM whose dynamical variables explicitly keep track of momentum exchanges between the electron and phonons. Our symmetry-adapted HEOM enable us to overcome the numerical instabilities inherent to the commonly used real-space HEOM. The HEOM method is then used to study the spectral function and thermodynamic quantities of chains containing up to ten sites. The HEOM results compare favorably to existing literature. To provide an independent assessment of the HEOM approach and to gain insight into the importance of finite-size effects, we devise a quantum Monte Carlo (QMC) procedure to evaluate finite-temperature single-particle correlation functions in imaginary time and apply it to chains containing up to twenty sites. QMC results reveal that finite-size effects are quite weak, so that the results on 5 to 10-site chains, depending on the parameter regime, are representative of larger systems. A detailed comparison between the HEOM and QMC data place our HEOM method among reliable methods to compute real-time finite-temperature correlation functions in parameter regimes ranging from low- to high-temperature, and weak- to strong-coupling regime.
\end{abstract}

\maketitle

\section{Introduction}
The coupling of electronic excitations (electrons or excitons) to quantum lattice vibrations fundamentally determines physical properties of a wide variety of systems. The examples include organic semiconductors~\cite{Baessler-Koehler-book,ChemRev.107.926,AdvFunctMater.25.1915} and photosynthetic pigment--protein complexes.~\cite{RevModPhys.90.035003,PhysRep.567.1,Photosynthetic-excitons-book} In such systems, the density of excitations, which are typically excited by light or introduced by doping, is low. An electronic excitation in the field of phonons is most simply modeled within the Holstein molecular-crystal model,~\cite{AnnPhys.8.325} in which the excitation is locally and linearly coupled to intramolecular vibrations. The distinctive feature of organic semiconductors and photosynthetic complexes is that the energy scales of the electronic (excitonic) bandwidth, phonon energy, and electron (exciton)--phonon couplings are all comparable to one another and to the thermal energy. In other words, there is no obvious small parameter in which a perturbation expansion can be performed, and the standard weak-coupling (Redfield-like~\cite{REDFIELD19651}) and strong-coupling (F\"{o}rster-like~\cite{Foerster1964} or Marcus-like~\cite{RevModPhys.65.599}) theories do not properly describe excitation dynamics.~\cite{JChemPhys.130.234110} Numerical studies are thus indispensable in tackling the most interesting intermediate-coupling regime.

Thermodynamic and dynamic properties of the Holstein model have been investigated using a host of numerical methods. The most popular ones are the exact diagonalization on finite lattices,~\cite{PhysRevB.45.7730,PhysLettA.180.280,PhysRevB.53.9666,PhysRevB.56.4484,PhysRevB.56.4513,PhysRevB.56.13634,PhysRevB.60.14092} quantum Monte Carlo (QMC) methods,~\cite{PhysRevB.26.5033,PhysRevLett.49.1522,PhysRevB.27.6097,PhysRevB.30.1671,PhysRevLett.81.2514,PhysRevLett.81.5382,PhysRevB.69.024301,PhysRevB.71.184310} the density matrix renormalization group,~\cite{PhysRevB.57.6376,PhysRevLett.80.2661,PhysRevLett.80.5607} variational techniques,~\cite{PhysRevB.48.3721,JChemPhys.109.6540,PhysRevB.58.6208,PhysRevB.60.1633,PhysRevB.65.174306} and the momentum-average approximation.~\cite{PhysRevLett.97.036402,PhysRevB.74.245104} The majority of these approaches were restricted to ground-state considerations, while the evaluation of finite-temperature (real-time) correlation functions has received limited attention. Early approaches to the finite-temperature single-particle spectral properties of the Holstein model were restricted to analytical~\cite{JPhysCondensMatter.18.7669} and numerical~\cite{PhysRevB.45.7730,PhysRevB.55.14872} studies on two-site systems. These were followed by the dynamical mean-field theory.~\cite{PhysRevB.56.4494} Recently, Bon\v ca and collaborators have examined the Holstein polaron's spectral function using the finite-temperature Lanczos method.~\cite{PhysRevB.100.094307,PhysRevB.103.054304} They have also investigated the thermodynamics and spectral functions of the Holstein polaron employing a finite-temperature time-dependent density-matrix renormalization group method.~\cite{PhysRevB.102.165155} Two-particle correlation functions, such as the conductivity, resistivity, mobility, or diffusion constant, were computed by performing different types of unitary transformations,~\cite{JChemPhys.54.4843,JChemPhys.67.5818,JChemPhys.72.2763,PhysRevB.69.075212,jcp.128.114713,PhysRevB.79.235206,PhysRevB.99.104304,PhysRevX.10.021062} or by resorting to the dynamical mean field theory,~\cite{PhysRevLett.91.256403,PhysRevB.74.075101} the momentum-average approximation,~\cite{PhysRevLett.107.076403} QMC,~\cite{PhysRevLett.114.146401} or finite-temperature time-dependent density matrix renormalization group.~\cite{JPhysChemLett.11.4930}  

On the other hand, within the chemical-physics community, the method of choice to investigate the dynamical properties of the Holstein model is the hierarchical equations of motion (HEOM) method.~\cite{JPhysSocJpn.75.082001,PhysRevE.75.031107,JChemPhys.130.234111,JChemPhys.153.020901} The HEOM method is a numerically exact density-matrix technique to calculate the dynamics of a quantum system of interest (here, electronic excitations) that is linearly coupled to a Gaussian bath (here, phonons). Starting from the exact result of the Feynman--Vernon influence functional theory,~\cite{AnnPhys.24.118} the problem is formulated as an infinite hierarchy of dynamical equations for the reduced density matrix (RDM), which completely describes the system of interest, and the so-called auxiliary density matrices (ADMs). While the systematic hierarchy truncation schemes do exist,~\cite{JChemPhys.130.084105,MolPhys.116.780} the number of ADMs that are necessary to obtain converged results can be quite high, which severely limits the applicability of the HEOM method. The HEOM method has been employed to examine the excitonic dynamics and absorption and emission spectra in Holstein-like models of photosynthetic molecular aggregates,~\cite{ProcNatlAcadSci.106.17255,JChemTheoryComput.8.2808,JChemTheoryComput.11.3411,JComputChem.39.1779,JChemPhys.153.244122,JChemPhys.153.244110} as well as to study the equilibrium (e.g., the exciton--polaron size) and dynamical (e.g., mobility) properties of the canonical Holstein model.~\cite{JChemPhys.132.081101,JChemPhys.142.174103,JPhysChemLett.6.3110,JChemPhys.143.194106} Nevertheless, two aspects have received limited attention in the HEOM-method investigations of the Holstein model. (i) The translational symmetry, which reduces the numerical complexity of standard density-matrix techniques,~\cite{kuhncontribution,Kirabook} has not been combined with the HEOM method. (ii) The HEOM method is usually applied to situations in which nuclear degrees of freedom constitute a genuine thermodynamic bath, i.e., the spectral density of the electron--phonon interaction is a continuous function. Its applications to the canonical Holstein model, for which the spectral density consists of delta-like peaks, may face serious numerical problems~\cite{JChemPhys.150.184109} and have treated relatively small systems~\cite{JChemPhys.140.134106,ChemPhys.515.129} and/or relatively short time scales.~\cite{JPhysChemLett.6.3110} Developing novel techniques to improve the numerical stability of the HEOM method for discrete oscillator modes is thus an active line of research.~\cite{JChemPhys.150.184109,JChemPhys.153.204109}

In this study, we devise a momentum-space HEOM method suitable to compute real-time single-particle correlation functions and thermodynamic properties of the Holstein Hamiltonian at finite temperature. Our momentum-space formulation enhances the numerical stability of the HEOM method with respect to its real-space formulation. We then compute the spectral function and various thermodynamic quantities of the one-dimensional Holstein model containing up to ten sites and find quite good agreement with literature results in a wide range of model parameters. To provide an independent check of our HEOM results and to understand the importance of finite-size effects, we also present QMC results for imaginary-time single-particle correlation functions at finite temperature, which are obtained on chains containing up to 20 sites. We conclude that finite-size effects are not pronounced, while the HEOM and QMC results in imaginary time agree exceptionally well.

The paper is organized as follows. Having specified the model (Sec.~\ref{SSec:Holstein_model}) and the correlation functions we evaluate (Sec.~\ref{SSec:corr_functs}), in Sec.~\ref{SSec:HEOM_method} we develop the HEOM approach, while in Sec.~\ref{SSec:QMC_method} we develop the QMC approach. Section~\ref{Sec:Relation_to_existing} discusses our HEOM method in light of the well-known weak-coupling (Sec.~\ref{SSec:Relation_weak}) and strong-coupling (Sec.~\ref{SSec:Relation_strong}) theories, whereas Sec.~\ref{SSec:Relation_broadening} provides an interesting perspective on the artificial broadening of spectral lines, which is widely used in graphic representations of spectral functions. Section~\ref{Sec:Numerical_results} is devoted to a detailed presentation of the numerical results for the one-dimensional Holstein model. We first present the results of the HEOM method in different parameter regimes (Secs.~\ref{SSec:vary-T}--\ref{SSec:vary-omega0}), and then transform them to imaginary time to enable a direct comparison with the QMC data and to discuss finite-size effects (Sec.~\ref{SSec:finite-size-effects}). Section~\ref{Sec:discussion} provides further discussion of the methodologies used in this work, while Sec.~\ref{Sec:conclusion} summarizes our main results.

\section{Model and method}
\subsection{Holstein model}
\label{SSec:Holstein_model}
We study the Holstein Hamiltonian on a one-dimensional lattice containing $N$ sites with periodic boundary conditions (PBC). We work in the limit of low excitation (electron or exciton) density, in which it is enough to consider only the subspaces containing no excitations and a single excitation. The zero-excitation subspace is spanned by the collective state $|\mathrm{vac}\rangle$, in which all units are unexcited. For the time being, we assume that the energy of the collective unexcited state sets the zero of our energy scale. An orthonormal basis in the single-excitation subspace is the so-called local basis $\{|j\rangle|j\}$, whose basis state $|j\rangle$ contains a single excitation localized on site $j$. We develop the theory in the momentum representation, in which the Holstein Hamiltonian reads as
\begin{equation}
 H=H_\mathrm{e}+H_\mathrm{ph}+H_\mathrm{e-ph}.
\end{equation}
$H_\mathrm{e}$ is the tight-binding Hamiltonian of free electronic excitations that can hop between nearest-neighboring sites with hopping amplitude $J$
\begin{equation}
\label{Eq:H-e}
 H_\mathrm{e}=\sum_{k}\varepsilon_{k} |k\rangle\langle k|.
\end{equation}
In Eq.~\eqref{Eq:H-e}, the dimensionless wave number $k$ (expressed in units of inverse lattice spacing) may assume any of the $N$ allowed values in the first Brillouin zone (IBZ) $-\pi<k\leq\pi$, while the free-electron dispersion is $\varepsilon_k=\varepsilon_e-2J\cos(k)$, where $\varepsilon_e$ is the on-site vertical excitation energy. The free-phonon Hamiltonian is
\begin{equation}
 H_\mathrm{ph}=\sum_q\hbar\omega_q b_q^\dagger b_q,
\end{equation}
where $\omega_q$ denotes the dispersion relation of optical phonons, for which $\omega_0\neq 0$ (in the center of the Brillouin zone), while $b_q$ and $b_q^\dagger$ are the phonon annihilation and creation operators. The electron--phonon interaction is conveniently written as
\begin{equation}
 H_\mathrm{e-ph}=\sum_q V_qB_q.
\end{equation}
The purely electronic operator $V_q$ increases the momentum of the electronic subsystem by $q$,
\begin{equation}
\label{Eq:def_V_q}
 V_q=\sum_k |k+q\rangle\langle k|,
\end{equation}
while the purely phononic operator $B_q$ decreases the momentum of phonons by $q$,
\begin{equation}
 B_q=\frac{\gamma}{\sqrt{N}}\left(b_q+b_{-q}^\dagger\right),
\end{equation}
where $\gamma$ is the constant of the local electron--phonon coupling.
The operator $V_q$ satisfies $V_{-q}=V_q^\dagger$, and the same holds for $B_q$.

\subsection{Single-particle correlation functions}
\label{SSec:corr_functs}
The single-particle spectral function, which contains complete information about the single-particle spectrum,
is proportional to the imaginary part of the retarded (causal) Green's function in the real-frequency domain. The corresponding expression in the real-time domain reads as~\cite{Mahanbook}
\begin{equation}
\label{Eq:def_G_R}
 G^R(k,t)=-i\theta(t)\left\langle\left\{c_k(t),c_k^\dagger\right\}\right\rangle.
\end{equation}
Here, $c_k$ and $c_k^\dagger$ are electron annihilation and creation  operators, which in the low-density limit should be replaced by
\begin{equation}
\label{Eq:low-density-substitutions}
 c_k\to|\mathrm{vac}\rangle\langle k|,\quad
 c_k^\dagger\to|k\rangle\langle\mathrm{vac}|,
\end{equation}
while $\theta(t)$ is the step function.
Temporal evolution of the annihilation operator $c_k(t)=e^{iHt/\hbar}c_ke^{-iHt/\hbar}$ entering Eq.~\eqref{Eq:def_G_R} is governed by the total Hamiltonian $H$. The averaging is performed in the thermal equilibrium of the coupled electron--phonon system at temperature $T=(k_B\beta)^{-1}$
\begin{equation}
 \langle\cdot\rangle=\mathrm{Tr}\left(\cdot\:\frac{e^{-\beta H}}{Z}\right),\quad Z=\mathrm{Tr}\:e^{-\beta H}.
\end{equation}

In HEOM computations, we will separately compute the time-dependent greater
\begin{equation}
\label{Eq:G-greater-general}
 G^>(k,t)=-i\left\langle c_k(t) c_k^\dagger\right\rangle
\end{equation}
and lesser
\begin{equation}
\label{Eq:G-lesser-general}
 G^<(k,t)=i\left\langle c_k^\dagger c_k(t)\right\rangle
\end{equation}
Green's functions for $t>0$. In QMC computations, we compute the imaginary-time counterpart of the greater Green's function
\begin{equation}
\label{Eq:C_k_tau_QMC}
    C(k,\tau)=\left\langle c_k(\tau)c_k^\dagger\right\rangle{_{\mathrm{p}}},
\end{equation}
where $0\leq\tau\leq\beta\hbar$ is the imaginary time, while $c_k(\tau)=e^{H\tau/\hbar}c_ke^{-H\tau/\hbar}$ is the imaginary time-dependent operator.
{The subscript $\mathrm{p}$ denotes that the average is taken over the space of purely phononic states $\expval{O}_{\mathrm{p}}=\frac{\mathrm{Tr}_{\mathrm{p}}\qty(Oe^{-\beta H})}{\mathrm{Tr}_{\mathrm{p}}\qty(e^{-\beta H})}$.}

Knowing $G^>$, the single-particle spectral function $A(k,\omega)$ is obtained using the fluctuation--dissipation theorem.~\cite{Mahanbook} In the following, we shift the zero of the energy scale to the on-site vertical excitation energy $\varepsilon_e$, as is commonly done in the literature. It is physically plausible to assume that $\beta\varepsilon_e\gg 1$, i.e., that thermal fluctuations alone cannot bridge the gap between the unexcited and singly-excited states. The fluctuation--dissipation theorem then reduces to
\begin{equation}
\label{Eq:A-from-G-greater}
 A(k,\omega)=-\frac{1}{2\pi}\mathrm{Im}\:G^>(k,\omega).
\end{equation}
A more detailed discussion is provided in Sec.~I of Ref.~\onlinecite{comment071021}.
While $A(k,\omega)$ describes the addition of an electron to the system with initially no electrons, the following quantity
\begin{equation}
\label{Eq:A-plus-k-omega}
    A^+(k,\omega)=\frac{1}{2\pi}\mathrm{Im}\:G^<(k,\omega)
\end{equation}
describes the removal of an electron from the system. The electron-addition and electron-removal spectral functions are not mutually independent, which is discussed in Sec.~I of Ref.~\onlinecite{comment071021} and Sec.~\ref{SSec:A-electron-removal}.

The comparison between HEOM and QMC results is most conveniently done by transforming HEOM results to the imaginary-time domain
\begin{equation}
\label{Eq:C_k_tau_HEOM}
    C_\mathrm{HEOM}(k,\tau)=\int_{-\infty}^{+\infty}d\omega\:e^{-\omega\tau}A(k,\omega)
\end{equation}
and comparing $C_\mathrm{HEOM}(k,\tau)$ with $C(k,\tau)$.

\subsection{HEOM method for single-particle correlation functions}
\label{SSec:HEOM_method}
\subsubsection{Preliminaries}
Employing the low-density replacements [Eq.~\eqref{Eq:low-density-substitutions}] in Eqs.~\eqref{Eq:G-greater-general} and~\eqref{Eq:G-lesser-general}, we obtain
\begin{eqnarray}
 G^>(k,t)=-i\left\langle k\left|\rho_k^>(t)\right|\mathrm{vac}\right\rangle\label{Eq:G-greater-1}\\
 \rho_k^{(I,>)}(t)=\mathrm{Tr}_\mathrm{ph}\left(U_\mathrm{e-ph}^{(I)}(t)|k\rangle\langle\mathrm{vac}|\frac{e^{-\beta H_\mathrm{ph}}}{Z_\mathrm{ph}}\right)\label{Eq:G-greater-2}
\end{eqnarray}

\begin{eqnarray}
 G^<(k,t)=i\left\langle k\left|\rho_k^<(t)\right|\mathrm{vac}\right\rangle\label{Eq:G-lesser-1}\\
 \rho_k^{(<,I)}(t)=\mathrm{Tr}_\mathrm{ph}\left(U_\mathrm{e-ph}^{(I)}(t)\frac{e^{-\beta H}}{Z}|k\rangle\langle\mathrm{vac}|\right)\label{Eq:G-lesser-2}
\end{eqnarray}

While Eqs.~\eqref{Eq:G-greater-1} and~\eqref{Eq:G-lesser-1} are formulated in the Schr\"{o}dinger picture, Eqs.~\eqref{Eq:G-greater-2} and~\eqref{Eq:G-lesser-2} are written in the interaction picture with respect to the noninteracting part $H_\mathrm{e}+H_\mathrm{ph}$ of the total Hamiltonian and the evolution operator $U_\mathrm{e-ph}^{(I)}(t)$ reads as ($T$ is the chronological time-ordering sign)
\begin{equation}
\label{Eq:U_e-ph_int_picture}
 U_\mathrm{e-ph}^{(I)}(t)=T\exp\left[-\frac{i}{\hbar}\int_0^t ds\:H_\mathrm{e-ph}^{(I)}(s)\right].
\end{equation}
In Eq.~\eqref{Eq:G-greater-2}, $Z_\mathrm{ph}=\mathrm{Tr}_\mathrm{ph}e^{-\beta H_\mathrm{ph}}$, while $\mathrm{Tr}_\mathrm{ph}$ denotes the partial trace over phonons. The computation of $G^>$ proceeds in the zero-excitation subspace, as signalized by the free-phonon thermal distribution that is characteristic for the unexcited system. On the other hand, the computation of $G^<$ requires a precomputation of the thermal-equilibrium state of the coupled electron--phonon system, and it thus proceeds in the single-excitation subspace. While the computation of $G^>$ is analogous to the computation of the absorption spectrum of a molecular aggregate, the computation of $G^<$ is analogous to the computation of the emission spectrum. In both cases, the HEOM computations are less demanding than they usually are because the objects propagated in time are not two-sided (density matrix-like), but one-sided (wave function-like). This is reflected in Eqs.~\eqref{Eq:G-greater-2} and~\eqref{Eq:G-lesser-2} by the absence of the backward evolution operator $U_\mathrm{e-ph}^{(I)\dagger}(t)$. Equation~\eqref{Eq:G-greater-2} represents temporal evolution starting from the initial state that is factorized, and the manipulations that are necessary to rewrite it as HEOM should be similar to the ones employed in the standard HEOM derivation.~\cite{AnnPhys.24.118,JChemPhys.130.234111} Recasting Eq.~\eqref{Eq:G-lesser-2} in form of HEOM is more complicated because the initial condition for time evolution cannot be factorized.

\subsubsection{HEOM method to compute $G^>(k,t)$}
According to the Feynman--Vernon influence functional theory,~\cite{AnnPhys.24.118} the only phonon quantity that enters the reduced electronic description is the thermal free-phonon correlation function
\begin{equation}
\label{Eq:thermal-free-phonon-cf}
 \mathcal{C}_{q_2q_1}(t)=\mathrm{Tr}_\mathrm{ph}\left\{B_{q_2}^{(I)}(t)B_{q_1}^{(I)}(0)\frac{e^{-\beta H_\mathrm{ph}}}{Z_\mathrm{ph}}\right\}.
\end{equation}
The time dependence of phonon operators is with respect to the free-phonon Hamiltonian $H_\mathrm{ph}$, which is emphasized by the superscript $(I)$.
The correlation function $\mathcal{C}_{q_2q_1}(t)$ is a sum of two complex exponential (oscillating) terms ($t>0$)
\begin{equation}
\label{Eq:C_q2_q1_t}
 \mathcal{C}_{q_2q_1}(t)=\delta_{q_2,-q_1}(\hbar\omega_0)^2\sum_{m=0}^1 c_{q_2m}e^{-\mu_{q_2m}t},
\end{equation}
where
\begin{equation}
 c_{q0}=\left(\frac{\gamma}{\hbar\omega_0\sqrt{N}}\right)^2\left[1+n_\mathrm{BE}(\omega_q,T)\right],\:\mu_{q0}=+i\omega_q,
\end{equation}
\begin{equation}
 c_{q1}=\left(\frac{\gamma}{\hbar\omega_0\sqrt{N}}\right)^2n_\mathrm{BE}(\omega_q,T),\:\mu_{q1}=-i\omega_q,
\end{equation}
while $n_\mathrm{BE}(\omega_q,T)=\left(e^{\beta\hbar\omega_q}-1\right)^{-1}$ is the number of phonons excited in mode $q$ at temperature $T$.

Integrating out phonons on the right-hand side of Eq.~\eqref{Eq:G-greater-2}, we obtain
\begin{equation}
\label{Eq:exact-sol-lesser}
 \rho_k^{(I,>)}(t)=\mathcal{U}^{(I,>)}(t)|k\rangle\langle\mathrm{vac}|,
\end{equation}
where the reduced evolution superoperator $\mathcal{U}^{(I,>)}(t)$ is given by
\begin{widetext}
 \begin{equation}
 \label{Eq:U_I_gtr_t}
    \mathcal{U}^{(I,>)}(t)=T\mathrm{exp}\left[-\omega_0^2\sum_{qm}\int_0^t ds_2\int_0^{s_2}ds_1\:V_q^{(I)}(s_2)^C\:c_{qm}e^{-\mu_{qm}(s_2-s_1)}\:V_{-q}^{(I)}(s_1)^C\right].
 \end{equation} 
\end{widetext}
The action of the hyperoperator $V^C$ on an arbitrary operator $O$ is defined as $V^CO=VO$. The hyperoperator notation is similar to that used in Ref.~\onlinecite{JChemPhys.153.244122}. The time-ordering sign $T$ enforces the chronological order of the arguments of hyperoperators $V^C$ (later times are moved to the left).
The structure of $\mathcal{U}^{(I,>)}(t)$ suggests that this superoperator conserves the momentum of the electronic subsystem. In other words, the only nontrivial matrix element of $\rho_k^>(t)$ is $\left\langle k\left|\rho_k^>(t)\right|\mathrm{vac}\right\rangle$, which is in agreement with Eq.~\eqref{Eq:G-greater-1}.

Starting from Eq.~\eqref{Eq:exact-sol-lesser}, the HEOM is derived in the standard manner.~\cite{JChemPhys.130.234111} The ADMs are characterized by the vector $\mathbf{n}=\{n_{qm}|q\in\mathrm{IBZ},m=0,1\}$ of nonnegative integers $n_{qm}\geq 0$ that count the order of phonon assistance. The ADM $\sigma_{k,\mathbf{n}}^{(I,>,n)}(t)$, which appears on the depth $n=\sum_{qm}n_{qm}$ of the hierarchy, is defined as
\begin{widetext}
 \begin{equation}
 \label{Eq:def-adm-greater}
  \sigma_{k,\mathbf{n}}^{(I,>,n)}(t)=T\left\{\prod_{qm}\left[i\omega_0^2\:c_{qm}\int_0^t ds\:e^{-\mu_{qm}(t-s)}\:V_{-q}^{(I)}(s)\right]^{n_{qm}}\mathcal{U}^{(I,>)}(t)\right\}|k\rangle\langle\mathrm{vac}|.
 \end{equation}
\end{widetext}
By virtue of the conservation of the electronic momentum, the only nontrivial matrix element of $\sigma_{k,\mathbf{n}}^{(>,n)}(t)$ will be $\left\langle k-k_\mathbf{n}\left|\sigma_{k,\mathbf{n}}^{(>,n)}(t)\right|\mathrm{vac}\right\rangle$, where
\begin{equation}
 k_\mathbf{n}=\sum_{qm}qn_{qm}
\end{equation}
is the electronic momentum carried by $\sigma_{k,\mathbf{n}}^{(>,n)}$. Defining the auxiliary Green's function (AGF) at depth $n$ characterized by vector $\mathbf{n}$
\begin{equation}
\label{Eq:def-aux-G-greater}
 G^{(>,n)}_\mathbf{n}(k-k_\mathbf{n},t)=-i\left\langle k-k_\mathbf{n}\left|\sigma_{k,\mathbf{n}}^{(>,n)}(t)\right|\mathrm{vac}\right\rangle,
\end{equation}
we obtain the following HEOM for Green's functions
\begin{equation}
\label{Eq:HEOM-greater-unscaled}
\begin{split}
 &\partial_t G^{(>,n)}_\mathbf{n}(k-k_\mathbf{n},t)=\\&-i\left(\Omega_{k-k_\mathbf{n}}+\mu_\mathbf{n}\right)G^{(>,n)}_\mathbf{n}(k-k_\mathbf{n},t)\\
 &+i\sum_{qm}G^{(>,n+1)}_{\mathbf{n}_{qm}^+}(k-k_\mathbf{n}-q,t)\\
 &+i\omega_0^2\sum_{qm}n_{qm}c_{qm}G^{(>,n-1)}_{\mathbf{n}_{qm}^-}(k-k_\mathbf{n}+q,t).
 \end{split}
\end{equation}
In the last equation, we introduce
\begin{equation}
\label{Eq:def-mu_n}
 \mu_\mathbf{n}=-i\sum_{qm}\mu_{qm}n_{qm}=\sum_{q}\omega_q\left(n_{q0}-n_{q1}\right),
\end{equation}
while $\Omega_{k}=\varepsilon_k/\hbar$ is the angular frequency of the free-electron state $|k\rangle$. The greater Green's function is then obtained as the root of the hierarchy, $G^{>}(k,t)=G_\mathbf{0}^{(>,0)}(k,t)$. The hierarchy presented in Eq.~\eqref{Eq:HEOM-greater-unscaled} is propagated separately for each allowed value of $k$ (there are $N$ mutually independent hierarchies) with the initial condition
\begin{equation}
\label{Eq:init_cond_greater}
 G^{(>,n)}_\mathbf{n}(k,t=0)=-i\delta_{n,0},
\end{equation}
which follows from Eqs.~\eqref{Eq:def-adm-greater} and~\eqref{Eq:def-aux-G-greater}.

In the zero-temperature limit, the hierarchy in Eq.~\eqref{Eq:HEOM-greater-unscaled} reduces to the hierarchy presented in Refs.~\onlinecite{PhysRevLett.97.036402,PhysRevB.74.245104}. Our HEOM may thus be regarded as an extension of the hierarchy in Refs.~\onlinecite{PhysRevLett.97.036402,PhysRevB.74.245104} to nonzero temperatures. Due to the zero-temperature assumption, the hierarchy in Refs.~\onlinecite{PhysRevLett.97.036402,PhysRevB.74.245104} has only a single branch (our $m=0$ branch). Since the thermal (or incoherent) phonon occupations are identically equal to zero, phonon-assisted processes proceed via virtual (or coherent) phonons, that are first created from the phonon vacuum and subsequently destroyed. On the other hand, our hierarchy has two branches describing phonon-assisted processes that proceed via thermal phonons and in which the order of phonon absorption and emission events is immaterial. One may attempt to solve Eq.~\eqref{Eq:HEOM-greater-unscaled} by Fourier or Laplace transform, as was done in Refs.~\onlinecite{PhysRevLett.97.036402,PhysRevB.74.245104}. This would transform the hierarchy of first-order linear differential equations into a (sparse) system of linear algebraic equations. While such a route is certainly possible, here, we opt for solving Eq.~\eqref{Eq:HEOM-greater-unscaled} directly in the time domain, after which the spectral properties are computed using the discrete Fourier transform.

\subsubsection{HEOM method to compute $G^<(k,t)$}
The main obstacle in the derivation of HEOM for the lesser Green's function is the fact that the density matrix $\rho_k^<(0)$ at the initial instant is not factorizable, see Eq.~\eqref{Eq:G-lesser-2}. Nevertheless, one may always write
\begin{equation}
\label{Eq:im-time-factorization}
\begin{split}
 &\frac{e^{-\beta H}}{Z}=\frac{e^{-\beta H_\mathrm{e}}}{Z_\mathrm{e}}\frac{e^{-\beta H_\mathrm{ph}}}{Z_\mathrm{ph}}\times\\&T\exp\left[-\frac{1}{\hbar}\int_0^{\beta\hbar}d\tau\overline{H}_\mathrm{e-ph}(\tau)\right],
 \end{split}
\end{equation}
where $Z_\mathrm{ph}$ is the free-phonon partition sum, $Z_\mathrm{e}=Z/Z_\mathrm{ph}$ is still unknown electronic partition sum (it contains all the effects due to the electron--phonon interaction), while $\overline{H}_\mathrm{e-ph}(\tau)$ denotes the electron--phonon coupling in the imaginary-time interaction picture
\begin{equation}
 \overline{H}_\mathrm{e-ph}(\tau)=e^{(H_e+H_\mathrm{ph})\tau/\hbar}H_\mathrm{e-ph}e^{-(H_e+H_\mathrm{ph})\tau/\hbar}.
\end{equation}
Inserting Eq.~\eqref{Eq:im-time-factorization} in Eq.~\eqref{Eq:G-lesser-2}, the partial trace over phonons can be straightforwardly computed to obtain
\begin{equation}
\label{Eq:rho-k-lesser}
 \rho_k^{(I,<)}(t)=\left(\mathcal{U}^{(I,<)}(t,\beta\hbar)\frac{e^{-\beta H_\mathrm{e}}}{Z_\mathrm{e}}\right)|k\rangle\langle\mathrm{vac}|.
\end{equation}
The exact evolution superoperator $\mathcal{U}^{(I,<)}(t,\beta\hbar)$ is the time-ordered exponential of the sum of three influence phases
\begin{widetext}
 \begin{eqnarray}
  \mathcal{U}^{(I,<)}(t,\beta\hbar)=\mathcal{T}\exp\left\{-\left[\Phi_1(t)+\Phi_2(\beta\hbar)+\Phi_3(t,\beta\hbar)\right]\right\}\label{Eq:G-greater-partial-trace-3-phis}\\
  \Phi_1(t)=\omega_0^2\sum_{qm}\int_0^t ds_2\int_0^{s_2}ds_1\:V_q^{(I)}(s_2)^C\:
  c_{qm}e^{-\mu_{qm}(s_2-s_1)}\:V_{-q}^{(I)}(s_1)^C\\
  \Phi_2(\beta\hbar)=-\omega_0^2\sum_{qm}\int_0^{\beta\hbar}d\tau_2\int_{0}^{\tau_2}d\tau_1\:^C\overline{V}_{-q}(\tau_1)\:
  c_{qm}e^{i\mu_{qm}(\tau_2-\tau_1)}
  \:^C\overline{V}_{q}(\tau_2)\\
  \Phi_3(t,\beta\hbar)=-i\omega_0^2\sum_{qm}\int_0^t ds\int_0^{\beta\hbar}d\tau\:V_q^{(I)}(s)^C\:
  c_{qm}e^{-\mu_{qm}s}e^{i\mu_{qm}(\beta\hbar-\tau)}
  \:^C\overline{V}_{-q}(\tau)
 \end{eqnarray}
\end{widetext}
The influence phase $\Phi_1(t)$ has already appeared in the derivation of HEOM for $G^>$ [see Eq.~\eqref{Eq:U_I_gtr_t}] and it represents the pure real-time evolution. The influence phase $\Phi_2(\beta\hbar)$ is representative of the pure imaginary-time evolution, while $\Phi_3(t,\beta\hbar)$ contains cross contributions among real- and imaginary-time evolutions. We note that similar expressions have been obtained in the course of the derivation of the HEOM for the initial states that cannot be factorized into a purely electronic and purely phononic part.~\cite{JChemPhys.141.044114,JChemPhys.143.194106} In contrast to Refs.~\onlinecite{JChemPhys.141.044114,JChemPhys.143.194106}, here, the backward evolution operator is missing from Eqs.~\eqref{Eq:G-greater-2} and~\eqref{Eq:G-lesser-2}, so that the hyperoperators $V^C$ and $^CV$ act only from one side.
The action of the hyperoperator $^CV$ on an arbitrary operator $O$ is defined as $^CVO=OV$. The parentheses on the right-hand side of Eq.~\eqref{Eq:rho-k-lesser} stress that all the hyperoperators $^C\overline{V}_{-q}(\tau)$ that act from the right should be applied before operator $|k\rangle\langle\mathrm{vac}|$. The time-ordering sign $\mathcal{T}$ in Eq.~\eqref{Eq:G-greater-partial-trace-3-phis} enforces the chronological order of the real-time arguments $s$ of hyperoperators $V^C$ (later times are moved to the left) and the anti-chronological order of the imaginary-time arguments $\tau$ of hyperoperators $^C V$ (later times are moved to the right).
There is no specific ordering of the real-time instants with respect to the imaginary-time instants because the hyperoperators $V^C$ and $^CV$ act on operator $e^{-\beta H_\mathrm{e}}/Z_\mathrm{e}$ from opposite sides. The anti-chronological ordering of the imaginary-time arguments of operators $^CV$ is necessary in order to maintain the correct chronological order in the general term of the expansion of Eq.~\eqref{Eq:rho-k-lesser} [see also Eqs.~\eqref{Eq:G-lesser-2},~\eqref{Eq:U_e-ph_int_picture}, and~\eqref{Eq:im-time-factorization}], whose operators are ordered as follows: $V_{q_n}^{(I)}(s_n)\dots V_{q_1}^{(I)}(s_1)\:[e^{-\beta H_\mathrm{e}}/Z_\mathrm{e}]\:\overline{V}_{p_m}(\tau_m)\dots \overline{V}_{p_1}(\tau_1)$, where $t \geq s_n \geq \dots \geq s_1 \geq 0$ and $\beta\hbar \geq \tau_m \geq \dots \geq \tau_1 \geq 0$, $n+m$ is even, and $q_n+\dots q_1+p_m+\dots p_1=0$. The ADM $\sigma_{k,\mathbf{n}}^{(I,<,n)}(t)$ at depth $n$ is defined as
\begin{widetext}
 \begin{equation}
 \label{Eq:adm-lesser}
 \begin{split}
  \sigma_{k,\mathbf{n}}^{(I,<,n)}(t,\beta\hbar)=\left\{\mathcal{T}\prod_{qm}\left[i\omega_0^2\int_0^t ds\:c_{qm}e^{-\mu_{qm}(t-s)}V_{-q}^{(I)}(s)^C+\right.\right. \\ \left.\left. \omega_0^2\:e^{-\mu_{qm}t}\int_0^{\beta\hbar}d\tau\:c_{qm}e^{i\mu_{qm}(\beta\hbar-\tau)}\:^C\overline{V}_{-q}(\tau)\right]^{n_{qm}} \mathcal{U}^{(I,<)}(t,\beta\hbar)\frac{e^{-\beta H_e}}{Z_e}\right\}|k\rangle\langle\mathrm{vac}|.
  \end{split}
 \end{equation}
\end{widetext}
The AGF on level $n$ that is characterized by vector $\mathbf{n}$ is then defined analogously to Eq.~\eqref{Eq:def-aux-G-greater}
\begin{equation}
 G^{(<,n)}_{\mathbf{n}}(k-k_\mathbf{n},t)=i\left\langle k-k_\mathbf{n}\left|\sigma_{k,\mathbf{n}}^{(<,n)}(t,\beta\hbar)\right|\mathrm{vac}\right\rangle
\end{equation}
and the HEOM for lesser Green's functions assume the form of Eq.~\eqref{Eq:HEOM-greater-unscaled}. The initial condition under which the hierarchy for $G^<$ is solved is obtained by setting $t=0$ in Eq.~\eqref{Eq:adm-lesser}, which results in
\begin{widetext}
\begin{equation}
\label{Eq:init-cond-lesser}
 \sigma_{k,\mathbf{n}}^{(n,<)}(0,\beta\hbar)=\left\{\mathcal{T}\prod_{qm}\left[\omega_0^2\int_0^{\beta\hbar}d\tau\:c_{qm}e^{i\mu_{qm}(\beta\hbar-\tau)}\:^C\overline{V}_{-q}(\tau)\right]^{n_{qm}}e^{-\Phi_2(\beta\hbar)}\:\frac{e^{-\beta H_e}}{Z_e}\right\}|k\rangle\langle\mathrm{vac}|.
\end{equation}
\end{widetext}
Setting $n=0$ and $\mathbf{n}=\mathbf{0}$ in Eq.~\eqref{Eq:init-cond-lesser}, we recognize that $\sigma_{k,\mathbf{0}}^{(0,<)}(0,\beta\hbar)$ is, up to the factor $|k\rangle\langle\mathrm{vac}|$, identical to the expression for the electronic RDM, see Eqs.~\eqref{Eq:rho-k-lesser} and~\eqref{Eq:adm-lesser}. The equilibrium state of the coupled electron--phonon system has been computed in different ways. The imaginary-time HEOM was formulated only for a specific spectral density of the electron--phonon coupling and its structure was not compatible with the structure of the real-time HEOM.~\cite{JChemPhys.141.044114} The correlated equilibrium state was also computed by the technique of stochastic unraveling~\cite{JChemPhys.143.194106,PhysRevB.85.115412} or by finding the steady state of the real-time HEOM.~\cite{JChemPhys.147.044105} In the following section, we derive the imaginary-time HEOM for the correlated electron--phonon equilibrium that is specifically suited for the single-mode Holstein model and whose form is fully compatible with the real-time HEOM derived in this section.

\subsubsection{HEOM method to compute the thermal equilibrium of coupled electrons and phonons}
\label{SSSec:thermal_eq_dm}
Taking the partial trace of Eq.~\eqref{Eq:im-time-factorization} over phonons we obtain the following expression for the (normalized) thermal-equilibrium RDM for an electron that is coupled to phonons
\begin{equation}
\label{Eq:normalized-th-eq}
 \rho^\mathrm{eq}(\beta\hbar)=\mathcal{T}\:e^{-\Phi_2(\beta\hbar)}\frac{e^{-\beta H_\mathrm{e}}}{Z_\mathrm{e}}.
\end{equation}
However, the normalization constant $Z_\mathrm{e}$ is not known. To compute $Z_\mathrm{e}$, let us consider the unnormalized and imaginary time-dependent analogue of Eq.~\eqref{Eq:normalized-th-eq}
\begin{equation}
 \rho^{\mathrm{un},\mathrm{eq}}(\tau)=\mathcal{T}\:e^{-\Phi_2(\tau)}\:e^{-H_\mathrm{e}\tau/\hbar}
\end{equation}
where $0\leq\tau\leq\beta\hbar$. The imaginary-time HEOM is then obtained as usually, while the definition of the unnormalized ADM at level $n$ reads as
\begin{widetext}
 \begin{equation}
 \label{Eq:imag-time-adms-un}
  \sigma_{\mathbf{n}}^{(n,\mathrm{un},\mathrm{eq})}(\tau)=\mathcal{T}\prod_{qm}\left[\omega_0^2\int_0^{\tau}d\tau'\:c_{qm}e^{i\mu_{qm}(\tau-\tau')}\:^C\overline{V}_{-q}(\tau')\right]^{n_{qm}}e^{-\Phi_2(\tau)}\:e^{-H_\mathrm{e}\tau/\hbar}.
 \end{equation}
\end{widetext}
Because of the momentum conservation, the ADM $\sigma_{\mathbf{n}}^{(n,\mathrm{un},\mathrm{eq})}(\tau)$ has only $N$ nonzero matrix elements, $\langle k-k_\mathbf{n}|\sigma_{\mathbf{n}}^{(n,\mathrm{un},\mathrm{eq})}(\tau)|k\rangle$, and it turns out that the imaginary-time HEOM can be solved independently for each of $N$ allowed values of $k$. The imaginary-time HEOM
\begin{equation}
\label{Eq:im-time-HEOM}
 \begin{split}
  &\partial_{\tau}\left\langle k-k_\mathbf{n}\left|\sigma_{\mathbf{n}}^{(n,\mathrm{un},\mathrm{eq})}(\tau)\right|k\right\rangle=\\
  &-\left(\Omega_{k-k_\mathbf{n}}+\mu_\mathbf{n}\right)\left\langle k-k_\mathbf{n}\left|\sigma_{\mathbf{n}}^{(n,\mathrm{un},\mathrm{eq})}(\tau)\right|k\right\rangle\\
  &+\sum_{qm}\left\langle k-k_\mathbf{n}-q\left|\sigma_{\mathbf{n}_{qm}^+}^{(n+1,\mathrm{un},\mathrm{eq})}(\tau)\right|k\right\rangle\\
  &+\omega_0^2\sum_{qm}n_{qm}c_{qm}\left\langle k-k_\mathbf{n}+q\left|\sigma_{\mathbf{n}_{qm}^-}^{(n-1,\mathrm{un},\mathrm{eq})}(\tau)\right|k\right\rangle
 \end{split}
\end{equation}
is then solved with the initial condition that can be inferred from Eq.~\eqref{Eq:imag-time-adms-un} and that reads as~\cite{JChemPhys.141.044114}
\begin{equation}
 \left\langle k-k_\mathbf{n}\left|\sigma_{\mathbf{n}}^{(n,\mathrm{un},\mathrm{eq})}(0)\right|k\right\rangle=\delta_{n,0}.
\end{equation}
This initial condition assumes that all free-electron levels are equally populated, which is in agreement with the fact that point $\tau=0$ corresponds to the infinite-temperature limit. Furthermore, thermal fluctuations are so strong that they completely suppress the effects due to the electron--phonon interaction, so that all ADMs are initially equal to zero. Once the imaginary-time propagation of Eq.~\eqref{Eq:imag-time-adms-un} is done, the normalization constant $Z_\mathrm{e}$ is obtained as
\begin{equation}
 Z_\mathrm{e}=\sum_k\left\langle k\left|\sigma_{\mathbf{0}}^{(0,\mathrm{un},\mathrm{eq})}(\beta\hbar)\right|k\right\rangle,
\end{equation}
so that the normalized ADMs in thermal equilibrium read as
\begin{equation}
 \sigma_{\mathbf{n}}^{(n,\mathrm{eq})}=Z_\mathrm{e}^{-1}\sigma_{\mathbf{n}}^{(n,\mathrm{un},\mathrm{eq})}(\beta\hbar).
\end{equation}
Finally, the initial condition for the real-time propagation of HEOM for $G^<$ reads as
\begin{equation}
 G^{(<,n)}_{\mathbf{n}}(k-k_\mathbf{n},0)=i\left\langle k-k_\mathbf{n}\left|\sigma_\mathbf{n}^{(n,\mathrm{eq})}\right|k\right\rangle.
\end{equation}

\subsubsection{Thermodynamic properties}
Thermodynamic properties of the Holstein model have been extensively investigated using the QMC method in combination with the Feynman's path-integral theory~\cite{PhysRevLett.49.1522,PhysRevB.27.6097,PhysRevB.30.1671} or the canonical Lang--Firsov transformation of the electron--phonon Hamiltonian~\cite{PhysRevB.69.024301} or the density-matrix renormalization group.~\cite{PhysRevB.102.165155} The algorithm presented in Refs.~\onlinecite{PhysRevLett.49.1522,PhysRevB.27.6097,PhysRevB.30.1671}, in which the purely electronic model resulting from the analytical elimination of the nuclear degrees of freedom is simulated using the QMC, is very close in spirit to the method developed here. It is mainly for this reason that we here compute the same thermodynamic quantities as in Refs.~\onlinecite{PhysRevLett.49.1522,PhysRevB.27.6097,PhysRevB.30.1671}.

The kinetic energy of the electron is a purely electronic quantity and it thus depends only on the electronic RDM $\sigma_\mathbf{0}^{(0,\mathrm{eq})}$ 
\begin{equation}
\label{Eq:E-kin}
 E_\mathrm{kin}=\sum_k\varepsilon_k f_k=\sum_k\varepsilon_k\left\langle k\left|\sigma_\mathbf{0}^{(0,\mathrm{eq})}\right|k\right\rangle,
\end{equation}
where $f_k=\left\langle k\left|\sigma_\mathbf{0}^{(0,\mathrm{eq})}\right|k\right\rangle\in[0,1]$ is the population of electronic state with momentum $k$. The electron--phonon interaction energy is a mixed quantity that is linear in phonon coordinates. Even though our approach mainly deals with reduced, i.e., purely electronic quantities, the results of Ref.~\onlinecite{JChemPhys.137.194106} suggest that the interaction energy should be extracted from ADMs at depth 1 as follows
\begin{equation}
\label{Eq:E-e-ph}
\begin{split}
 E_\mathrm{e-ph}&=\frac{\gamma}{\sqrt{N}}\sum_{kq}\left\langle |k+q\rangle\langle k|\left(b_q+b_{-q}^\dagger\right)\right\rangle\\
 &=\hbar\omega_0\sum_q\sum_{m=0}^1\sqrt{c_{qm}}\sum_k\left\langle k-q\left|\sigma_{\mathbf{0}_{qm}^+}^{(1,\mathrm{eq})}\right|k\right\rangle.
 \end{split}
\end{equation}
The fermion--boson correlation function contains information on how the electron distorts its surroundings by following how the presence of an electron on a site affects the nuclear motion on sites that are $r$ lattice constants away 
\begin{equation}
\label{Eq:normalized-fbcf}
\begin{split}
 \varphi_r&=\frac{\gamma}{E_\mathrm{e-ph}}\sum_j\left\langle|j\rangle\langle j|\left(b_{j+r}+b_{j+r}^\dagger\right)\right\rangle\\
 &=\frac{\hbar\omega_0}{E_\mathrm{e-ph}}\sum_{q}e^{iqr}\sum_{m=0}^1\sqrt{c_{qm}}\sum_k\left\langle k-q\left|\sigma_{\mathbf{0}_{qm}^+}^{(1,\mathrm{eq})}\right|k\right\rangle.
 \end{split}
\end{equation}
This quantity has been used to gain insight into the spatial extent of the polaron.~\cite{PhysRevB.27.6097,PhysRevB.30.1671,JChemPhys.142.174103}

\subsection{QMC method for single-particle correlation function}
\label{SSec:QMC_method}

We used a QMC method to calculate the single-particle correlation function in imaginary time. The method is based on the path-integral representation of the correlation function. In this representation, it is possible to integrate out the phononic degrees of freedom. For this reason, one ends up with the sum over electronic coordinates that is performed using the Monte Carlo method. Our approach essentially follows the path of early works in Refs.~\onlinecite{PhysRevLett.49.1522,PhysRevB.27.6097,PhysRevB.30.1671} with a difference that we extend these works (where only thermodynamic quantities were calculated) to calculate the imaginary-time correlation function. In addition, we formulate the method in such a way that phase space of electronic coordinates can be directly sampled. This is significantly more convenient than the sampling based on the Metropolis algorithm which needs initial equilibration of the system and which suffers from the possibility of trapping the system in local minima. The QMC method that we used is also similar to the stochastic path integral method developed in the chemical physics community.\cite{PhysRevB.85.115412} Within that method, phonon degrees of freedom are integrated out and the problem is eventually reduced to the problem of dynamics of the electronic subsystem in the presence of noise.

In more detail, to evaluate the imaginary time correlation function
\begin{equation}
 C\qty(
 k,\tau)= \frac{\mathrm{Tr}_{\mathrm{p}}
 \qty(e^{-\qty(\beta-\frac{\tau}{\hbar})H_{\mathrm{ph}}} 
 c_{k}
 e^{-\frac{\tau}{\hbar}H}
 c_{k}^\dagger )
 }
 {\mathrm{Tr}_{\mathrm{p}} e^{-\beta H_{\mathrm{ph}}}}
\end{equation}
we calculate it in the site basis
\begin{equation}
 C_{ab}\qty(\tau)= \frac{\mathrm{Tr}_{\mathrm{p}}\qty(e^{-\qty(\beta-\frac{\tau}{\hbar})H_{\mathrm{ph}}} 
 c_{a} 
 e^{-\frac{\tau}{\hbar}H}
 c_{b}^\dagger )
 }
 {\mathrm{Tr}_{\mathrm{p}} e^{-\beta H_{\mathrm{ph}}}} 
\end{equation}
and then we perform the transformation
\begin{equation}
 C\qty(
 k,\tau)=\frac{1}{N}\sum_{a=0}^{N-1}\sum_{b=0}^{N-1}
 e^{\mathrm{i}k\qty(b-a)} C_{ab}\qty(\tau).
\end{equation}
We calculate $C_{ab}\qty(\tau)$ starting from
\begin{equation}\label{eq:cab528}
\begin{split}
 & C_{ab}\qty(\tau)=\\ 
 & \frac
 {\int \dd{\qty{x}} \dd{\qty{y}} 
 \mel{\qty{x}}{e^{-\qty(\beta-\frac{\tau}{\hbar})H_{\mathrm{ph}}}}{\qty{y}}  
 \mel{a\qty{y}}{e^{-\frac{\tau}{\hbar}H}}{b\qty{x}} 
 }
 {\int \dd{\qty{x}} \dd{\qty{y}}
 \mel{\qty{x}}{e^{-\qty(\beta-\frac{\tau}{\hbar})H_{\mathrm{ph}}}}{\qty{y}}  
 \mel{\qty{y}}{e^{-\frac{\tau}{\hbar}H_{\mathrm{ph}}}}{\qty{x}} 
 }
 \end{split}
\end{equation}
where $\qty{x}$ and $\qty{y}$ denote the vectors containing the coordinates of all phonon modes.
The term $\mel{\qty{x}}{e^{-\qty(\beta-\frac{\tau}{\hbar})H_{\mathrm{ph}}}}{\qty{y}}$ is expressed in an analytical form, since the following identity holds for a single phonon mode
\begin{equation}\label{eq:lho491}
\begin{split}
 \mel{x}{e^{-\alpha H_{\mathrm{ph}}}}{y}=&
 \sqrt{\frac{m\omega_0}{2\pi\hbar\sinh\qty(\alpha\hbar\omega_0)}} \cdot \\
& e^{-\frac{m\omega_0}{2\hbar\sinh\qty(\alpha\hbar\omega_0)}
 \qty[\qty(x^2+y^2)\cosh\qty(\alpha\hbar\omega_0)-2xy]},
 \end{split}
\end{equation}
while the analogous term for multiple phonon modes is a simple product of the terms for each mode ($m$ denotes the mass of the oscillator that can be set to 1 without the loss of generality). 
To evaluate the matrix elements $\mel{a\qty{y}}{e^{-\frac{\tau}{\hbar}H}}{b\qty{x}} $ and $\mel{\qty{y}}{e^{-\frac{\tau}{\hbar}H_{\mathrm{ph}}}}{\qty{x}}$ we discretize the imaginary time $\tau$ by dividing it into $K$ steps of duration $\Delta \tau=\frac{\tau}{K}$. We make use of the Suzuki-Trotter expansion of first order which has a small controllable error of order $\qty(\Delta \tau)^2$
\begin{equation}
 e^{- \frac{\Delta \tau}{\hbar} H} \approx 
 e^{- \frac{\Delta \tau}{\hbar} H_{\mathrm{ph}}}
 e^{- \frac{\Delta \tau}{\hbar} H_{\mathrm{e-ph}}}
 e^{- \frac{\Delta \tau}{\hbar} H_{\mathrm{e}}}.
\end{equation}
We then have
\begin{equation}
\begin{split}
 \mel{a\qty{y}}{e^{-\frac{\Delta\tau}{\hbar}H}}{b\qty{x}} &=
 \int \dd{\qty{z}} \sum_c 
 \mel{\qty{y}}{e^{-\frac{\Delta\tau}{\hbar}H_{\mathrm{ph}}}}{\qty{z}} \times \\
 &\mel{a\qty{z}}{e^{-\frac{\Delta\tau}{\hbar}H_{\mathrm{e-ph}}}}{c\qty{x}}
 \mel{c}{e^{-\frac{\Delta\tau}{\hbar}H_{\mathrm{e}}}}{b}.
 \end{split}
\end{equation}
The term involving the electron-phonon interaction can be evaluated as
\begin{equation}
 \mel{a\qty{z}}{e^{-\frac{\Delta\tau}{\hbar}H_{\mathrm{e-ph}}}}{c\qty{x}}
 =\delta_{ac}
 e^{-\frac{\Delta\tau}{\hbar}Cx_a}
 \delta\qty(\qty{z}-\qty{x}), 
\end{equation}
where $C=\frac{\gamma}{\hbar\omega_0}\sqrt{2\hbar m\omega_0^3}$ and $x_a$ denotes the coordinate of the phonon at site $a$. The term containing $H_{\mathrm{e}}$ reads
\begin{equation}
 \mel{c}{e^{-\frac{\Delta\tau}{\hbar}H_e}}{b}=
 \frac{1}{N}\sum_k e^{-\Delta\tau \varepsilon_k}
 e^{\mathrm{i}k\qty(c-b)}\equiv f\qty(c-b)
\end{equation}
and we denote it as $f\qty(c-b)$ in what follows.
We thus obtain
\begin{equation}\label{eq:matel1-295}
\begin{split}
&\mel{a\qty{y}}{e^{-\frac{\tau}{\hbar}H}}{b\qty{x}} =
\sum_{i_1\ldots i_{K-1}} \\
& f\qty(a-i_1)
f\qty(i_1-i_2)
\ldots
f\qty(i_{K-1}-b)  \\
& \int \dd{\qty{x}^{(1)}}
\ldots
\dd{\qty{x}^{(K-1)}} 
\mel{\qty{y}}{e^{-\frac{\Delta\tau}{\hbar}H_{\mathrm{ph}}}}{\qty{x}^{(1)}}\\
&\mel{\qty{x}^{(1)}}{e^{-\frac{\Delta\tau}{\hbar}H_{\mathrm{ph}}}}{\qty{x}^{(2)}}
\dots
\mel{\qty{x}^{(K-1)}}{e^{-\frac{\Delta\tau}{\hbar}H_{\mathrm{ph}}}}{\qty{x}}
\\
& e^{-\frac{\Delta \tau}{\hbar} C \qty(x_a^{(1)}+x_{i_1}^{(2)}+\ldots+x_{i_{K-2}}^{(K-1)}+x_{i_{K-1}})}
\end{split}
\end{equation}
and
\begin{equation}\label{eq:matel1-296}
\begin{split}
&\mel{\qty{y}}{e^{-\frac{\tau}{\hbar}H_{\mathrm{ph}}}}{\qty{x}}=\\
& \int \dd{\qty{x}^{(1)}}
\ldots
\dd{\qty{x}^{(K-1)}} 
\mel{\qty{y}}{e^{-\frac{\Delta\tau}{\hbar}H_{\mathrm{ph}}}}{\qty{x}^{(1)}}\\
&\mel{\qty{x}^{(1)}}{e^{-\frac{\Delta\tau}{\hbar}H_{\mathrm{ph}}}}{\qty{x}^{(2)}}
\dots
\mel{\qty{x}^{(K-1)}}{e^{-\frac{\Delta\tau}{\hbar}H_{\mathrm{ph}}}}{\qty{x}}.
\end{split}
\end{equation}
After substitution of expressions \eqref{eq:lho491}, \eqref{eq:matel1-295} and \eqref{eq:matel1-296} to Eq.~\eqref{eq:cab528} we obtain
\begin{equation}
C_{ab}\qty(\tau)=\frac{\mathcal{N}}{\mathcal{D}}, 
\end{equation}
with 
\begin{equation}\label{eq:D904}
\begin{split}
 \mathcal{N}=&\sum_{i_1\ldots i_{K-1}}
f\qty(a-i_1)
f\qty(i_1-i_2)
\ldots
f\qty(i_{K-1}-b)  \\
\int & 
\dd{\qty{x}}
\dd{\qty{y}}
\dd{\qty{x}^{(1)}}
\ldots
\dd{\qty{x}^{(K-1)}}\\
& e^{-Q\qty(\qty{x},\qty{y},\qty{x}^{(1)},\ldots,\qty{x}^{(K-1)})}\\
& e^{-L\qty(\qty{x},\qty{y},\qty{x}^{(1)},\ldots,\qty{x}^{(K-1)})}
\end{split}
\end{equation}
and
\begin{equation}\label{eq:D905}
\begin{split}
 \mathcal{D}=\int 
&\dd{\qty{x}}
\dd{\qty{y}}
\dd{\qty{x}^{(1)}}
\ldots
\dd{\qty{x}^{(K-1)}}\\
& e^{-Q\qty(\qty{x},\qty{y},\qty{x}^{(1)},\ldots,\qty{x}^{(K-1)})}.
\end{split}
\end{equation}
The term
$Q\qty(\qty{x},\qty{y},\qty{x}^{(1)},\ldots,\qty{x}^{(K-1)})$
in Eqs. \eqref{eq:D904} and \eqref{eq:D905} 
denotes the quadratic form of phonon coordinates, that contains the terms of second order only (i.e. the square of a phonon coordinate or the product of two phonon coordinates), while the term
$L\qty(\qty{x},\qty{y},\qty{x}^{(1)},\ldots,\qty{x}^{(K-1)})$
denotes the linear term that contains a linear combination of phonon coordinates. The integrals in Eqs. \eqref{eq:D904} and \eqref{eq:D905} are the Gaussian integrals that can be evaluated analytically using the formula for the Gaussian integral
\begin{equation}\label{eq:gaus3690}
 \int \dd[n]{\vb{z}} e^{-\frac{1}{2}\vb{z}\cdot\vb{A}\cdot\vb{z}}
 e^{\vb{b}\cdot \vb{z}}=\qty(2\pi)^{n/2} \qty(\det\vb{A})^{-1/2}
 e^{\frac{1}{2}\vb{b}\cdot \vb{A}^{-1} \cdot \vb{b}},
 \end{equation}
 where $\vb{z}$ is a real vector of dimension $n$, $\vb{A}$ is a real symmetric positive definite matrix of dimension $n\times n$, $\vb{b}$ is a vector of dimension $n$, $\det\vb{A}$ is the determinant of matrix $\vb{A}$ and $\vb{A}^{-1}$ is its inverse. We then obtain
 \begin{equation}
 \begin{split}
C_{ab}\qty(\tau)=  
\sum_{i_1\ldots i_{K-1}} &
f\qty(a-i_1)
f\qty(i_1-i_2)
\ldots
f\qty(i_{K-1}-b)\times  \\
& e^{\frac{1}{2}\vb{b}\cdot \vb{A}^{-1} \cdot \vb{b}},
\end{split}
 \end{equation}
 where $\vb{A}$ is the matrix representing the quadratic form $Q$ and $\vb{b}$ is the vector representing the linear form $L$.
To evaluate the last sum, we introduce the replacement of variables
\begin{equation}\label{eq:rvar391}
a-i_1=j_1, i_1-i_2=j_2, \ldots, i_{K-2}-i_{K-1}=j_{K-1}.
\end{equation} 
The sum then takes the form
\begin{equation}
\begin{split}
 C_{ab}\qty(\tau)=  
 \sum_{j_1\ldots j_{K-1}}&
f\qty(j_1)
f\qty(j_2)
\ldots
f\qty(j_{K-1})\\
& f\qty(a-b-\sum_{i=1}^{K-1}j_i)
e^{\frac{1}{2}\vb{b}\cdot \vb{A}^{-1} \cdot \vb{b}}.
\end{split}
\end{equation}
To evaluate the last sum, we treat the non-negative function 
\begin{equation}
 g\qty(j_1,\ldots,j_{K-1})=
f\qty(j_1)
f\qty(j_2)
\ldots
f\qty(j_{K-1})
\end{equation}
as the probability density.  
We perform Monte Carlo summation by choosing random integers $j_1$, $j_2$, $\dots$, $j_{K-1}$ that follow the distribution $h(j)=\frac{f\qty(j)}{\sum_j f\qty(j)}$ and perform the summation of the term $f\qty(a-b-\sum_{i=1}^{K-1}j_i) e^{\frac{1}{2}\vb{b}\cdot \vb{A}^{-1} \cdot \vb{b}}$. The replacement of variables introduced in Eq.~\eqref{eq:rvar391} enabled us to directly sample the phase space of electronic coordinates rather than to perform sampling based on the Metropolis algorithm. The results presented in this work were obtained from $10^5$ samples of electronic coordinates, unless otherwise stated. The integer $K$ that defines the discretization of imaginary time was chosen so that $\Delta\tau\approx 0.05\frac{\hbar}{J}$.

We used the same ideas to calculate the thermodynamic expectation values using the QMC method. The details of the method for these cases are given in Sec.~II of Ref.~\onlinecite{comment071021}.

\normalcolor

\section{Relation of the HEOM method to existing results}
\label{Sec:Relation_to_existing}
\subsection{Weak-coupling limit}
\label{SSec:Relation_weak}
In the weak-coupling limit, the electron--phonon interaction constant is the smallest energy scale in the problem. It is then reasonable to treat the interaction within the second-order perturbation theory. In the HEOM approach, the second-order approximation is performed by truncating the hierarchy at depth 1 and solving the resulting equations for ADMs at depth 1 in the Markov and adiabatic approximations.~\cite{JChemPhys.122.041103,JChemPhys.130.234111} One eventually obtains the Rayleigh--Schr\"{o}dinger perturbation theory expression for the self energy.~\cite{Mahanbook} More details are provided in Sec.~III of Ref.~\onlinecite{comment071021}.

\subsection{Strong-coupling limit}
\label{SSec:Relation_strong}
In the opposite, strong-coupling limit, the electronic coupling $J$ is the smallest energy in the problem. The Holstein Hamiltonian then reduces to $N$ independent single-site problems (the so-called independent-boson models), whose analytic solution is known.~\cite{Mahanbook} In the following discussion, we assume for simplicity that phonons are dispersionless, i.e., $\omega_q=\omega_0$ for all $q$. In Sec.~IV of Ref.~\onlinecite{comment071021}, we demonstrate how the exact solution presented here is recast as the Lang--Firsov formula~\cite{SovPhysJETP.16.1301} for the single-site Green's function. The crux of the derivation is to note that, in the single-site limit, the hyperoperators appearing in the evolution superoperator become time independent, which renders the time-ordering sign ineffective.~\cite{JChemPhys.153.244122} Here, we repeat the zero-temperature result
\begin{equation}
\label{Eq:Lang-Firsov-T-0}
\begin{split}
 G^>(\omega)=&-2\pi i\: e^{-\gamma^2/(\hbar\omega_0)^2}\times\\&\sum_{n=0}^{+\infty}\frac{1}{n!}\left(\frac{\gamma}{\hbar\omega_0}\right)^{2n}\delta\left(\omega+\frac{\gamma^2}{\hbar^2\omega_0}-n\omega_0\right).
\end{split}
\end{equation}

\subsection{Artificial broadening of spectral lines}
\label{SSec:Relation_broadening}
For graphic representations of the spectral function [Eq.~\eqref{Eq:A-from-G-greater}], the $\delta$ functions entering Eq.~\eqref{Eq:Lang-Firsov-T-0} are commonly replaced by their Lorentzian representation, $\delta(x)\to\eta/[\pi(x^2+\eta^2)]$, where the small parameter $\eta$ is the artificial broadening of the peaks. The discussion in Sec.~IV of Ref.~\onlinecite{comment071021} suggests that the artificial broadening can be introduced in the exact result embodied in Eq.~\eqref{Eq:U_I_gtr_t} by simply replacing $\mu_{qm}\to\mu_{qm}+\eta$. On the level of the HEOM, this replacement is reflected as an artificial damping at rate $n\eta$ of the AGFs at depth $n$, see Eqs.~\eqref{Eq:HEOM-greater-unscaled} and~\eqref{Eq:def-mu_n}. The excellent agreement between the analytical and numerical results in the single-site limit with different levels of artificial broadening is demonstrated in Figs.~1 and~2 in Sec.~IV Ref.~\onlinecite{comment071021}. The replacement $\mu_{qm}\to\mu_{qm}+\eta$, however, changes the physical situation we are dealing with by effectively replacing the discrete phonon spectrum by a continuous one.

To elaborate on the last point, we recall that the thermal free-phonon correlation function $\mathcal{C}_{q_2q_1}(t)$ [Eq.~\eqref{Eq:thermal-free-phonon-cf}] can be written as follows~\cite{May-Kuhn-book}
\begin{equation}
\label{Eq:fluct-diss-C-phon}
\mathcal{C}_{q_2q_1}(t)=\delta_{q_2,-q_1}\frac{\hbar}{\pi}\int_{-\infty}^{+\infty}d\omega\:\mathcal{J}(\omega)\:\frac{e^{i\omega t}}{e^{\beta\hbar\omega}-1},
\end{equation}
where the spectral density of the electron--phonon interaction for the (single-mode) Holstein model
\begin{equation}
\label{Eq:SD_single_mode}
\mathcal{J}(\omega)=\frac{\pi}{\hbar}\left(\frac{\gamma}{\sqrt{N}}\right)^2\left[\delta(\omega-\omega_0)-\delta(\omega+\omega_0\right)]
\end{equation}
has two discrete peaks at $\pm\omega_0$. In Sec.~V of Ref.~\onlinecite{comment071021}, we demonstrate that the replacement $\mu_{qm}\to\mu_{qm}+\eta$ means that, on the level of the spectral density, delta peaks should be replaced by Lorentizans characterized by the full width at half maximum $\eta$. Equation~\eqref{Eq:SD_single_mode} is then replaced by the so-called underdamped Brownian oscillator spectral density~\cite{Mukamel-book}
\begin{equation}
\label{Eq:sd-underdamped-bo}
 \mathcal{J}(\omega)=\Lambda\left[\frac{\omega\eta}{(\omega-\omega_0)^2+\eta^2}+\frac{\omega\eta}{(\omega+\omega_0)^2+\eta^2}\right],
\end{equation}
where $\Lambda=\gamma^2/(N\hbar\omega_0)$ is the appropriate polaron binding energy. The demonstration in Sec.~V of Ref.~\onlinecite{comment071021} leans on the fact that the artificial broadening is the smallest energy scale in the problem, i.e., $\eta\ll\omega_0$ and $\hbar\eta\ll k_BT$. These assumptions are commonly satisfied whenever spectral lines are artificially broadened.~\cite{PhysRevB.74.245104,PhysRevB.100.094307}

\begin{figure}
    \centering
    \includegraphics[scale=1.05]{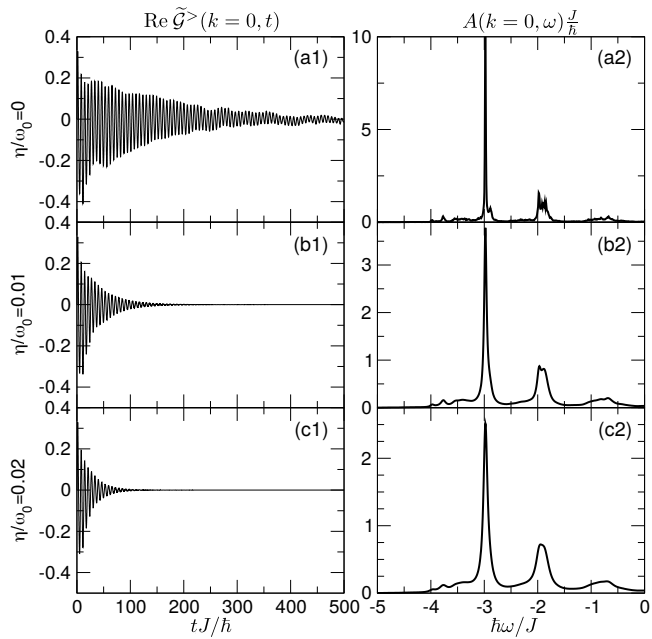}
    \caption{(a1), (b1), and (c1) Time dependence of the real part of the envelope $\widetilde{\mathcal{G}}^>(k=0,t)$ of the greater Green's function in the zone center for three different values of the broadening parameter $\eta$. For a precise definition of $\widetilde{\mathcal{G}}^>(k,t)$, see Eq.~\eqref{Eq:def-envelope}. (a2), (b2), and (c2) Spectral function $A(k=0,\omega)$ in the zone center for three different values of the broadening parameter $\eta$. The computation is performed on a $N=8$-site chain with $k_BT/J=0.4$, $\hbar\omega_0/J=1$, and $\gamma/J=\sqrt{2}$, while the maximum hierarchy depth is set to $D=7$.}
    \label{Fig:fig_finite_eta}
\end{figure}

As a numerical example beyond the single-site limit, in Fig.~\ref{Fig:fig_finite_eta} we discuss how different levels of artificial broadening $\eta$ affect the Green's function in real time [panels (a1)--(c1)] and the spectral function [panels (a2)--(c2)] in the intermediate-coupling regime and at relatively low temperature.
The broadening renders our HEOM method more numerically stable by ensuring that all dynamical quantities decay to zero on a time scale $\sim\eta^{-1}$, compare Figs.~\ref{Fig:fig_finite_eta}(b1) and~\ref{Fig:fig_finite_eta}(c1) to Fig.~\ref{Fig:fig_finite_eta}(a1), in which the Green's function at long times oscillates around zero. However, we see that already $\eta/\omega_0=0.01$ greatly underestimates the decay time of the Green's function, which means that the width of the quasiparticle peak and its first phonon replica is greatly overestimated, compare Figs.~\ref{Fig:fig_finite_eta}(b2) and~\ref{Fig:fig_finite_eta}(c2) to Fig.~\ref{Fig:fig_finite_eta}(a2). In the following, we will not introduce any artificial broadening to our HEOM.

\section{Numerical results}
\label{Sec:Numerical_results}
We assume that the phonons are dispersionless, $\omega_q=\omega_0$, and focus on the so-called extreme quantum regime, $\hbar\omega_0/J=1$, in which the approximate treatments specifically developed for the adiabatic and antiadiabatic regimes are not appropriate. Furthermore, we limit our investigations to the most challenging crossover regime $\gamma/J=\sqrt{2}$, in which the dimensionless electron--phonon coupling constant is $\lambda=\gamma^2/(2J\hbar\omega_0)=1$. As the default value of the temperature, we choose $k_BT/J=1$. All of these dimensionless ratios ($\hbar\omega_0/J$, $\gamma/J$, and $k_BT/J$) will be changed, one at a time, to examine the applicability of the HEOM method in various parameter regimes (low-/high-temperature, weak-/strong-coupling, adiabatic/antiadiabatic regime).
\subsection{Hierarchy rescaling, rotating-wave system, and twisted boundary conditions}
The HEOM for $G^>$ and $G^<$ are both of the same form, see Eq.~\eqref{Eq:HEOM-greater-unscaled}. It is known that propagating such a form of HEOM produces numerical instabilities, especially for very strong electron--phonon couplings.~\cite{JChemPhys.130.084105} Equation~\eqref{Eq:def-adm-greater} suggests that the ADM at depth $n$ scales as the $(2n)$-th power of the interaction strength, meaning that, for sufficiently strong interaction, the magnitude of ADMs increases with $n$. However, it is desirable that the magnitude of ADMs at deeper hierarchy levels be smaller than at shallower levels. This is accomplished by performing the hierarchy rescaling introduced in Ref.~\onlinecite{JChemPhys.130.084105}. We define the dimensionless time $\widetilde{t}=\omega_0 t$ and rescale the AGFs as follows
\begin{equation}
 \widetilde{G}^{(n)}_\mathbf{n}(k-k_\mathbf{n},\widetilde{t})=\omega_0^{-n}f(\mathbf{n})G^{(n)}_\mathbf{n}\left(k-k_\mathbf{n},\omega_0^{-1}\widetilde{t}\right),
\end{equation}
where the rescaling factor $f(\mathbf{n})$ reads as
\begin{equation}
 f(\mathbf{n})=\prod_{qm}\left[c_{qm}^{n_{qm}}n_{qm}!\right]^{-1/2},
\end{equation}
while the factor $\omega_0^{-n}$ renders $\widetilde{G}^{(n)}_\mathbf{n}$ dimensionless.
The dimensionless and rescaled HEOM then reads as
\begin{equation}
\label{Eq:HEOM-greater-scaled}
 \begin{split}
 &\partial_{\widetilde{t}} \widetilde{G}^{(n)}_\mathbf{n}\left(k-k_\mathbf{n},\widetilde{t}\right)=\\&-i\frac{\Omega_{k-k_\mathbf{n}}+\mu_\mathbf{n}}{\omega_0}\widetilde{G}^{(n)}_\mathbf{n}\left(k-k_\mathbf{n},\widetilde{t}\right)\\
 &+i\sum_{qm}\sqrt{1+n_{qm}}\sqrt{c_{qm}}\:\widetilde{G}^{(n+1)}_{\mathbf{n}_{qm}^+}\left(k-k_\mathbf{n}-q,\widetilde{t}\right)\\
 &+i\sum_{qm}\sqrt{n_{qm}}\sqrt{c_{qm}}\:\widetilde{G}^{(n-1)}_{\mathbf{n}_{qm}^-}\left(k-k_\mathbf{n}+q,\widetilde{t}\right).
 \end{split}
\end{equation}

The numerical stability of the HEOM method is further enhanced by performing the transition to the so-called rotating-wave system by separating out the rapidly oscillating from the slowly changing component (the envelope) of $\widetilde{G}_\mathbf{n}^{(n)}$ as follows
\begin{equation}
\label{Eq:def-envelope}
\begin{split}
 \widetilde{G}^{(n)}_\mathbf{n}(k-k_\mathbf{n},\widetilde{t})=&\exp\left(-i\frac{\Omega_{k-k_\mathbf{n}}+\mu_\mathbf{n}}{\omega_0}\widetilde{t}\right)\times\\&\widetilde{\mathcal{G}}^{(n)}_\mathbf{n}(k-k_\mathbf{n},\widetilde{t}).
\end{split}
\end{equation}
We propagate the HEOM for the envelope $\widetilde{\mathcal{G}}^{(n)}_\mathbf{n}(k-k_\mathbf{n},\widetilde{t})$ using the fourth-order Runge--Kutta method. We use the time step $\Delta\widetilde{t}=\omega_0\Delta t=0.02$ in all the computations to be presented.

Having computed $G^{>/<}(k,\widetilde{t})$ for $0\leq\widetilde{t}\leq\widetilde{t}_\mathrm{max}$, the spectral analysis is performed by first continuing the Green's function symmetrically for $-\widetilde{t}_\mathrm{max}\leq\widetilde{t}\leq 0$ according to $G^{>/<}(k,-\widetilde{t})=-G^{>/<}(k,\widetilde{t})^*$ and then multiplying it with the Hann window function
\begin{equation}
\label{Eq:hann_window}
    w(\widetilde{t})=\cos^2\left(\frac{\pi\widetilde{t}}{2\widetilde{t}_\mathrm{max}}\right),\:-\widetilde{t}_\mathrm{max}\leq\widetilde{t}\leq\widetilde{t}_\mathrm{max}.
\end{equation}
The maximum time $\widetilde{t}_\mathrm{max}$ is always long enough so that the signal windowing accurately captures the spectral content of the short-time dynamics and yet eliminates the long-time oscillations of $G^{>/<}$ around zero, which originate from finite-size effects. In most cases, we take $\widetilde{t}_\mathrm{max}=500$.  

The imaginary-time HEOM [Eq.~\eqref{Eq:im-time-HEOM}] is rescaled in a similar manner. Its propagation in dimensionless imaginary time $\widetilde{\tau}=\omega_0\tau$ is performed from 0 to $\beta\hbar\omega_0$. Typical number of imaginary-time steps that is sufficient to obtain converged results for the electron--phonon equilibrium is of the order of 100.

The number of dynamical variables $G^{(n)}_{\mathbf{n}}(k-k_\mathbf{n},t)$ entering the HEOM is in principle infinite. Nevertheless, converged results are obtained by truncating the hierarchy at certain maximum depth $D$. The number of active variables at depth $n$, $0\leq n\leq D$, is $\binom{n+2N-1}{n}$, while their total number in the hierarchy is
\begin{equation}
\label{Eq:n_active_dms}
    n_\mathrm{active}=\sum_{n=0}^D\binom{n+2N-1}{n}=\binom{2N+D}{D}.
\end{equation}
The dimensionality of the hierarchy grows very fast with increasing the maximum depth $D$ and even faster with increasing the system size $N$. This limits the applicability of the HEOM method to relatively small systems and moderate maximum depths.

To reduce the influence of the finite-size effects on the thermodynamic quantities and to obtain $A(k,\omega)$ with a decent $k$-space resolution, we employ the twisted boundary conditions (TBC),~\cite{PhysRevE.64.016702} which are widely used to minimize one-body finite-size effects both at vanishing~\cite{PhysRevB.100.094307,PhysRevB.103.054304} and finite~\cite{PhysRevX.5.041041,PhysRevB.96.205145} electron densities. In practice, the free-electron dispersion relation $\varepsilon_k$ appearing in Eq.~\eqref{Eq:H-e} is replaced by $\varepsilon_{k\theta}=\varepsilon_e-2J\cos(k+\theta)$, where $\theta$ is the twist angle. Even though the computations are performed on a finite chain of length $N$, the $N$ allowed $k$ points can be continuously connected by varying the twist angle through the range $[0,2\pi/N)$. A typical choice for the values of the twist angle is $\theta_m=2\pi m/(NN_\theta)$, where $m=0,\dots,N_\theta-1$.

\subsection{Estimate of the maximum hierarchy depth}
In computations based on HEOM it is always necessary to study if the depth $D$ is sufficiently large so that the relevant quantities do not significantly change when $D$ is increased further.

\begin{figure}
 \centering
 \includegraphics[scale=1.02]{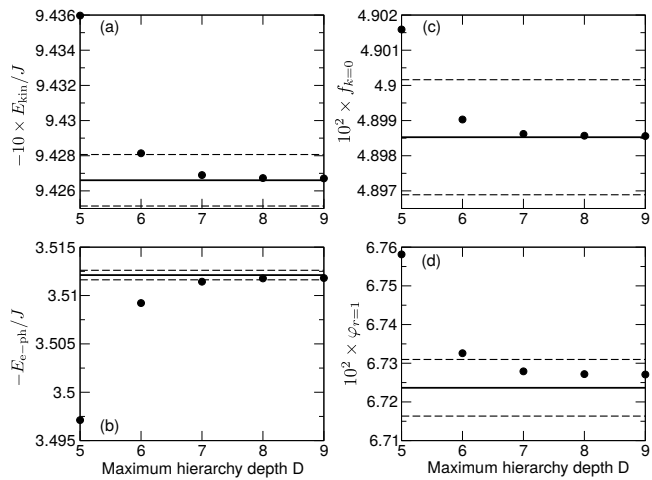}
 \caption{Dependence of various thermodynamic quantities on the maximum hierarchy depth $D$ (full dots) and comparison with QMC results (solid line). Vertical distance between the solid line and dashed lines is equal to the standard deviation of the sample containing 10 different realizations of the QMC algorithm. The model parameters assume the following values: $k_BT/J=1$, $\hbar\omega_0/J=1$, $\gamma/J=\sqrt{2}$, $N=8$, $N_\theta=6$. (a) Negative kinetic energy of the electron [Eq.~\eqref{Eq:E-kin}] in units of $J$ multiplied by a factor of 10 (averaging over TBC is performed). (b) Negative electron--phonon interaction energy [Eq.~\eqref{Eq:E-e-ph}] in units of $J$ (averaging over TBC is performed). (c) Population of the zero-momentum electronic state multiplied by a factor of 100. TBC are used to generate $f_k$ for $NN_\theta=48$ values of $k$ in the IBZ and the overall normalization is such that the sum of these $NN_\theta$ values of $f_k$ is equal to 1. (d) Fermion--boson correlation function [Eq.~\eqref{Eq:normalized-fbcf}] at distance of a single lattice spacing multiplied by a factor of 100 (PBC are used).}
 \label{Fig:Fig1}
\end{figure}

We show an example of such study for different thermodynamic quantities in Figs.~\ref{Fig:Fig1}(a)--\ref{Fig:Fig1}(d). Their dependence on $D$ is shown in full dots in these figures. The computations are performed on a lattice with $N=8$ sites, the twist angle $\theta$ may assume $N_\theta=6$ different values, while $D$ is varied between 5 and 9. The results of the HEOM method are compared with QMC results represented by solid lines in Figs.~\ref{Fig:Fig1}(a)--\ref{Fig:Fig1}(d), while the width of the interval delimited by dashed lines is a measure of the statistical noise in QMC data. It is estimated by performing 10 different realizations of the QMC algorithm and computing the standard deviation of thus generated sample. The agreement between the HEOM and QMC results is remarkable. As a good compromise between the numerical effort and accuracy, we opt in this case for the maximum depth $D=7$, so that the number of active variables is $n_\mathrm{active}=245,157$.

An example of the estimate of $D$ when the results for $G^>(k,t)$ and $A(k,\omega)$ are concerned is presented in Figures~3 and~4 in Sec.~VI of Ref.~\onlinecite{comment071021}. The figures demonstrate that the results in both time and frequency domains do not change significantly when $D$ is increased from 7 to 8. At the same time, the late-time revival of oscillations in $\mathrm{Re}\:\mathcal{G}^>(k=0,t)$ for $D=6$ indicates that $D=6$ may not be sufficient to obtain convergent results. These observations suggest that taking $D=7$ produces meaningful results for both $G^>(k,t)$ and $A(k,\omega)$ in this case.

\subsection{Variations in temperature}
\label{SSec:vary-T}
\begin{figure}
 \centering
 \includegraphics{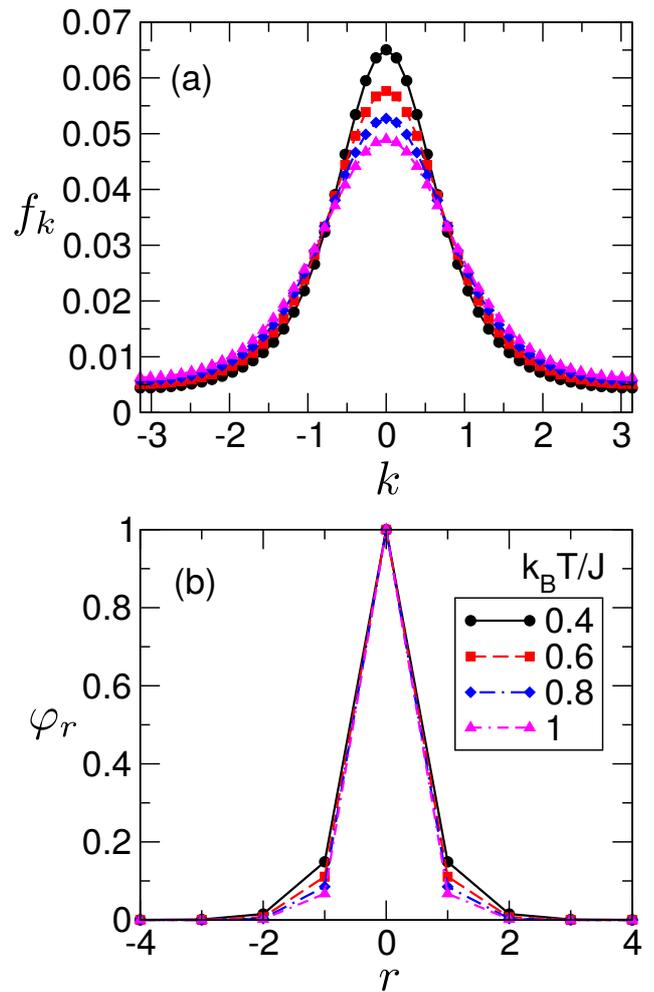}
 \caption{(a) Electronic distribution over momenta $f_k$ for different temperatures. TBC are used to generate $f_k$ in $NN_\theta$ points in the IBZ and the overall normalization is such that the sum of these $NN_\theta$ values of $f_k$ is equal to 1. (b) Fermion--boson correlation function $\varphi_r$ for different temperatures (PBC are observed). Other model parameters assume the following values: $\hbar\omega_0/J=1$, $\gamma/J=\sqrt{2}$, $N=8$, $N_\theta=6$, $D=7$.}
 \label{Fig:Fig2}
\end{figure}

The effects of the temperature variations on the electronic momentum distribution and the fermion--boson correlation function are studied in Figs.~\ref{Fig:Fig2}(a) and~\ref{Fig:Fig2}(b), respectively. The TBC are employed to obtain the electronic momentum distribution on a relatively dense grid in the IBZ, as specified in greater detail in the caption of Fig.~\ref{Fig:Fig2}. As the temperature is increased, the momentum distribution flattens. At the same time, the corresponding real-space distribution $f_r=\frac{1}{N}\sum_k e^{ikr} f_k$ shrinks and develops a more pronounced peak at $r=0$, which suggests that the spatial extent of the polaron is decreased. This is also apparent from Fig.~\ref{Fig:Fig2}(b). These observations are in agreement with earlier results.~\cite{jcp.128.114713} The results presented in Fig.~\ref{Fig:Fig2}(a) are further supported by their remarkable agreement with the corresponding QMC results, see Fig.~5 in Sec.~VII of Ref.~\onlinecite{comment071021}.

\begin{figure}
 \centering
 \includegraphics[scale=1.0]{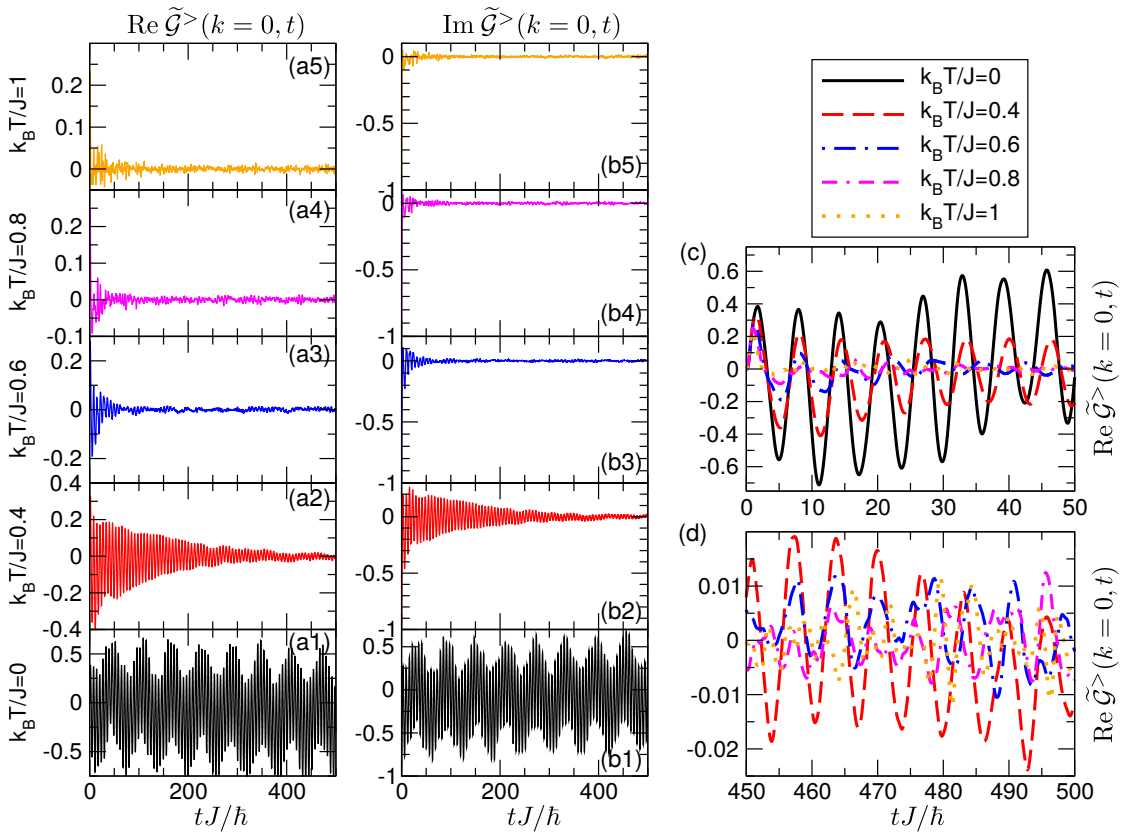}
 \caption{Time dependence of the real [(a1)--(a5)] and imaginary part [(b1)--(b5)] of the envelope $\widetilde{\mathcal{G}}^>(k=0,t)$ of the greater Green's function at $k=0$ for temperatures ranging from $k_BT/J=0$ [in (a1) and (b1)] to $k_BT/J=1$ [in (a5) and (b5)]. (c) and (d) Time dependence of the envelope $\widetilde{\mathcal{G}}^>(k=0,t)$ during the initial [(c)] and later [(d)] stages of temporal evolution for different temperatures. Other model parameters assume the following values: $\hbar\omega_0/J=1$, $\gamma/J=\sqrt{2}$, $N=8$, $N_\theta=6$, $D=7$.}
 \label{Fig:Fig3}
\end{figure}

In Fig.~\ref{Fig:Fig3} we present how temperature variations affect the time dependence of the envelope $\widetilde{\mathcal{G}}^>(k=0,t)$ of the greater Green's function in the zone center. At zero temperature, the phonon-assisted processes proceed via virtual (coherent) phonons. There is no net energy exchange between the electron and the phonon subsystem, which is seen as the persistent oscillations in both the real and imaginary parts of $\widetilde{\mathcal{G}}^>(k=0,t)$, see Figs.~\ref{Fig:Fig3}(a1) and~\ref{Fig:Fig3}(b1). As the temperature is increased, the oscillations in $\widetilde{\mathcal{G}}^>(k=0,t)$ become damped, see Figs.~\ref{Fig:Fig3}(a2)--\ref{Fig:Fig3}(b5), which is a direct consequence of the energy exchange between the electron and thermally excited (incoherent) phonons. The higher the temperature, the more pronounced the damping of $\widetilde{\mathcal{G}}^>(k=0,t)$. According to Fig.~\ref{Fig:Fig3}(c), the damping sets in already at the very initial stages of time evolution. Because of the finite-size effects, the envelopes at higher $T$ do not decay to zero, but rather oscillate around it, see Fig.~\ref{Fig:Fig3}(d). At the latest instants we examine, the oscillation amplitude of $\mathrm{Re}\:\widetilde{\mathcal{G}}(k=0,t)$ drops to less than 10\% of its value around $t=0$, compare the vertical-axis scales of Figs.~\ref{Fig:Fig3}(a2)--\ref{Fig:Fig3}(a5) and Fig.~\ref{Fig:Fig3}(d). At the same time, the amplitude of the imaginary part drops to less than 1\% of its initial value, which is equal to unity, see also Eq.~\eqref{Eq:init_cond_greater}.

\begin{figure}
 \centering
 \includegraphics[scale=0.5]{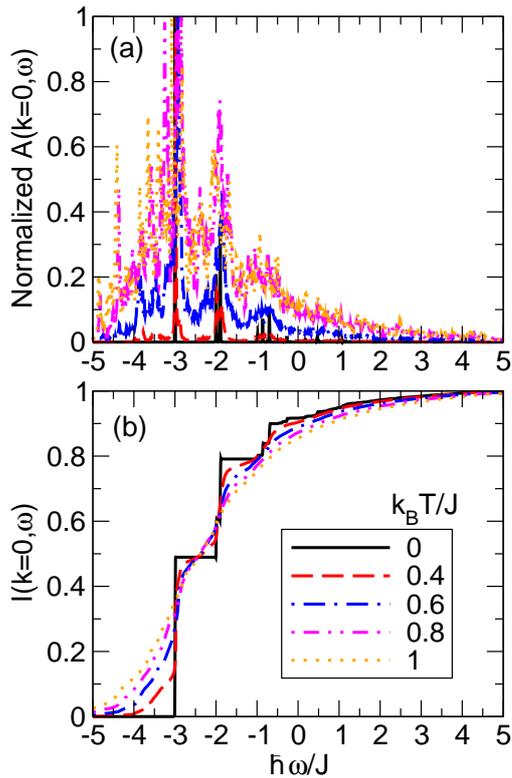}
 \caption{(a) Spectral functions $A(k=0,\omega)$ for different temperatures. For the sake of graphical presentation, the spectral functions are renormalized so that the intensity of the QP peak at $\hbar\omega_\mathrm{QP}/J\approx -3$ is equal to unity for all the temperatures examined. (b) Integrated spectral weight $I(k=0,\omega)$ [Eq.~\eqref{Eq:I_k_omega}] at different temperatures. The sum rule $\displaystyle{\lim_{\omega\to+\infty}I(k=0,\omega)=1}$ is satisfied. Other model parameters assume the following values: $\hbar\omega_0/J=1$, $\gamma/J=\sqrt{2}$, $N=8$, $N_\theta=6$, $D=7$.}
 \label{Fig:Fig4}
\end{figure}

Transforming the data in Fig.~\ref{Fig:Fig3} to the frequency domain, the corresponding spectral functions for $k=0$ at different temperatures are shown in Fig.~\ref{Fig:Fig4}(a). For the sake of graphical presentation, the spectral functions in Fig.~\ref{Fig:Fig4}(a) are normalized so that the quasiparticle (QP) peak, which is located around $\hbar\omega_\mathrm{QP}/J\approx -3$, is of unit intensity. The position of the QP peak agrees very well with the results from the variational ground-state study,~\cite{PhysRevB.60.1633} the cluster perturbation theory,~\cite{PhysRevB.68.184304} and the momentum average approximation.~\cite{PhysRevB.74.245104} Figure~\ref{Fig:Fig4}(b) presents the integrated spectral weight
\begin{equation}
\label{Eq:I_k_omega}
 I(k,\omega)=\int_{-\infty}^\omega d\omega'\:A(k,\omega').
\end{equation}
at $k=0$ for different temperatures. At zero temperature, the peaks of $A(k=0,\omega)$ are quite narrow and their structure closely resembles the single-site vibronic progression, see the narrowest peaks in Fig.~\ref{Fig:Fig4}(a) and the step-like increments in $I(k=0,\omega)$ in Fig.~\ref{Fig:Fig4}(b). The small, but finite, peak width is the consequence of performing time propagation on a finite chain and up to finite maximum time $t_\mathrm{max}$ and employing the Hann windowing procedure [Eq.~\eqref{Eq:hann_window}]. As the temperature is increased, the peaks become broadened, and the intensity of the QP peak is redistributed towards lower energies, see the bottom left parts of Figs.~\ref{Fig:Fig4}(a) and~\ref{Fig:Fig4}(b). The peak broadening is determined by the decay time of the envelope $\widetilde{\mathcal{G}}^>(k=0,t)$, see Fig.~\ref{Fig:Fig3}, and is larger at higher temperature. The spectral-intensity shift originates from the process in which the electron destroys one or more thermally excited phonons. The higher is the temperature, the larger is the number of thermal phonons and the more pronounced is the shift, compare $I(k=0,\omega)$ curves for different temperatures in the region $\omega<\omega_\mathrm{QP}$ in Fig.~\ref{Fig:Fig4}(b). The integrated spectral density satisfies the sum rule for the spectral function $\int d\omega\:A(k,\omega)=1$, see the upper right part of Fig.~\ref{Fig:Fig4}(b).

\onecolumngrid

\twocolumngrid

\begin{figure}
 \centering
 \includegraphics[scale=0.75]{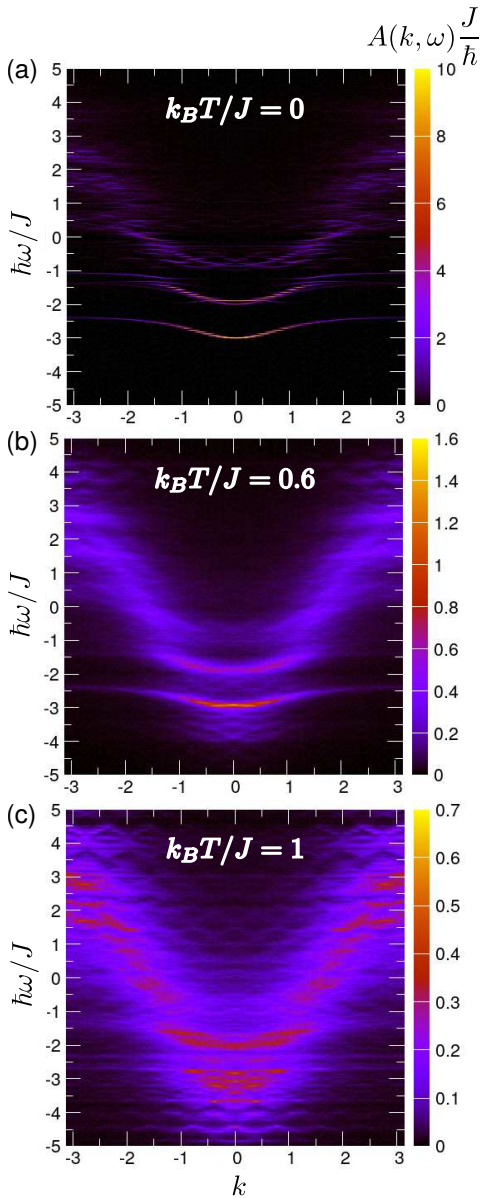}
 \caption{Spectral function $A(k,\omega)$ at temperatures (a) $T=0$, (b) $k_BT/J=0.6$, and (c) $k_BT/J=1$. Other model parameters assume the following values: $\hbar\omega_0/J=1$, $\gamma/J=\sqrt{2}$, $N=8$, $N_\theta=6$, $D=7$.}
 \label{Fig:Fig5}
\end{figure}

The momentum- and energy-resolved spectral functions $A(k,\omega)$ for three different temperatures are compared in Figs.~\ref{Fig:Fig5}(a)--\ref{Fig:Fig5}(c). At zero temperature, our result for $A(k,\omega)$, see Fig.~\ref{Fig:Fig5}(a), agrees quite well with the zero-temperature results of the cluster perturbation theory~\cite{PhysRevB.68.184304} and the momentum average approximation,~\cite{PhysRevB.74.245104} as well as with the low-temperature results of the finite-temperature Lanczos method.~\cite{PhysRevB.100.094307} For a single electron on an infinite chain, the QP peak at $\omega_\mathrm{QP}(k)$ is a delta function, while satellite peaks are of nonzero width. In that case, the lowest-energy polaronic band, which is energetically narrow enough so that it is entirely below the minimum energy for inelastic scattering $\hbar\omega_\mathrm{QP}(k=0)+\hbar\omega_0$, is perfectly coherent, while higher-energy bands formed by satellite peaks are incoherent.~\cite{PhysRevB.56.4494} On finite chains studied here, the width of the QP peak is essentially determined by the frequency resolution $\pi/t_\mathrm{max}$ [we continue $G^{>}(k,t)$ to negative times $-t_\mathrm{max}\leq t\leq 0$], so that the lowest-lying band in Fig.~\ref{Fig:Fig5}(a) may be regarded as perfectly coherent. The satellite peaks at each $k$ display some structure, mainly due to the finite chain length $N$, and are thus somewhat wider, see the band whose minimum lies at $\hbar\omega/J\approx -2$ in Fig.~\ref{Fig:Fig5}(a). This band may be regarded as incoherent. As the temperature is increased, in addition to the broadening of the bands, the spectral intensity redistributes towards the region $\omega<\omega_\mathrm{QP}(k)$. This intensity redistribution is most pronounced in the vicinity of the zone center, where the QP weight is appreciable. At $k_BT/J=0.6$, due to the moderate spectral-intensity redistribution, the (broadened) polaronic band together with the associated one-phonon replica can still be recognized, see Fig.~\ref{Fig:Fig5}(b). On the other hand, at $k_BT/J=1$, the spectral-intensity redistribution and the broadening are so pronounced that all the features of the zero-temperature spectrum merge into a continuum-like structure, see Fig.~\ref{Fig:Fig5}(c).

\subsection{Variations in the electron--phonon coupling constant}
\label{SSec:vary-gamma}
The stronger the electron--phonon coupling, the larger the maximum depth of the hierarchy needed to obtain meaningful results. To keep the numerical effort within reasonable limits, varying the electron--phonon coupling we varied both the maximum depth $D$ and the number of sites $N$. The precise values of the model parameters that are changed together with the electron--phonon coupling are summarized in Table~\ref{Tab:Tab1}.

\begin{table}[]
    \centering
    \begin{tabular}{c|c|c|c|c|c|c}
        regime & $\gamma/J$ & $\lambda$ & $N$ & $N_\theta$ & $D$ & $n_\mathrm{active}$ \\
        \hline
        weak-coupling & 1 & 0.5 & 10 & 5 & 6 & 230,230\\
        intermediate-coupling & $\sqrt{2}$ & 1 & 8 & 6 & 7 & 245,157\\
        strong-coupling & 2 & 2 & 6 & 8 & 9 & 293,930
    \end{tabular}
    \caption{Values of model parameters that are changed together with the electron--phonon coupling constant.}
    \label{Tab:Tab1}
\end{table}

\begin{figure}
 \centering
 \includegraphics[scale=0.9]{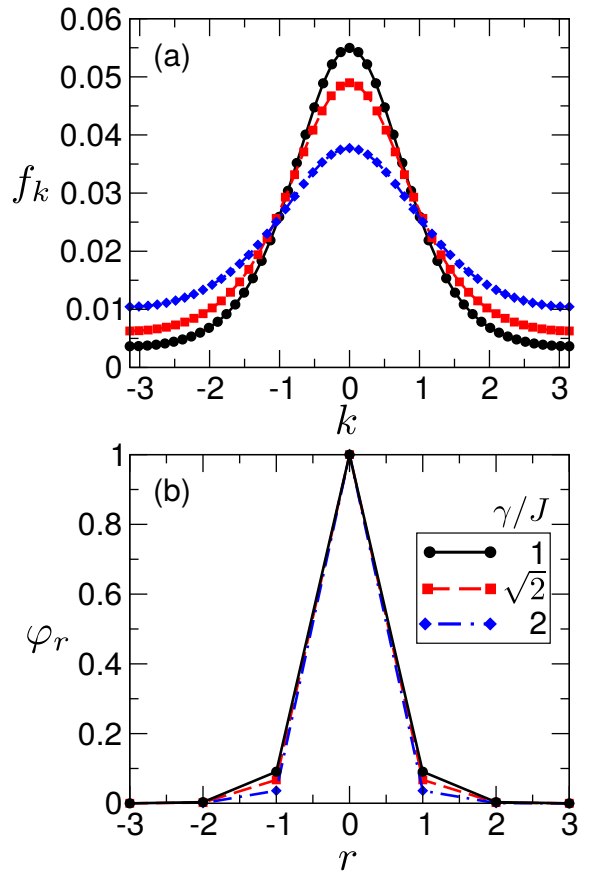}
 \caption{(a) Electronic distribution in the momentum space for different values of the electron--phonon coupling constant $\gamma$ (or $\lambda$). TBC are used to generate $f_k$ in $NN_\theta$ points in the IBZ and the overall normalization is such that the sum of these $NN_\theta$ values of $f_k$ is equal to 1. (b) Fermion--boson correlation function (PBC are observed) for different values of the electron--phonon coupling constant $\gamma$ (or $\lambda$). Other model parameters assume the following values: $k_BT/J=1$, $\hbar\omega_0/J=1$, while the values of $N,N_\theta$, and $D$ are summarized in Table~\ref{Tab:Tab1}.}
 \label{Fig:Fig6}
\end{figure}

As the dimensionless electron--phonon coupling constant $\lambda$ is increased from 0.5 to 1 and 2 ($\gamma/J$ is increased from 1 to $\sqrt{2}$ and 2), the electronic momentum distribution flattens, see Fig.~\ref{Fig:Fig6}(a), which together with the narrowing of the peak of the fermion--boson correlation function displayed in Fig.~\ref{Fig:Fig6}(b) suggests that the polaron becomes more localized. Increasing the electron--phonon coupling strength thus has similar effects on the equilibrium polaron properties as increasing the temperature, compare Figs.~\ref{Fig:Fig2} and~\ref{Fig:Fig6}, which is in line with existing studies.~\cite{jcp.128.114713} The results presented in Fig.~\ref{Fig:Fig6}(a) are further supported by their remarkable agreement with the corresponding QMC results, see Fig.~6 in Sec.~VII of Ref.~\onlinecite{comment071021}.

\begin{figure}
    \centering
    \includegraphics[scale=0.65]{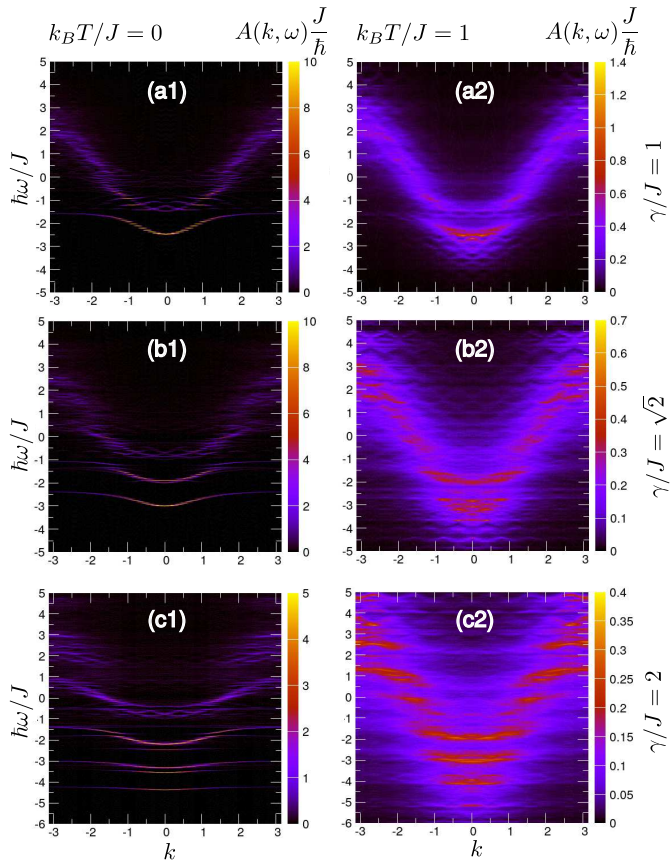}
    \caption{Spectral function $A(k,\omega)$ (measured in units $\hbar/J$) for different electron--phonon interaction strengths [$\gamma/J=1$ in (a1) and (a2), $\gamma/J=\sqrt{2}$ in (b1) and (b2), $\gamma/J=2$ in (c1) and (c2)] and different temperatures [$k_BT/J=0$ in (a1)--(c1) and $k_BT/J=1$ in (a2)--(c2)]. The optical-phonon energy is $\hbar\omega_0/J=1$, while the values of $N,N_\theta$, and $D$ are summarized in Table~\ref{Tab:Tab1}.}
    \label{Fig:Fig11}
\end{figure}

Spectral properties for different interaction strengths and temperatures are summarized in Figs.~\ref{Fig:Fig11}(a1)--\ref{Fig:Fig11}(c2). The zero-temperature results presented in Figs.~\ref{Fig:Fig11}(a1) and~\ref{Fig:Fig11}(c1) compare favorably to available zero-temperature~\cite{PhysRevB.68.184304,PhysRevB.74.245104} and low-temperature~\cite{PhysRevB.100.094307,PhysRevB.102.165155} results for $A(k,\omega)$. Moreover, our result for the polaronic band $\omega_\mathrm{QP}(k)$ (under TABC) in the weak-coupling regime, see Fig.~\ref{Fig:Fig11}(a1), is in good agreement with the ground-state polaronic dispersion relation obtained in Ref.~\onlinecite{PhysRevB.60.1633}. In the strong-coupling regime and at $T=0$, see Fig.~\ref{Fig:Fig11}(c1), a clearly observable peak appears above the QP peak $\hbar\omega_\mathrm{QP}/J\sim -4.4$ and below its one-phonon replica, in our case at around $\hbar\omega_\mathrm{BP}/J\sim -3.55$. This peak corresponds to the so-called bound polaron state, in which a phonon is bound to the polaron.~\cite{PhysRevB.60.1633,PhysRevB.100.094307} At $k_BT/J=1$, our data presented in Fig.~\ref{Fig:Fig11}(b2) overall agree with the results of the finite-temperature Lanczos method, see Fig.~2(d) of Ref.~\onlinecite{PhysRevB.100.094307}. The spectral-intensity shift below the polaronic band becomes both larger and more intensive as the electron--phonon coupling is increased. In the weak- and intermediate-coupling regimes, the smearing of the (shifted) polaronic band makes it barely recognizable at finite temperature, see Figs.~\ref{Fig:Fig11}(a2) and~\ref{Fig:Fig11}(b2). On the other hand, in the strong-coupling regime, see Fig.~\ref{Fig:Fig11}(c2), the peaks at finite $T$ are clearly recognizable, mutually separated by $\hbar\omega_0/J$, and resemble the vibronic progression in the single-site limit. Previous studies~\cite{PhysRevB.68.184304} suggested that the physical effects in the strong-coupling regime are predominantly local, which is also reflected in the very narrow polaronic band observed in Fig.~\ref{Fig:Fig11}(c1).

\subsection{Variations in the optical phonon frequency}
\label{SSec:vary-omega0}
Here, we vary both the phonon energy $\hbar\omega_0/J$ and the coupling constant $\gamma/J$ in such a way that we remain in the strong-coupling regime $\lambda=2$, which limits our investigations to $N=6$-site chains. The temperature is fixed to its default value $k_BT/J=1$, while the precise values of the model parameters varied are summarized in Table~\ref{Tab:Tab2}.
\begin{table}
\begin{tabular}{c|c|c|c|c|c|c}
    regime & $\hbar\omega_0/J$ & $\gamma/J$ & $N$ & $N_\theta$ & $D$ & $n_\mathrm{active}$ \\
    \hline
    adiabatic & 0.2 & $\sqrt{0.8}$ & 6 & 8 & 11 & 1,352,078\\
    extreme quantum & 1 & 2 & 6 & 8 & 9 & 293,930\\
    anti-adiabatic & 3 & $\sqrt{12}$ & 6 & 8 & 9 & 293,930
\end{tabular}
\caption{Values of model parameters that are changed together with the optical phonon frequency.}
\label{Tab:Tab2}
\end{table}

\begin{figure}
    \centering
    \includegraphics[scale=0.63]{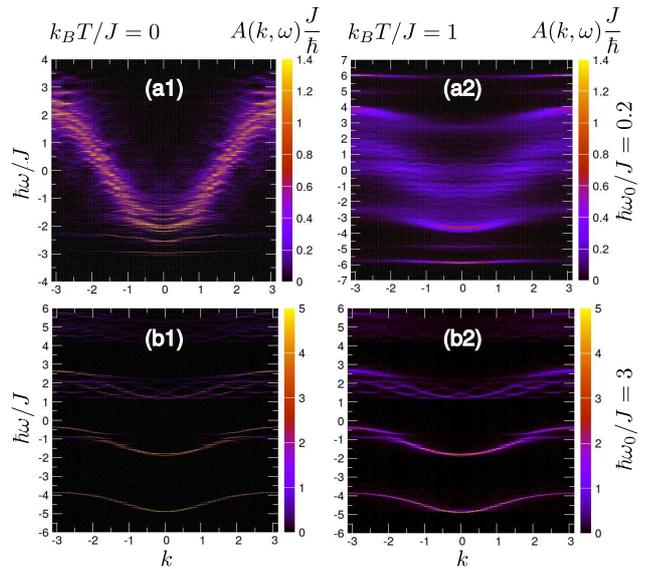}
    \caption{Spectral function $A(k,\omega)$ (measured in units $\hbar/J$) for different values of the adiabaticity ratio [$\hbar\omega_0/J=0.2$ in (a1) and (a2), $\hbar\omega_0/J=3$ in (b1) and (b2)] and temperature [$T=0$ in (a1) and (b1), $k_BT/J=1$ in (a2) and (b2)]. The changes in the electron--phonon interaction strength that are necessary to remain in the strong-coupling regime ($\lambda=2$) are summarized in Table~\ref{Tab:Tab2}. The vertical-axis ranges in (a1) and (a2) are different.}
    \label{Fig:Fig_vary_omega0}
\end{figure}

The adiabatic regime, $\hbar\omega_0/J=0.2$, is numerically most challenging for the HEOM method.
At the temperature we consider, $k_BT/J=1$, the number of thermally excited phonons is large [$k_BT/(\hbar\omega_0)=5$], and the hierarchical couplings are strong because they are determined by the product of the large electron--phonon coupling constant and the large number of thermally excited phonons. Therefore, one should study in more detail how the results for $G^>(k,t)$ and $A(k,\omega)$ change with the maximum hierarchy depth $D$. In Figs.~7 and~8 in Sec.~VIII of Ref.~\onlinecite{comment071021} we compare the real-time and real-frequency data in the adiabatic regime for maximum hierarchy depths $D=9,10,$ and $11$. While the real-time data for $D=10$ and $11$ are to some extent similar, their spectral contents seems to be very different. A comparison between the QMC and HEOM results, which is performed in Fig.~9 in Sec.~VIII of Ref.~\onlinecite{comment071021}, reveals that, as $D$ is increased from 9 to 11, the agreement between the QMC and HEOM results in imaginary time becomes better. This favorable comparison suggests that the HEOM results for $A(k,\omega)$ obtained using $D=11$, which are shown in Figs.~\ref{Fig:Fig_vary_omega0}(a1) (at $T=0$) and~\ref{Fig:Fig_vary_omega0}(a2) (at $k_BT/J=1$), may be representative of the true result.

In the antiadiabatic regime, $\hbar\omega_0/J=3$, the temperature we study is relatively low, $k_BT/(\hbar\omega_0)=1/3$, so that the zero-temperature spectrum in Fig.~\ref{Fig:Fig_vary_omega0}(b1) is quite similar to the finite-temperature spectrum in Fig.~\ref{Fig:Fig_vary_omega0}(b2). The spectrum consists of the polaronic band, the QP peak being located at $\hbar\omega_\mathrm{QP}/J\sim -4.9$, which is accompanied by its single-phonon and two-phonon replicas, whose minima lie at approximately $\hbar(\omega_\mathrm{QP}+\omega_0)/J$ and $\hbar(\omega_\mathrm{QP}+2\omega_0)/J$, respectively. In contrast to the zero-temperature bands in Fig.~\ref{Fig:Fig_vary_omega0}(b1), the finite-temperature bands in Fig.~\ref{Fig:Fig_vary_omega0}(b2) are somewhat broadened. The bandwidth $W_\mathrm{pol}(T=0)$ of the polaronic band at $T=0$ agrees reasonably well with the Lang--Firsov prediction $W_\mathrm{pol}(T=0)/W_0=\exp(-\gamma^2/(\hbar\omega_0)^2)$ ($W_0=4J$ is the bare bandwidth), while the corresponding finite-temperature expression $W_\mathrm{pol}(T)/W_0=\exp(-\gamma^2\coth(\beta\hbar\omega_0/2)/(\hbar\omega_0)^2)$ underestimates the width of the lowest-lying band in Fig.~\ref{Fig:Fig_vary_omega0}(b2).

\subsection{Electron-removal spectral function}
\label{SSec:A-electron-removal}
\begin{figure}[htbp!]
    \centering
    \includegraphics[scale=1.1]{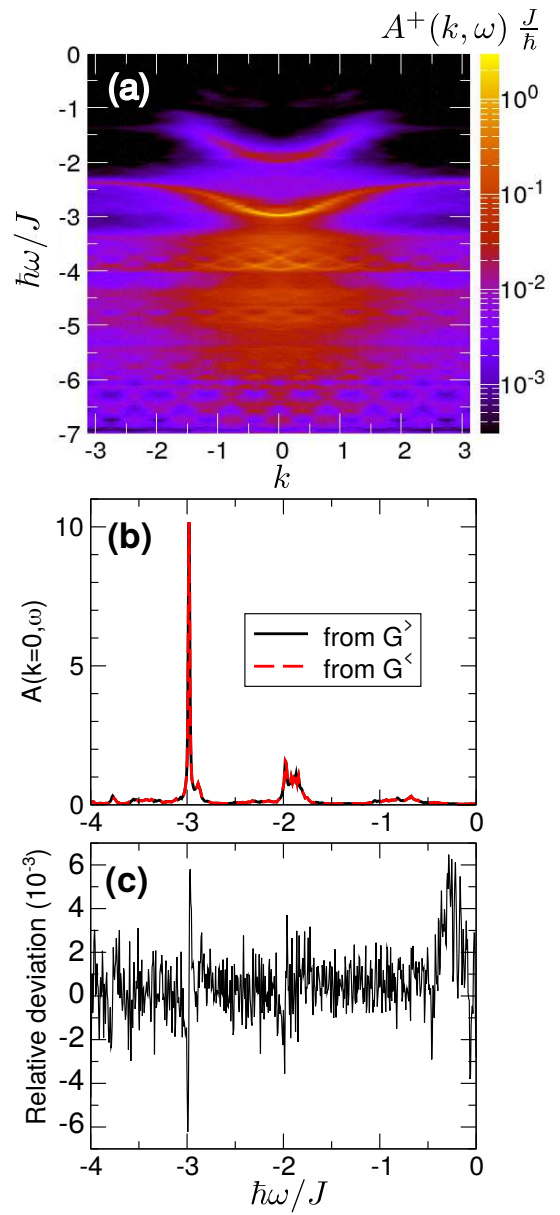}
    \caption{(a) Momentum- and energy-resolved electron-removal spectral function $A^+(k,\omega)$ for $\gamma/J=\sqrt{2}$, $\hbar\omega_0/J=1$, and $k_BT/J=0.4$. Note the logarithmic scale of the color bar. (b) and (c): Numerical verification of the fluctuation--dissipation theorem for $A$ and $A^+$. (b) Spectral function $A(k=0,\omega)$ in the zone center obtained using Eq.~\eqref{Eq:A-from-G-greater} (solid line, labeled "from $G^>$") and using Eqs.~\eqref{Eq:A-plus-k-omega},~\eqref{Eq:relate_A_and_A_plus}, and~\eqref{Eq:normalization_A_A_plus} (dashed line, labeled "from $G^<$"). (c) Relative deviation (in $10^{-3}$) of $A(k=0,\omega)$ obtained from $G^>$ with respect to $A(k=0,\omega)$ obtained from $G^<$.}
    \label{Fig:fig_lesser_overall_280921}
\end{figure}

Figure~\ref{Fig:fig_lesser_overall_280921}(a) presents the electron-removal spectral function $A^+(k,\omega)$ defined in Eq.~\eqref{Eq:A-plus-k-omega} for $\gamma/J=\sqrt{2}$, $\hbar\omega_0/J=1$, and $k_BT/J=0.4$.
We observe a prominent band whose minimum is situated at the quasiparticle peak $\hbar\omega_\mathrm{QP}/J\sim -3$ along with its single-phonon replica, which is much less intense and starts at $\hbar(\omega_\mathrm{QP}+\omega_0)/J$. The bands at frequencies $\omega<\omega_\mathrm{QP}$ are not well resolved, but one can still glimpse certain wide structures whose minima lie at $\hbar(\omega_\mathrm{QP}-n\omega_0)/J$, $n=1,2,3$, and whose intensity diminishes as $n$ is increased. The results summarized in Fig.~\ref{Fig:fig_lesser_overall_280921}(a) overall agree with the results for $A^+(k,\omega)$ presented in Ref.~\onlinecite{PhysRevB.102.165155} for the same parameter regime. $A^+(k,\omega)$ satisfies the sum rule
\begin{equation}
    \int_{-\infty}^{+\infty} d\omega\:A^+(k,\omega)=f_k
\end{equation}
where $f_k$ is the population of electronic state with momentum $k$.

According to the fluctuation--dissipation theorem,~\cite{Mahanbook} $A(k,\omega)$ and $A^+(k,\omega)$ are not mutually independent and are related by
\begin{equation}
\label{Eq:relate_A_and_A_plus}
    A(k,\omega)=\mathcal{N}e^{\beta\hbar\omega}A^+(k,\omega).
\end{equation}
The normalization constant $\mathcal{N}$ enters Eq.~\eqref{Eq:relate_A_and_A_plus} because $A$ and $A^+$ satisfy different sum rules, while its value
\begin{equation}
\label{Eq:normalization_A_A_plus}
    \mathcal{N}=\sum_k\int_{-\infty}^{+\infty} d\omega\:A(k,\omega)e^{-\beta\hbar\omega}
\end{equation}
is determined by the requirement that the number of electrons is equal to unity. To numerically check the validity of Eq.~\eqref{Eq:relate_A_and_A_plus}, in Fig.~\ref{Fig:fig_lesser_overall_280921}(b) we plot the spectral function $A(k=0,\omega)$ in the zone center computed using $G^>(k=0,t)$ [Eq.~\eqref{Eq:A-from-G-greater}, solid line] and $G^<(k=0,t)$ [Eqs.~\eqref{Eq:A-plus-k-omega},~\eqref{Eq:relate_A_and_A_plus} and~\eqref{Eq:normalization_A_A_plus}, dashed line]. Figure~\ref{Fig:fig_lesser_overall_280921}(b) presents the relative deviation between the results obtained in these two manners. While already Fig.~\ref{Fig:fig_lesser_overall_280921}(a) indicates that the agreement is good, Fig.~\ref{Fig:fig_lesser_overall_280921}(b) reveals that the relative difference between the data originating from $G^>$ and $G^<$ oscillates around zero and is smaller than 1\%. These results offer further support for the correctness of our numerical implementation of the HEOM method. 

\subsection{Finite-size effects}
\label{SSec:finite-size-effects}
The numerical cost of the HEOM method significantly increases with the chain length $N$, see Eq.~\eqref{Eq:n_active_dms}. In all previous computations, $N$ was restricted to relatively small values (between 6 and 10). It is therefore highly important to understand how the finite-size effects influence the results presented so far. To that end, we compare the imaginary-time correlation functions obtained from the HEOM spectral function [Eq.~\eqref{Eq:C_k_tau_HEOM}] and the QMC computations [Eq.~\eqref{Eq:C_k_tau_QMC}]. In addition to examining finite-size effects, the comparison of the HEOM and QMC results provides an independent check of the HEOM results, which makes this study self-contained.

\begin{figure}
 \centering
 \includegraphics[scale=0.44]{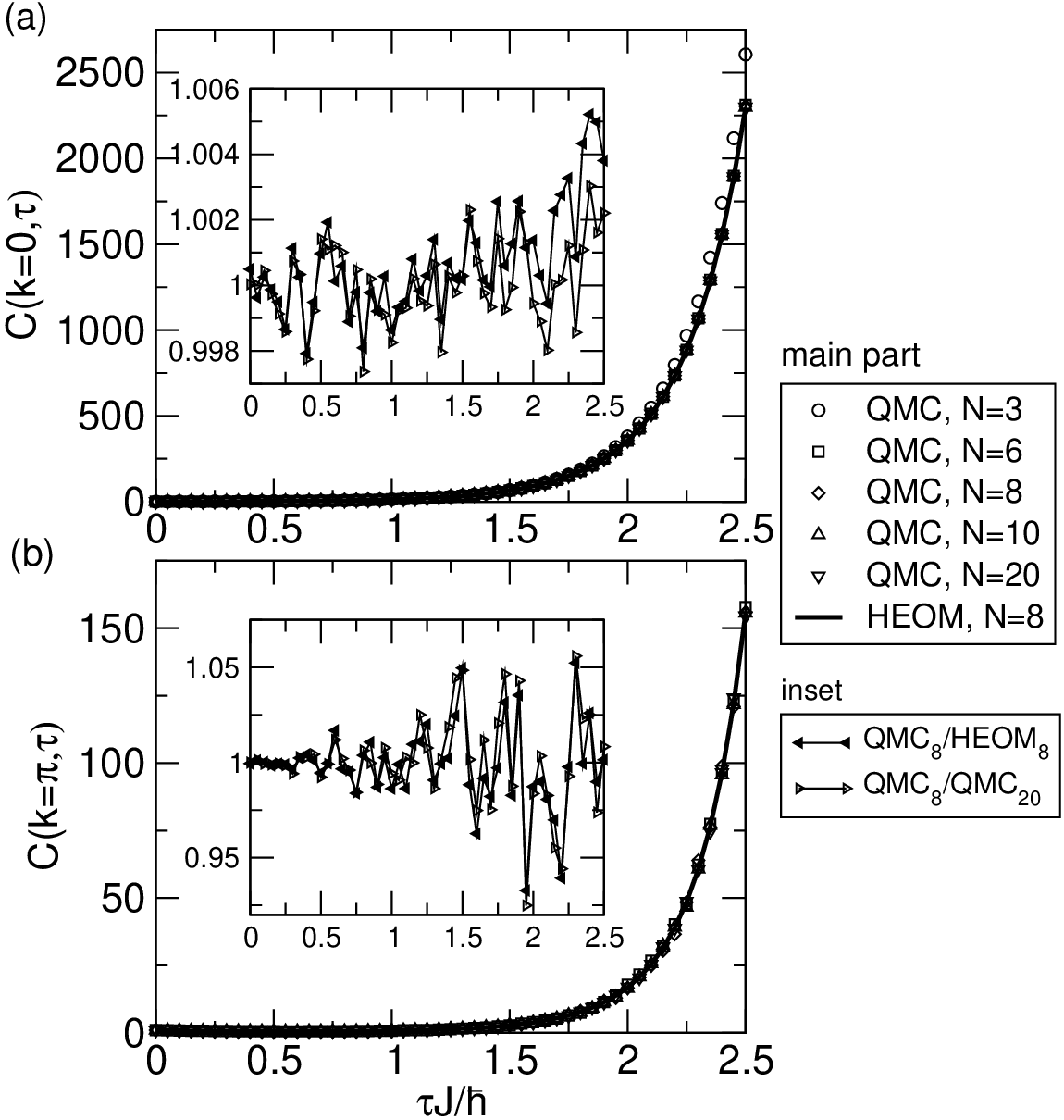}
 \caption{Imaginary-time correlation function $C(k,\tau)$ [Eqs.~\eqref{Eq:C_k_tau_QMC} and~\eqref{Eq:C_k_tau_HEOM}] (a) in the zone center $k=0$ and (b) at the zone edge $k=\pi$ computed using QMC with different chain lengths (empty symbols) and HEOM (solid line). Insets present the ratio of QMC and HEOM results for $N=8$ (full left-triangles) and the ratio of QMC results for $N=8$ and $N=20$ (empty right-triangles). The model parameters assume the following values: $k_BT/J=0.4$, $\hbar\omega_0/J=1$, and $\gamma/J=\sqrt{2}$.}
 \label{Fig:Fig7}
\end{figure}

In Figs.~\ref{Fig:Fig7}(a) and~\ref{Fig:Fig7}(b) we compare the QMC results for different chain lengths (empty symbols) with the HEOM result (solid line) for $k_BT/J=0.4$, $\hbar\omega_0/J=1$, and $\gamma/J=\sqrt{2}$.
A more detailed comparison between the QMC and HEOM results for the fixed chain length is performed by calculating their mutual ratio, see the full left-triangles in the insets. The influence of the chain length on the imaginary-time results can be inferred from the ratio of QMC results for two very different chain lengths, see the empty right-triangles in the insets. The QMC results in Fig.~\ref{Fig:Fig7}(a) are virtually independent of $N$ as long as $N\geq 8$, and the agreement between the QMC results for an 8-site and a 20-site chain is up to 0.3\%, see the empty right-triangles in the inset of Fig.~\ref{Fig:Fig7}(a). The oscillations of the ratio QMC$_8$/QMC$_{20}$ reflect the statistical noise in the QMC data. The HEOM result (for $N=8$) agrees with the corresponding QMC result up to 0.6\%, see the full left-triangles in the inset of Fig.~\ref{Fig:Fig7}(a). Apart from the oscillations that reflect QMC statistical noise, there is also a systematic deviation of the QMC from the HEOM data that increases with $\tau$. Keeping in mind that $C(k,\beta J)$ is proportional to the equilibrium population of state $|k\rangle$, we may conclude that the aforementioned systematic deviation reflects small differences between the momentum distribution function obtained within QMC and HEOM methods, see also Fig.~ 5(b) in Sec.~VII of Ref.~\onlinecite{comment071021}. At the zone edge, the QMC and HEOM results exhibit somewhat worse agreement, see the full left-triangles in the inset of Fig.~\ref{Fig:Fig7}(b), yet their relative difference is smaller than 5\% for all $\tau$. The systematic deviation between the HEOM and QMC data is hidden by the very strong noise in the QMC data. The statistical noise is much stronger at the zone boundary than in the zone center, compare the ranges of the vertical axes in the insets of Figs.~\ref{Fig:Fig7}(a) and~\ref{Fig:Fig7}(b), and, in principle, it could be eliminated by increasing the statistical sample of the QMC computation.
We have checked that this is indeed the case by performing QMC calculations for $N=8$ sites with the statistical sample that is 10 and 100 times larger than the one used in Figs.~\ref{Fig:Fig7}(a) and~\ref{Fig:Fig7}(b). The results of these calculations are presented in Figs.~10(a)--10(d) in Sec.~IX of Ref.~\onlinecite{comment071021}. There, we observe that increasing sample size reduces the statistical noise in the QMC data and ultimately reveals a small systematic deviation between the HEOM and QMC data whose magnitude is consistent with the small differences between the HEOM and QMC momentum distribution functions, see also Fig.~\ref{Fig:Fig7}(a) and Fig.~5 of Ref.~\onlinecite{comment071021}. This relatively low-temperature regime is challenging for both the QMC and HEOM methods. While QMC necessitates finer discretization of the quite long imaginary-time interval over which the algorithm is performed, the HEOM may encounter problems with the long-time weakly damped oscillations in the Green's functions, see Figs.~\ref{Fig:Fig3}(a2) and~\ref{Fig:Fig3}(b2).

\begin{figure}
 \centering
 \includegraphics[scale=0.44]{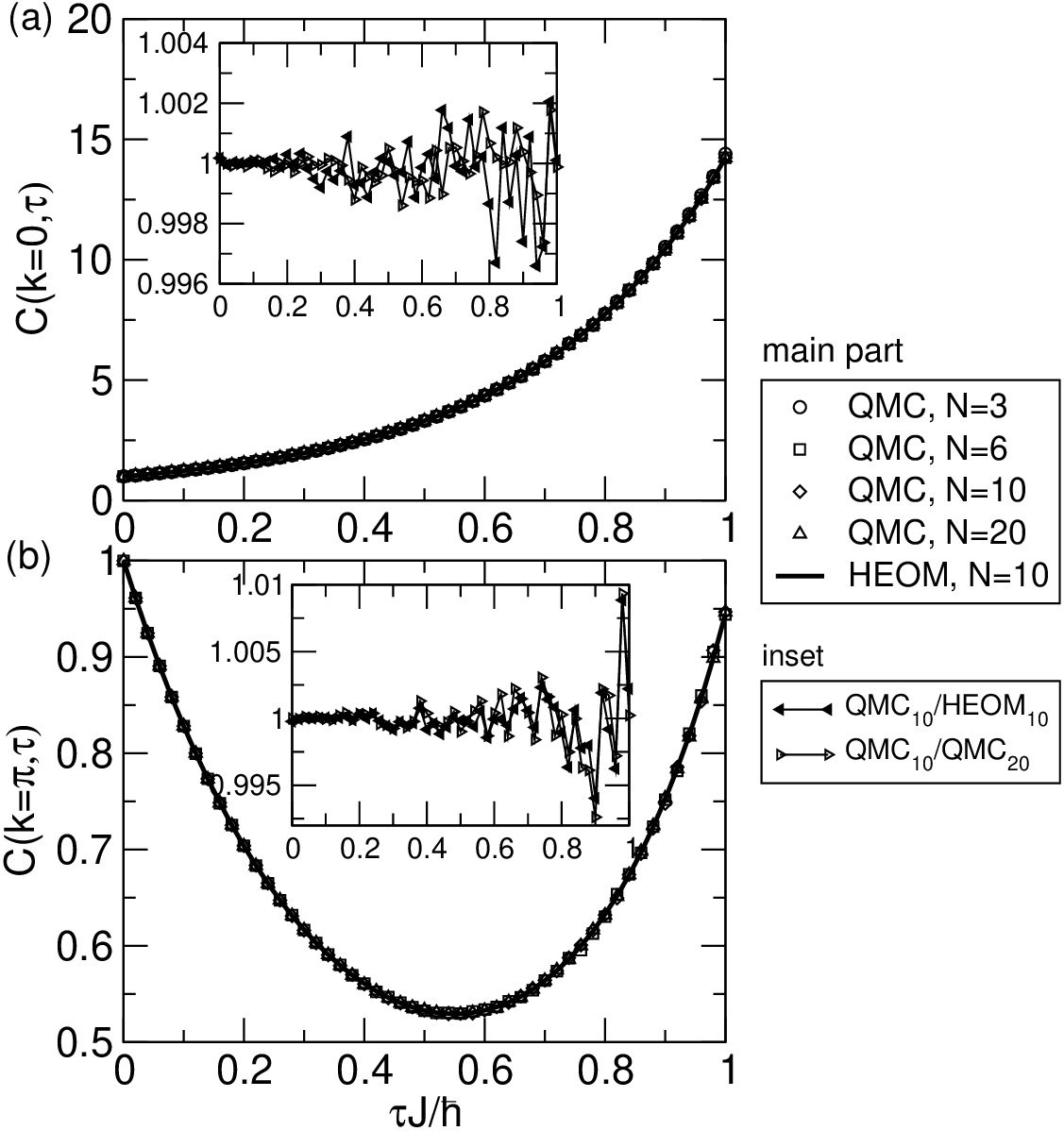}
 \caption{Imaginary-time correlation function $C(k,\tau)$ [Eqs.~\eqref{Eq:C_k_tau_QMC} and~\eqref{Eq:C_k_tau_HEOM}] (a) in the zone center $k=0$ and (b) at the zone edge $k=\pi$ computed using QMC with different chain lengths (empty symbols) and HEOM (solid line). Insets present the ratio of QMC and HEOM results for $N=10$ (full left-triangles) and the ratio of QMC results for $N=10$ and $N=20$ (empty right-triangles). The model parameters assume the following values: $k_BT/J=1$, $\hbar\omega_0/J=1$, and $\gamma/J=1$.}
 \label{Fig:Fig8}
\end{figure}

As the temperature is increased, the finite-size effects become less pronounced, even for reduced interaction strengths. This may be concluded from Figs.~\ref{Fig:Fig8}(a) and~\ref{Fig:Fig8}(b), which is obtained for $k_BT/J=1$, $\hbar\omega_0/J=1$, and $\gamma/J=1$. There, both in the zone center [Fig.~\ref{Fig:Fig8}(a)] and at the zone edges [Fig.~\ref{Fig:Fig8}(b)], QMC results depend very weakly on $N$ (empty right-triangles in the insets) and agree quite well (up to 1\%) with the HEOM results (full left-triangles in the insets).

\begin{figure}
 \centering
 \includegraphics[scale=0.44]{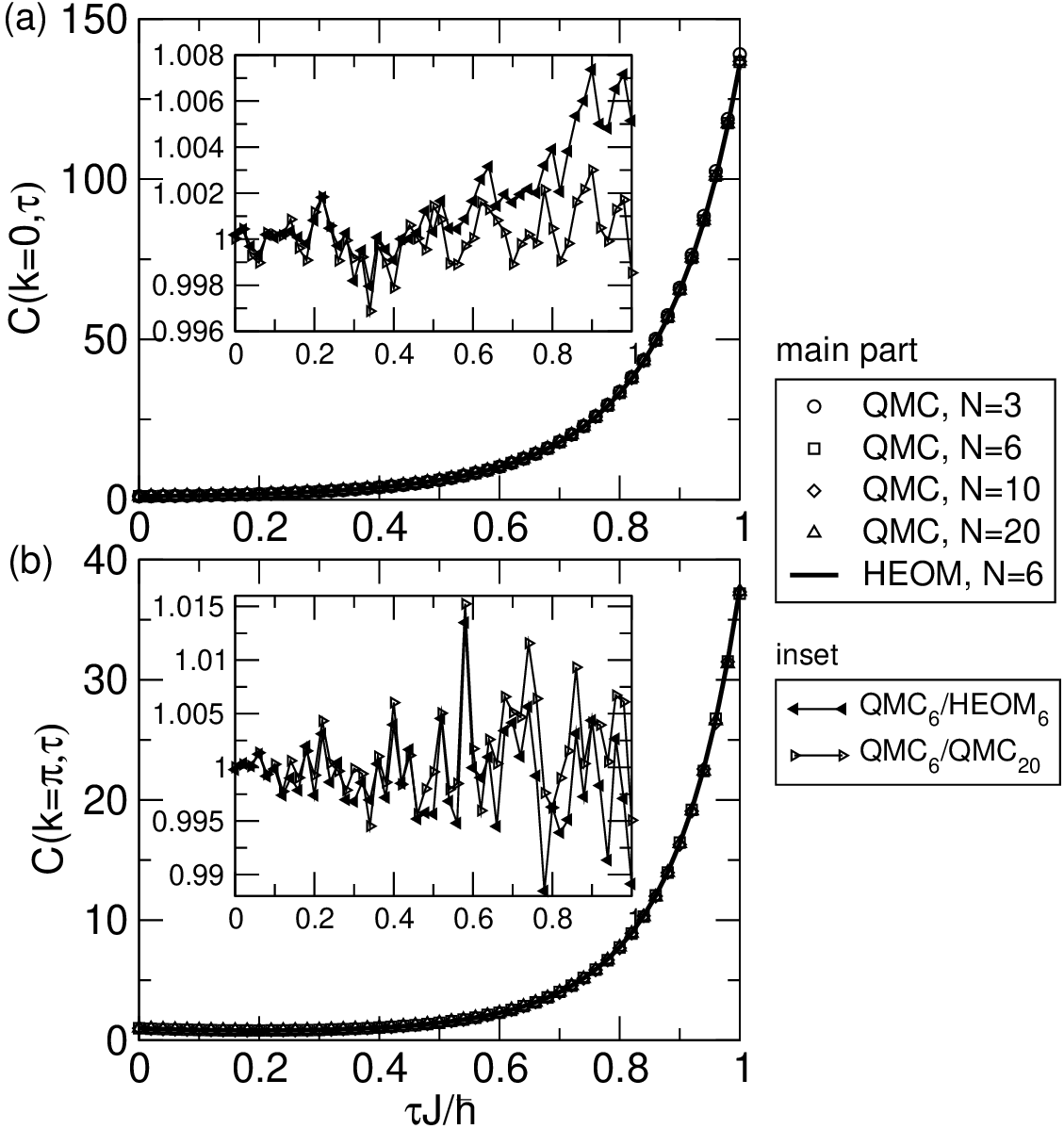}
 \caption{Imaginary-time correlation function $C(k,\tau)$ [Eqs.~\eqref{Eq:C_k_tau_QMC} and~\eqref{Eq:C_k_tau_HEOM}] (a) in the zone center $k=0$ and (b) at the zone edge $k=\pi$ computed using QMC with different chain lengths (empty symbols) and HEOM (solid line). Insets present the ratio of QMC and HEOM results for $N=6$ (full left-triangles) and the ratio of QMC results for $N=6$ and $N=20$ (empty right-triangles). The model parameters assume the following values: $k_BT/J=1$, $\hbar\omega_0/J=1$, and $\gamma/J=2$.}
 \label{Fig:Fig9}
\end{figure}

For strong electron--phonon coupling and at elevated temperatures, the finite-size effects are also not very pronounced and the HEOM results agree very well with the QMC results, see Figs.~\ref{Fig:Fig9}(a) and~\ref{Fig:Fig9}(b) that are plotted for $k_BT/J=1$, $\hbar\omega_0/J=1$, and $\gamma/J=2$.
Similar conclusions may be drawn in the adiabatic and antiadiabatic regime, see Figs.~11 and~12 in Sec.~X of Ref.~\onlinecite{comment071021}, which display quite a good agreement between the QMC and HEOM imaginary-time correlation functions and suggest that the results obtained on $N=6$-site chains are representative of larger systems.

The imaginary-time results presented in this section do establish that the HEOM-method results on short chains are representative of larger systems and strongly suggest that the spectral functions presented in Secs.~\ref{SSec:vary-T}--\ref{SSec:vary-omega0} are close to the true result.
We put some caution on this claim by noting that the operator that transforms the real-frequency spectral function to imaginary-time correlation function has a nontrivial null-space. For this reason, several different spectral functions may yield the same correlation function in imaginary time within the given small uncertainty. Therefore the agreement of imaginary-time correlation functions obtained by QMC and by transforming the HEOM spectral function does not provide a definite proof that HEOM spectral functions are accurate but it does provide a strong evidence. In our most recent work,\cite{2112.15542} we compared the HEOM spectral functions with literature results for spectral functions obtained using the real-time/frequency numerically exact methods such as the finite-temperature Lanczos method~\cite{PhysRevB.100.094307,PhysRevB.103.054304} and the density-matrix renormalization group method.~\cite{PhysRevB.102.165155} Excellent agreement between our results and the literature results gives a definite proof of accuracy of HEOM for parameter values where literature results were available.

We note however that the number of sites when finite-size effects become negligible depends both on the parameters of the model and the physical quantity of interest. For example, in our previous work on the mobility in the Holstein model\cite{PhysRevB.99.104304} we found that for small electron-phonon interaction when the mean free path is large, several tens of sites are required to obtain the converged result. Large number of sites is also necessary to obtain the long-time result if one is interested in the transport at finite bias.\cite{2111.06137}

\section{Discussion}
\label{Sec:discussion}
Despite the unfavorable scaling of the amount of computational resources with system size [Eq.~\eqref{Eq:n_active_dms}], our HEOM approach to evaluate real-time correlation functions at finite temperatures occupies a special place in the gamut of numerical tools for interacting electron--phonon systems.

The exact diagonalization is a method of choice to accurately compute ground-state and finite-temperature spectral properties of relatively small clusters. However, obtaining real-time correlation functions is quite challenging. Furthermore, strictly speaking, the spectral function is a set of delta peaks on the frequency axis, and to get smooth spectra, one has to introduce artificial broadening. As discussed in Sec.~\ref{SSec:Relation_broadening}, this artificial broadening actually changes the physical system we are dealing with by replacing a finite number of phonon modes by a thermodynamic reservoir consisting of infinitely many phonon modes. Even though the spectra become representative of the thermodynamic limit, the value of the broadening is to be chosen carefully.~\cite{PhysRevLett.123.036601} On the one hand, it should be sufficiently large to remove finite-size effects, i.e., to damp the long-time oscillations in $G^>(k,t)$, see Fig.~\ref{Fig:Fig3}, to zero. On the other hand, it should be sufficiently small not to overbroaden the genuine features of spectra, i.e., not to affect the short-time decay of $G^{>}(k,t)$, see Fig.~\ref{Fig:Fig3}. In other words, even though the results are exact, the arbitrariness in the broadening casts doubts on the exactness of the physical quantities that may be obtained from the broadened data (e.g., the conductivity in the so-called bubble or independent-particle approximation,~\cite{RepProgPhys.83.036501} in which the current--current (two-particle) correlation function is approximated as the product of two single-particle correlation functions). QMC may treat larger clusters, as we demonstrated, and thus minimize finite-size effects without introducing artificial broadenings. However, QMC methods to evaluate dynamic correlation functions may face severe problems. On the one hand, to obtain correlation functions on the real-frequency axis, one may perform the analytic continuation from the imaginary-frequency axis, which is in principle an ill-posed problem. On the other hand, one may attempt to directly evaluate correlation functions on the real-time axis, in which case the infamous sign problem is encountered. The difficulties in extracting real-time or real-frequency data from the results on the imaginary axis may be appreciated by contrasting Fig.~\ref{Fig:Fig9} of the main text and Figs.~11 and~12 in Sec.~X of Ref.~\onlinecite{comment071021}, which are obtained in very different parameter regimes. The imaginary-time data in these cases are both qualitatively (the overall shape of the curves) and quantitatively (the vertical-axis ranges) similar to one another. Nevertheless, the spectral functions presented in Figs.~\ref{Fig:Fig11}(c2),~\ref{Fig:Fig_vary_omega0}(a2) and~\ref{Fig:Fig_vary_omega0}(b2) are considerably different from one another. In contrast to the above-discussed approaches, the HEOM method presented here deals directly with the real-time correlation functions and provides numerically exact results for the cluster of given size. HEOM data can be safely used to further compute physical quantities of interest, and the results will be numerically exact for the given system size. Even though we employed HEOM on relatively short chains, artificial broadening parameters are not necessary to obtain results representative of larger systems, which is demonstrated by the very good agreement between the imaginary-time correlation functions emerging from HEOM and QMC computations.

Another quantity relevant for the single-polaron problem is the phonon spectral function, which can be obtained from the greater Green's function in the phonon sector $D^>(q,t)=-i\langle [b_q(t)+b_{-q}^\dagger(t)][b_{-q}+b_q^\dagger]\rangle$. Its computation using the HEOM formalism developed here is complicated by at least two factors. Firstly, the partial trace over phonons lies at the heart of the HEOM formalism, meaning that correlation functions of mixed electron--phonon operators are not straightforwardly accessible from the purely electronic ADMs. Some other formalisms, such as the dissipaton equations of motion~\cite{MolPhys.116.780} or the generalized hierarchical equations,~\cite{JChemPhys.148.014103,JChemPhys.148.014104} provide prescriptions to evaluate correlation functions of mixed electron--phonon operators using the ADMs. Secondly, while the hierarchy for $G^{>/<}(k,t)$ is single-sided [there is no backward evolution operator in Eqs.~\eqref{Eq:G-greater-1}--\eqref{Eq:G-lesser-2}], the hierarchy for $D^>(q,t)$ would be double-sided. The numerical effort to obtain $D^>(q,t)$ would thus be greater than in the case of $G^{>/<}(k,t)$.

Strictly speaking, the phonon spectral function in the zero-density limit, i.e., for one electron on an infinite chain, would be equal to the free-phonon spectral function directly accessible from Eq.~\eqref{Eq:C_q2_q1_t}.~\cite{JPhysCondensMatter.18.7299}
Even though our formulation is restricted to the single-electron case, the numerical results, which would be obtained on finite chains, would actually correspond to finite electron densities.~\cite{JPhysCondensMatter.18.7299} At low temperatures, the phonon spectra of the Holstein model on a finite chain feature the free-phonon line (bare, unrenormalized phonon line) at $\omega=\omega_0$ and a band that replicates the polaron band and whose minimum is at $(q,\omega)=(0,0)$.~\cite{PhysRevB.102.165155,JPhysCondensMatter.18.7299,PhysRevB.82.104304} The minimum of the polaron band is at $\omega=0$ because the zero-momentum phonon couples to the (constant) electron density, see Eq.~\eqref{Eq:def_V_q}. These features may be identified already in the single-site limit, when the phonon spectral function may be directly evaluated
\begin{equation*}
\begin{split}
    iD^>(t)&=[1+n_\mathrm{BE}(\omega_0,T)]e^{-i\omega_0t}+4\left(\frac{\gamma}{\hbar\omega_0}\right)^2\\
    &+n_\mathrm{BE}(\omega_0,T)e^{i\omega_0t}
\end{split}
\end{equation*}
As the temperature is increased, the free-phonon line at negative frequency $\omega=-\omega_0$ acquires appreciable intensity, the reflected polaron band appears at $\omega<0$, while the polaron band broadens at larger momenta.~\cite{PhysRevB.102.165155}

Generally speaking, the extension of the developed method beyond the single-excitation case is not straightforward. This is most directly seen on the example of the equilibrium RDM (Sec.~\ref{SSSec:thermal_eq_dm}) within the Holstein model of mutually noninteracting spinless fermions (the many-fermion version of the present model). The expression for the RDM [Eq.~\eqref{Eq:normalized-th-eq}] and the definition of the ADM $\sigma_\mathbf{n}^{(n,\mathrm{un,eq})}(\tau)$ [Eq.~\eqref{Eq:imag-time-adms-un}] remain the same as in the single-electron case, the only difference being that $V_q=\sum_k c_{k+q}^\dagger c_k$, cf. Eq.~\eqref{Eq:def_V_q}. However, the number of their nonzero matrix elements is significantly increased so that $\mathrm{Tr}\left(c_{q_1}^\dagger\dots c_{q_l}^\dagger c_{p_1}\dots c_{p_m}\sigma_\mathrm{n}^{(n,\mathrm{un,eq})}\right)\neq 0$ only if $l=m$ and $q_1+\dots+q_l-(p_1+\dots+p_m)=k_\mathbf{n}$. The evolution of single-particle quantities (singlets) depends on the evolution of two-particle quantities (doublets) etc., and one has to truncate the hierarchy induced by many-body effects and yet keep the hierarchy stemming from the electron--phonon interaction. One possibility is to employ the cluster-expansion method, see e.g. Ref.~\onlinecite{Kirabook}. Another possibility, particularly appealing when studying time-dependent (exciton--)polaron formation triggered by a laser excitation,~\cite{PhysChemChemPhys.18.7966} is the dynamics controlled truncation scheme, see Refs.~\onlinecite{JChemPhys.153.244122,RevModPhys.70.145}. For the Holstein--Hubbard model, an extension of the present approach is even more complicated, and the HEOM formalism has been successfully applied only to the Holstein--Hubbard dimer,~\cite{PhysRevB.75.081101,JChemPhys.155.064106} where a direct enumeration of many-body states and the separation of different spin sectors are feasible.

Another aspect of our HEOM approach that is worth discussing is its symmetry-adapted formulation in the momentum space. While the application of the HEOM method to single-mode situations is not new,~\cite{JChemPhys.142.174103,JPhysChemLett.6.3110,ChemPhys.515.129} previous studies implemented the method in the real space, without exploiting its translational symmetry. In contrast to perturbative theories, which lean on approximations that have to be carried out in a specific basis (e.g., the momentum eigenbasis for the Redfield theory or the coordinate eigenbasis for the Marcus/F\"{o}rster theory), the HEOM permits us to calculate the properties of interest in any basis. Furthermore, the equilibrium RDM of the electronic subsystem [Eq.~\eqref{Eq:normalized-th-eq}], as well as the Green's functions [Eq.~\eqref{Eq:G-greater-general}], can also be represented in any basis. However, the most convenient representation is the one in which the electronic RDM is diagonal. In that representation, the effects of the electron--phonon interaction on the reduced electronic properties (i.e., the polaronic effects) are taken into account in the simplest possible way. The corresponding basis is known as the global basis,~\cite{PhysRevB.84.245430} or pointer/preferred basis~\cite{JChemPhys.153.244110} in the framework of decoherence theory.~\cite{RevModPhys.75.715,Schlosshauer-book} According to the momentum-conservation law, the electronic single-particle density matrix is diagonal in the eigenbasis $\{|k\rangle\}$ of the electronic momentum. We thus conclude that our momentum-space HEOM is indeed formulated in the preferred basis for the Holstein Hamiltonian.

Our optimal formulation of the problem comes with another advantage. The authors of Ref.~\onlinecite{JChemPhys.150.184109} pointed out that the temporal propagation of the real-space HEOM method developed in Ref.~\onlinecite{JPhysChemLett.6.3110} exhibits an exponential-instability wall. Having hit that wall, the observables diverge, and advanced techniques have to be employed to extract long-time dynamics of the single-mode Holstein model. All the results presented here unambiguously demonstrate that exploiting the translational symmetry of the model, i.e., transferring to the preferred basis of the problem, renders the equations numerically stable and thus obviates the need for advanced tools to mitigate potential instabilities. We propagated our symmetry-adapted HEOM in various parameter regimes up to quite long times $\omega_0t_\mathrm{max}\sim 500$ with a moderate time step $\omega_0\Delta t=0.02$, see Fig.~\ref{Fig:Fig3}, and observed no sign of numerical instabilities in our data.

\section{Conclusion}
\label{Sec:conclusion}
We develop a novel HEOM-based approach that is specifically suited for evaluation of real-time single-particle correlation functions and thermodynamic properties of the Holstein Hamiltonian at finite temperature. The conservation of the total momentum enables us to formulate the hierarchy in the momentum space, so that its dynamical variables describe multi-phonon absorption and emission processes in which the momentum is exchanged between the electron and quantum vibrations. Our momentum-space formulation is superior to the commonly used real-space formulation because it circumvents known numerical problems that arise during the propagation of the real-space HEOM.

We use our HEOM approach to compute the spectral function and thermodynamic quantities of the one-dimensional Holstein model containing up to 10 sites. Our results agree quite well with the available results of the finite-temperature Lanczos method, the coupled-cluster theory, and the momentum average approximation. A further support to the HEOM approach comes from the comparison with imaginary-time data obtained using our QMC approach that provides access to the properties of larger chains. QMC results demonstrate that finite size effects are not pronounced and that relatively short chains, containing 5--10 sites, capture the behavior characteristic for longer chains.

All these results suggest that our HEOM approach may greatly contribute to future studies of finite-temperature Holstein polaron dynamics. Using advanced propagation techniques,~\cite{JChemPhys.154.194104} our approach may become viable for larger or higher-dimensional systems and in multimode situations. It is interesting to note that our study is concurrent with other works aiming at extending the methods commonly applied to molecular systems (chemical-physics realm) to band-like situations (condensed-matter realm).~\cite{JChemPhys.156.024105} Our method may contribute to revealing the (exciton-)polaron formation dynamics, which is highly relevant for a proper interpretation of ultrafast experimental signals in molecular aggregates,~\cite{PhysRevB.84.245430} organic semiconductors,~\cite{Science.323.369} and photosynthetic pigment--protein complexes.~\cite{PhysChemChemPhys.18.7966} Finally, the HEOM developed here may be readily used to study finite-temperature transport properties of the Holstein model by approximating a two-particle correlation function as the product of two single-particle correlation functions.~\cite{RepProgPhys.83.036501}

\acknowledgments
We acknowledge funding provided by the Institute of Physics Belgrade, through the grant by the Ministry of Education, Science, and Technological Development of the Republic of Serbia. Numerical computations were performed on the PARADOX-IV supercomputing facility at the Scientific Computing Laboratory, National Center of Excellence for the Study of Complex Systems, Institute of Physics Belgrade. We thank Darko Tanaskovi\'c and Petar Mitri\'c for insightful and stimulating discussions.

\newpage
\bibliography{refs}

\begin{thebibliography}{110}%
\makeatletter
\providecommand \@ifxundefined [1]{%
 \@ifx{#1\undefined}
}%
\providecommand \@ifnum [1]{%
 \ifnum #1\expandafter \@firstoftwo
 \else \expandafter \@secondoftwo
 \fi
}%
\providecommand \@ifx [1]{%
 \ifx #1\expandafter \@firstoftwo
 \else \expandafter \@secondoftwo
 \fi
}%
\providecommand \natexlab [1]{#1}%
\providecommand \enquote  [1]{``#1''}%
\providecommand \bibnamefont  [1]{#1}%
\providecommand \bibfnamefont [1]{#1}%
\providecommand \citenamefont [1]{#1}%
\providecommand \href@noop [0]{\@secondoftwo}%
\providecommand \href [0]{\begingroup \@sanitize@url \@href}%
\providecommand \@href[1]{\@@startlink{#1}\@@href}%
\providecommand \@@href[1]{\endgroup#1\@@endlink}%
\providecommand \@sanitize@url [0]{\catcode `\\12\catcode `\$12\catcode
  `\&12\catcode `\#12\catcode `\^12\catcode `\_12\catcode `\%12\relax}%
\providecommand \@@startlink[1]{}%
\providecommand \@@endlink[0]{}%
\providecommand \url  [0]{\begingroup\@sanitize@url \@url }%
\providecommand \@url [1]{\endgroup\@href {#1}{\urlprefix }}%
\providecommand \urlprefix  [0]{URL }%
\providecommand \Eprint [0]{\href }%
\providecommand \doibase [0]{http://dx.doi.org/}%
\providecommand \selectlanguage [0]{\@gobble}%
\providecommand \bibinfo  [0]{\@secondoftwo}%
\providecommand \bibfield  [0]{\@secondoftwo}%
\providecommand \translation [1]{[#1]}%
\providecommand \BibitemOpen [0]{}%
\providecommand \bibitemStop [0]{}%
\providecommand \bibitemNoStop [0]{.\EOS\space}%
\providecommand \EOS [0]{\spacefactor3000\relax}%
\providecommand \BibitemShut  [1]{\csname bibitem#1\endcsname}%
\let\auto@bib@innerbib\@empty
\bibitem [{\citenamefont {K\"ohler}\ and\ \citenamefont
  {B\"asller}(2015)}]{Baessler-Koehler-book}%
  \BibitemOpen
  \bibfield  {author} {\bibinfo {author} {\bibfnamefont {A.}~\bibnamefont
  {K\"ohler}}\ and\ \bibinfo {author} {\bibfnamefont {H.}~\bibnamefont
  {B\"asller}},\ }\href@noop {} {\emph {\bibinfo {title} {Electronic processes
  in organic semiconductors}}}\ (\bibinfo  {publisher} {Wiley-VCH Verlag GmbH
  \& Co. KGaA},\ \bibinfo {year} {2015})\BibitemShut {NoStop}%
\bibitem [{\citenamefont {Coropceanu}\ \emph {et~al.}(2007)\citenamefont
  {Coropceanu}, \citenamefont {Cornil}, \citenamefont {da~Silva~Filho},
  \citenamefont {Olivier}, \citenamefont {Silbey},\ and\ \citenamefont
  {Br{\'e}das}}]{ChemRev.107.926}%
  \BibitemOpen
  \bibfield  {author} {\bibinfo {author} {\bibfnamefont {V.}~\bibnamefont
  {Coropceanu}}, \bibinfo {author} {\bibfnamefont {J.}~\bibnamefont {Cornil}},
  \bibinfo {author} {\bibfnamefont {D.~A.}\ \bibnamefont {da~Silva~Filho}},
  \bibinfo {author} {\bibfnamefont {Y.}~\bibnamefont {Olivier}}, \bibinfo
  {author} {\bibfnamefont {R.}~\bibnamefont {Silbey}}, \ and\ \bibinfo {author}
  {\bibfnamefont {J.-L.}\ \bibnamefont {Br{\'e}das}},\ }\bibfield  {title}
  {\enquote {\bibinfo {title} {Charge transport in organic semiconductors},}\
  }\href {\doibase 10.1021/cr050140x} {\bibfield  {journal} {\bibinfo
  {journal} {Chem. Rev.}\ }\textbf {\bibinfo {volume} {107}},\ \bibinfo {pages}
  {926--952} (\bibinfo {year} {2007})}\BibitemShut {NoStop}%
\bibitem [{\citenamefont {Mladenovi{\'{c}}}\ and\ \citenamefont
  {Vukmirovi{\'{c}}}(2015)}]{AdvFunctMater.25.1915}%
  \BibitemOpen
  \bibfield  {author} {\bibinfo {author} {\bibfnamefont {M.}~\bibnamefont
  {Mladenovi{\'{c}}}}\ and\ \bibinfo {author} {\bibfnamefont {N.}~\bibnamefont
  {Vukmirovi{\'{c}}}},\ }\bibfield  {title} {\enquote {\bibinfo {title} {Charge
  carrier localization and transport in organic semiconductors: Insights from
  atomistic multiscale simulations},}\ }\href {\doibase
  https://doi.org/10.1002/adfm.201402435} {\bibfield  {journal} {\bibinfo
  {journal} {Adv. Funct. Mater.}\ }\textbf {\bibinfo {volume} {25}},\ \bibinfo
  {pages} {1915--1932} (\bibinfo {year} {2015})}\BibitemShut {NoStop}%
\bibitem [{\citenamefont {Jang}\ and\ \citenamefont
  {Mennucci}(2018)}]{RevModPhys.90.035003}%
  \BibitemOpen
  \bibfield  {author} {\bibinfo {author} {\bibfnamefont {S.~J.}\ \bibnamefont
  {Jang}}\ and\ \bibinfo {author} {\bibfnamefont {B.}~\bibnamefont
  {Mennucci}},\ }\bibfield  {title} {\enquote {\bibinfo {title} {Delocalized
  excitons in natural light-harvesting complexes},}\ }\href {\doibase
  10.1103/RevModPhys.90.035003} {\bibfield  {journal} {\bibinfo  {journal}
  {Rev. Mod. Phys.}\ }\textbf {\bibinfo {volume} {90}},\ \bibinfo {pages}
  {035003} (\bibinfo {year} {2018})}\BibitemShut {NoStop}%
\bibitem [{\citenamefont {Schr\"oter}\ \emph {et~al.}(2015)\citenamefont
  {Schr\"oter}, \citenamefont {Ivanov}, \citenamefont {Schulze}, \citenamefont
  {Polyutov}, \citenamefont {Yan}, \citenamefont {Pullerits},\ and\
  \citenamefont {K\"uhn}}]{PhysRep.567.1}%
  \BibitemOpen
  \bibfield  {author} {\bibinfo {author} {\bibfnamefont {M.}~\bibnamefont
  {Schr\"oter}}, \bibinfo {author} {\bibfnamefont {S.D.}\ \bibnamefont
  {Ivanov}}, \bibinfo {author} {\bibfnamefont {J.}~\bibnamefont {Schulze}},
  \bibinfo {author} {\bibfnamefont {S.P.}\ \bibnamefont {Polyutov}}, \bibinfo
  {author} {\bibfnamefont {Y.}~\bibnamefont {Yan}}, \bibinfo {author}
  {\bibfnamefont {T.}~\bibnamefont {Pullerits}}, \ and\ \bibinfo {author}
  {\bibfnamefont {O.}~\bibnamefont {K\"uhn}},\ }\bibfield  {title} {\enquote
  {\bibinfo {title} {Exciton--vibrational coupling in the dynamics and
  spectroscopy of {F}renkel excitons in molecular aggregates},}\ }\href
  {\doibase https://doi.org/10.1016/j.physrep.2014.12.001} {\bibfield
  {journal} {\bibinfo  {journal} {Phys. Rep.}\ }\textbf {\bibinfo {volume}
  {567}},\ \bibinfo {pages} {1--78} (\bibinfo {year} {2015})}\BibitemShut
  {NoStop}%
\bibitem [{\citenamefont {van Amerongen}\ \emph {et~al.}(2000)\citenamefont
  {van Amerongen}, \citenamefont {Valkunas},\ and\ \citenamefont {van
  Grondelle}}]{Photosynthetic-excitons-book}%
  \BibitemOpen
  \bibfield  {author} {\bibinfo {author} {\bibfnamefont {V.}~\bibnamefont {van
  Amerongen}}, \bibinfo {author} {\bibfnamefont {L.}~\bibnamefont {Valkunas}},
  \ and\ \bibinfo {author} {\bibfnamefont {R.}~\bibnamefont {van Grondelle}},\
  }\href@noop {} {\emph {\bibinfo {title} {Photosynthetic Excitons}}}\
  (\bibinfo  {publisher} {World Scientific Publishing Co. Pte. Ltd.},\ \bibinfo
  {year} {2000})\BibitemShut {NoStop}%
\bibitem [{\citenamefont {Holstein}(1959)}]{AnnPhys.8.325}%
  \BibitemOpen
  \bibfield  {author} {\bibinfo {author} {\bibfnamefont {T.}~\bibnamefont
  {Holstein}},\ }\bibfield  {title} {\enquote {\bibinfo {title} {Studies of
  polaron motion: Part {I}. {T}he molecular-crystal model},}\ }\href {\doibase
  10.1006/aphy.2000.6020} {\bibfield  {journal} {\bibinfo  {journal} {Ann.
  Phys.}\ }\textbf {\bibinfo {volume} {8}},\ \bibinfo {pages} {325--342}
  (\bibinfo {year} {1959})}\BibitemShut {NoStop}%
\bibitem [{\citenamefont {Redfield}(1965)}]{REDFIELD19651}%
  \BibitemOpen
  \bibfield  {author} {\bibinfo {author} {\bibfnamefont {A.~G.}\ \bibnamefont
  {Redfield}},\ }\bibfield  {title} {\enquote {\bibinfo {title} {The theory of
  relaxation processes},}\ }in\ \href {\doibase
  https://doi.org/10.1016/B978-1-4832-3114-3.50007-6} {\emph {\bibinfo
  {booktitle} {Advances in Magnetic Resonance}}},\ \bibinfo {series} {Advances
  in Magnetic and Optical Resonance}, Vol.~\bibinfo {volume} {1},\ \bibinfo
  {editor} {edited by\ \bibinfo {editor} {\bibfnamefont {J.~S.}\ \bibnamefont
  {Waugh}}}\ (\bibinfo  {publisher} {Academic Press},\ \bibinfo {year} {1965})\
  pp.\ \bibinfo {pages} {1--32}\BibitemShut {NoStop}%
\bibitem [{\citenamefont {F{\"{o}}rster}(1964)}]{Foerster1964}%
  \BibitemOpen
  \bibfield  {author} {\bibinfo {author} {\bibfnamefont {Th.}\ \bibnamefont
  {F{\"{o}}rster}},\ }\href {\doibase 10.2172/4626886} {\emph {\bibinfo {title}
  {DELOCALIZED EXCITATION AND EXCITATION TRANSFER. Bulletin No. 18}}},\
  \bibinfo {type} {Tech. Rep.}\ \bibinfo {number} {FSU-2690-18}\ (\bibinfo
  {institution} {Florida State Univ., Tallahassee. Dept. of Chemistry},\
  \bibinfo {year} {1964})\BibitemShut {NoStop}%
\bibitem [{\citenamefont {Marcus}(1993)}]{RevModPhys.65.599}%
  \BibitemOpen
  \bibfield  {author} {\bibinfo {author} {\bibfnamefont {R.~A.}\ \bibnamefont
  {Marcus}},\ }\bibfield  {title} {\enquote {\bibinfo {title} {Electron
  transfer reactions in chemistry. {T}heory and experiment},}\ }\href {\doibase
  10.1103/RevModPhys.65.599} {\bibfield  {journal} {\bibinfo  {journal} {Rev.
  Mod. Phys.}\ }\textbf {\bibinfo {volume} {65}},\ \bibinfo {pages} {599--610}
  (\bibinfo {year} {1993})}\BibitemShut {NoStop}%
\bibitem [{\citenamefont {Ishizaki}\ and\ \citenamefont
  {Fleming}(2009{\natexlab{a}})}]{JChemPhys.130.234110}%
  \BibitemOpen
  \bibfield  {author} {\bibinfo {author} {\bibfnamefont {A.}~\bibnamefont
  {Ishizaki}}\ and\ \bibinfo {author} {\bibfnamefont {G.~R.}\ \bibnamefont
  {Fleming}},\ }\bibfield  {title} {\enquote {\bibinfo {title} {On the adequacy
  of the {R}edfield equation and related approaches to the study of quantum
  dynamics in electronic energy transfer},}\ }\href {\doibase
  10.1063/1.3155214} {\bibfield  {journal} {\bibinfo  {journal} {J. Chem.
  Phys.}\ }\textbf {\bibinfo {volume} {130}},\ \bibinfo {pages} {234110}
  (\bibinfo {year} {2009}{\natexlab{a}})}\BibitemShut {NoStop}%
\bibitem [{\citenamefont {Ranninger}\ and\ \citenamefont
  {Thibblin}(1992)}]{PhysRevB.45.7730}%
  \BibitemOpen
  \bibfield  {author} {\bibinfo {author} {\bibfnamefont {J.}~\bibnamefont
  {Ranninger}}\ and\ \bibinfo {author} {\bibfnamefont {U.}~\bibnamefont
  {Thibblin}},\ }\bibfield  {title} {\enquote {\bibinfo {title} {Two-site
  polaron problem: Electronic and vibrational properties},}\ }\href {\doibase
  10.1103/PhysRevB.45.7730} {\bibfield  {journal} {\bibinfo  {journal} {Phys.
  Rev. B}\ }\textbf {\bibinfo {volume} {45}},\ \bibinfo {pages} {7730--7738}
  (\bibinfo {year} {1992})}\BibitemShut {NoStop}%
\bibitem [{\citenamefont {Marsiglio}(1993)}]{PhysLettA.180.280}%
  \BibitemOpen
  \bibfield  {author} {\bibinfo {author} {\bibfnamefont {F.}~\bibnamefont
  {Marsiglio}},\ }\bibfield  {title} {\enquote {\bibinfo {title} {The spectral
  function of a one-dimensional {H}olstein polaron},}\ }\href {\doibase
  10.1016/0375-9601(93)90711-8} {\bibfield  {journal} {\bibinfo  {journal}
  {Phys. Lett. A}\ }\textbf {\bibinfo {volume} {180}},\ \bibinfo {pages}
  {280--284} (\bibinfo {year} {1993})}\BibitemShut {NoStop}%
\bibitem [{\citenamefont {Wellein}\ \emph {et~al.}(1996)\citenamefont
  {Wellein}, \citenamefont {R\"oder},\ and\ \citenamefont
  {Fehske}}]{PhysRevB.53.9666}%
  \BibitemOpen
  \bibfield  {author} {\bibinfo {author} {\bibfnamefont {G.}~\bibnamefont
  {Wellein}}, \bibinfo {author} {\bibfnamefont {H.}~\bibnamefont {R\"oder}}, \
  and\ \bibinfo {author} {\bibfnamefont {H.}~\bibnamefont {Fehske}},\
  }\bibfield  {title} {\enquote {\bibinfo {title} {Polarons and bipolarons in
  strongly interacting electron-phonon systems},}\ }\href {\doibase
  10.1103/PhysRevB.53.9666} {\bibfield  {journal} {\bibinfo  {journal} {Phys.
  Rev. B}\ }\textbf {\bibinfo {volume} {53}},\ \bibinfo {pages} {9666--9675}
  (\bibinfo {year} {1996})}\BibitemShut {NoStop}%
\bibitem [{\citenamefont {Capone}\ \emph {et~al.}(1997)\citenamefont {Capone},
  \citenamefont {Stephan},\ and\ \citenamefont {Grilli}}]{PhysRevB.56.4484}%
  \BibitemOpen
  \bibfield  {author} {\bibinfo {author} {\bibfnamefont {M.}~\bibnamefont
  {Capone}}, \bibinfo {author} {\bibfnamefont {W.}~\bibnamefont {Stephan}}, \
  and\ \bibinfo {author} {\bibfnamefont {M.}~\bibnamefont {Grilli}},\
  }\bibfield  {title} {\enquote {\bibinfo {title} {Small-polaron formation and
  optical absorption in {S}u-{S}chrieffer-{H}eeger and {H}olstein models},}\
  }\href {\doibase 10.1103/PhysRevB.56.4484} {\bibfield  {journal} {\bibinfo
  {journal} {Phys. Rev. B}\ }\textbf {\bibinfo {volume} {56}},\ \bibinfo
  {pages} {4484--4493} (\bibinfo {year} {1997})}\BibitemShut {NoStop}%
\bibitem [{\citenamefont {Wellein}\ and\ \citenamefont
  {Fehske}(1997)}]{PhysRevB.56.4513}%
  \BibitemOpen
  \bibfield  {author} {\bibinfo {author} {\bibfnamefont {G.}~\bibnamefont
  {Wellein}}\ and\ \bibinfo {author} {\bibfnamefont {H.}~\bibnamefont
  {Fehske}},\ }\bibfield  {title} {\enquote {\bibinfo {title} {Polaron band
  formation in the {H}olstein model},}\ }\href {\doibase
  10.1103/PhysRevB.56.4513} {\bibfield  {journal} {\bibinfo  {journal} {Phys.
  Rev. B}\ }\textbf {\bibinfo {volume} {56}},\ \bibinfo {pages} {4513--4517}
  (\bibinfo {year} {1997})}\BibitemShut {NoStop}%
\bibitem [{\citenamefont {Robin}(1997)}]{PhysRevB.56.13634}%
  \BibitemOpen
  \bibfield  {author} {\bibinfo {author} {\bibfnamefont {J.~M.}\ \bibnamefont
  {Robin}},\ }\bibfield  {title} {\enquote {\bibinfo {title} {Spectral
  properties of the small polaron},}\ }\href {\doibase
  10.1103/PhysRevB.56.13634} {\bibfield  {journal} {\bibinfo  {journal} {Phys.
  Rev. B}\ }\textbf {\bibinfo {volume} {56}},\ \bibinfo {pages} {13634--13637}
  (\bibinfo {year} {1997})}\BibitemShut {NoStop}%
\bibitem [{\citenamefont {Zhang}\ \emph {et~al.}(1999)\citenamefont {Zhang},
  \citenamefont {Jeckelmann},\ and\ \citenamefont {White}}]{PhysRevB.60.14092}%
  \BibitemOpen
  \bibfield  {author} {\bibinfo {author} {\bibfnamefont {C.}~\bibnamefont
  {Zhang}}, \bibinfo {author} {\bibfnamefont {E.}~\bibnamefont {Jeckelmann}}, \
  and\ \bibinfo {author} {\bibfnamefont {S.~R.}\ \bibnamefont {White}},\
  }\bibfield  {title} {\enquote {\bibinfo {title} {Dynamical properties of the
  one-dimensional {H}olstein model},}\ }\href {\doibase
  10.1103/PhysRevB.60.14092} {\bibfield  {journal} {\bibinfo  {journal} {Phys.
  Rev. B}\ }\textbf {\bibinfo {volume} {60}},\ \bibinfo {pages} {14092--14104}
  (\bibinfo {year} {1999})}\BibitemShut {NoStop}%
\bibitem [{\citenamefont {Hirsch}\ \emph {et~al.}(1982)\citenamefont {Hirsch},
  \citenamefont {Sugar}, \citenamefont {Scalapino},\ and\ \citenamefont
  {Blankenbecler}}]{PhysRevB.26.5033}%
  \BibitemOpen
  \bibfield  {author} {\bibinfo {author} {\bibfnamefont {J.~E.}\ \bibnamefont
  {Hirsch}}, \bibinfo {author} {\bibfnamefont {R.~L.}\ \bibnamefont {Sugar}},
  \bibinfo {author} {\bibfnamefont {D.~J.}\ \bibnamefont {Scalapino}}, \ and\
  \bibinfo {author} {\bibfnamefont {R.}~\bibnamefont {Blankenbecler}},\
  }\bibfield  {title} {\enquote {\bibinfo {title} {Monte {C}arlo simulations of
  one-dimensional fermion systems},}\ }\href {\doibase
  10.1103/PhysRevB.26.5033} {\bibfield  {journal} {\bibinfo  {journal} {Phys.
  Rev. B}\ }\textbf {\bibinfo {volume} {26}},\ \bibinfo {pages} {5033--5055}
  (\bibinfo {year} {1982})}\BibitemShut {NoStop}%
\bibitem [{\citenamefont {Raedt}\ and\ \citenamefont
  {Lagendijk}(1982)}]{PhysRevLett.49.1522}%
  \BibitemOpen
  \bibfield  {author} {\bibinfo {author} {\bibfnamefont {H.~De}\ \bibnamefont
  {Raedt}}\ and\ \bibinfo {author} {\bibfnamefont {A.}~\bibnamefont
  {Lagendijk}},\ }\bibfield  {title} {\enquote {\bibinfo {title} {Critical
  quantum fluctuations and localization of the small polaron},}\ }\href
  {\doibase 10.1103/PhysRevLett.49.1522} {\bibfield  {journal} {\bibinfo
  {journal} {Phys. Rev. Lett.}\ }\textbf {\bibinfo {volume} {49}},\ \bibinfo
  {pages} {1522--1525} (\bibinfo {year} {1982})}\BibitemShut {NoStop}%
\bibitem [{\citenamefont {De~Raedt}\ and\ \citenamefont
  {Lagendijk}(1983)}]{PhysRevB.27.6097}%
  \BibitemOpen
  \bibfield  {author} {\bibinfo {author} {\bibfnamefont {H.}~\bibnamefont
  {De~Raedt}}\ and\ \bibinfo {author} {\bibfnamefont {A.}~\bibnamefont
  {Lagendijk}},\ }\bibfield  {title} {\enquote {\bibinfo {title} {Numerical
  calculation of path integrals: The small-polaron model},}\ }\href {\doibase
  10.1103/PhysRevB.27.6097} {\bibfield  {journal} {\bibinfo  {journal} {Phys.
  Rev. B}\ }\textbf {\bibinfo {volume} {27}},\ \bibinfo {pages} {6097--6109}
  (\bibinfo {year} {1983})}\BibitemShut {NoStop}%
\bibitem [{\citenamefont {De~Raedt}\ and\ \citenamefont
  {Lagendijk}(1984)}]{PhysRevB.30.1671}%
  \BibitemOpen
  \bibfield  {author} {\bibinfo {author} {\bibfnamefont {H.}~\bibnamefont
  {De~Raedt}}\ and\ \bibinfo {author} {\bibfnamefont {A.}~\bibnamefont
  {Lagendijk}},\ }\bibfield  {title} {\enquote {\bibinfo {title} {Numerical
  study of {H}olstein's molecular-crystal model: Adiabatic limit and influence
  of phonon dispersion},}\ }\href {\doibase 10.1103/PhysRevB.30.1671}
  {\bibfield  {journal} {\bibinfo  {journal} {Phys. Rev. B}\ }\textbf {\bibinfo
  {volume} {30}},\ \bibinfo {pages} {1671--1678} (\bibinfo {year}
  {1984})}\BibitemShut {NoStop}%
\bibitem [{\citenamefont {Prokof'ev}\ and\ \citenamefont
  {Svistunov}(1998)}]{PhysRevLett.81.2514}%
  \BibitemOpen
  \bibfield  {author} {\bibinfo {author} {\bibfnamefont {N.~V.}\ \bibnamefont
  {Prokof'ev}}\ and\ \bibinfo {author} {\bibfnamefont {B.~V.}\ \bibnamefont
  {Svistunov}},\ }\bibfield  {title} {\enquote {\bibinfo {title} {Polaron
  problem by diagrammatic quantum {M}onte {C}arlo},}\ }\href {\doibase
  10.1103/PhysRevLett.81.2514} {\bibfield  {journal} {\bibinfo  {journal}
  {Phys. Rev. Lett.}\ }\textbf {\bibinfo {volume} {81}},\ \bibinfo {pages}
  {2514--2517} (\bibinfo {year} {1998})}\BibitemShut {NoStop}%
\bibitem [{\citenamefont {Kornilovitch}(1998)}]{PhysRevLett.81.5382}%
  \BibitemOpen
  \bibfield  {author} {\bibinfo {author} {\bibfnamefont {P.~E.}\ \bibnamefont
  {Kornilovitch}},\ }\bibfield  {title} {\enquote {\bibinfo {title}
  {Continuous-time quantum {M}onte {C}arlo algorithm for the lattice
  polaron},}\ }\href {\doibase 10.1103/PhysRevLett.81.5382} {\bibfield
  {journal} {\bibinfo  {journal} {Phys. Rev. Lett.}\ }\textbf {\bibinfo
  {volume} {81}},\ \bibinfo {pages} {5382--5385} (\bibinfo {year}
  {1998})}\BibitemShut {NoStop}%
\bibitem [{\citenamefont {Hohenadler}\ \emph {et~al.}(2004)\citenamefont
  {Hohenadler}, \citenamefont {Evertz},\ and\ \citenamefont {von~der
  Linden}}]{PhysRevB.69.024301}%
  \BibitemOpen
  \bibfield  {author} {\bibinfo {author} {\bibfnamefont {M.}~\bibnamefont
  {Hohenadler}}, \bibinfo {author} {\bibfnamefont {H.~G.}\ \bibnamefont
  {Evertz}}, \ and\ \bibinfo {author} {\bibfnamefont {W.}~\bibnamefont {von~der
  Linden}},\ }\bibfield  {title} {\enquote {\bibinfo {title} {Quantum {M}onte
  {C}arlo and variational approaches to the {H}olstein model},}\ }\href
  {\doibase 10.1103/PhysRevB.69.024301} {\bibfield  {journal} {\bibinfo
  {journal} {Phys. Rev. B}\ }\textbf {\bibinfo {volume} {69}},\ \bibinfo
  {pages} {024301} (\bibinfo {year} {2004})}\BibitemShut {NoStop}%
\bibitem [{\citenamefont {Spencer}\ \emph {et~al.}(2005)\citenamefont
  {Spencer}, \citenamefont {Samson}, \citenamefont {Kornilovitch},\ and\
  \citenamefont {Alexandrov}}]{PhysRevB.71.184310}%
  \BibitemOpen
  \bibfield  {author} {\bibinfo {author} {\bibfnamefont {P.~E.}\ \bibnamefont
  {Spencer}}, \bibinfo {author} {\bibfnamefont {J.~H.}\ \bibnamefont {Samson}},
  \bibinfo {author} {\bibfnamefont {P.~E.}\ \bibnamefont {Kornilovitch}}, \
  and\ \bibinfo {author} {\bibfnamefont {A.~S.}\ \bibnamefont {Alexandrov}},\
  }\bibfield  {title} {\enquote {\bibinfo {title} {Effect of electron-phonon
  interaction range on lattice polaron dynamics: A continuous-time quantum
  {M}onte {C}arlo study},}\ }\href {\doibase 10.1103/PhysRevB.71.184310}
  {\bibfield  {journal} {\bibinfo  {journal} {Phys. Rev. B}\ }\textbf {\bibinfo
  {volume} {71}},\ \bibinfo {pages} {184310} (\bibinfo {year}
  {2005})}\BibitemShut {NoStop}%
\bibitem [{\citenamefont {Jeckelmann}\ and\ \citenamefont
  {White}(1998)}]{PhysRevB.57.6376}%
  \BibitemOpen
  \bibfield  {author} {\bibinfo {author} {\bibfnamefont {E.}~\bibnamefont
  {Jeckelmann}}\ and\ \bibinfo {author} {\bibfnamefont {S.~R.}\ \bibnamefont
  {White}},\ }\bibfield  {title} {\enquote {\bibinfo {title} {Density-matrix
  renormalization-group study of the polaron problem in the {H}olstein
  model},}\ }\href {\doibase 10.1103/PhysRevB.57.6376} {\bibfield  {journal}
  {\bibinfo  {journal} {Phys. Rev. B}\ }\textbf {\bibinfo {volume} {57}},\
  \bibinfo {pages} {6376--6385} (\bibinfo {year} {1998})}\BibitemShut {NoStop}%
\bibitem [{\citenamefont {Zhang}\ \emph {et~al.}(1998)\citenamefont {Zhang},
  \citenamefont {Jeckelmann},\ and\ \citenamefont
  {White}}]{PhysRevLett.80.2661}%
  \BibitemOpen
  \bibfield  {author} {\bibinfo {author} {\bibfnamefont {C.}~\bibnamefont
  {Zhang}}, \bibinfo {author} {\bibfnamefont {E.}~\bibnamefont {Jeckelmann}}, \
  and\ \bibinfo {author} {\bibfnamefont {S.~R.}\ \bibnamefont {White}},\
  }\bibfield  {title} {\enquote {\bibinfo {title} {Density matrix approach to
  local {H}ilbert space reduction},}\ }\href {\doibase
  10.1103/PhysRevLett.80.2661} {\bibfield  {journal} {\bibinfo  {journal}
  {Phys. Rev. Lett.}\ }\textbf {\bibinfo {volume} {80}},\ \bibinfo {pages}
  {2661--2664} (\bibinfo {year} {1998})}\BibitemShut {NoStop}%
\bibitem [{\citenamefont {Bursill}\ \emph {et~al.}(1998)\citenamefont
  {Bursill}, \citenamefont {McKenzie},\ and\ \citenamefont
  {Hamer}}]{PhysRevLett.80.5607}%
  \BibitemOpen
  \bibfield  {author} {\bibinfo {author} {\bibfnamefont {R.~J.}\ \bibnamefont
  {Bursill}}, \bibinfo {author} {\bibfnamefont {R.~H.}\ \bibnamefont
  {McKenzie}}, \ and\ \bibinfo {author} {\bibfnamefont {C.~J.}\ \bibnamefont
  {Hamer}},\ }\bibfield  {title} {\enquote {\bibinfo {title} {Phase diagram of
  the one-dimensional {H}olstein model of spinless fermions},}\ }\href
  {\doibase 10.1103/PhysRevLett.80.5607} {\bibfield  {journal} {\bibinfo
  {journal} {Phys. Rev. Lett.}\ }\textbf {\bibinfo {volume} {80}},\ \bibinfo
  {pages} {5607--5610} (\bibinfo {year} {1998})}\BibitemShut {NoStop}%
\bibitem [{\citenamefont {Ivi\ifmmode~\acute{c}\else \'{c}\fi{}}\ \emph
  {et~al.}(1993)\citenamefont {Ivi\ifmmode~\acute{c}\else \'{c}\fi{}},
  \citenamefont {Kapor}, \citenamefont {\ifmmode~\check{S}\else
  \v{S}\fi{}krinjar},\ and\ \citenamefont {Popovi\ifmmode~\acute{c}\else
  \'{c}\fi{}}}]{PhysRevB.48.3721}%
  \BibitemOpen
  \bibfield  {author} {\bibinfo {author} {\bibfnamefont {Z.}~\bibnamefont
  {Ivi\ifmmode~\acute{c}\else \'{c}\fi{}}}, \bibinfo {author} {\bibfnamefont
  {D.}~\bibnamefont {Kapor}}, \bibinfo {author} {\bibfnamefont
  {M.}~\bibnamefont {\ifmmode~\check{S}\else \v{S}\fi{}krinjar}}, \ and\
  \bibinfo {author} {\bibfnamefont {Z.}~\bibnamefont
  {Popovi\ifmmode~\acute{c}\else \'{c}\fi{}}},\ }\bibfield  {title} {\enquote
  {\bibinfo {title} {Self-trapping in quasi-one-dimensional electron- and
  exciton-phonon systems},}\ }\href {\doibase 10.1103/PhysRevB.48.3721}
  {\bibfield  {journal} {\bibinfo  {journal} {Phys. Rev. B}\ }\textbf {\bibinfo
  {volume} {48}},\ \bibinfo {pages} {3721--3733} (\bibinfo {year}
  {1993})}\BibitemShut {NoStop}%
\bibitem [{\citenamefont {Romero}\ \emph {et~al.}(1998)\citenamefont {Romero},
  \citenamefont {Brown},\ and\ \citenamefont
  {Lindenberg}}]{JChemPhys.109.6540}%
  \BibitemOpen
  \bibfield  {author} {\bibinfo {author} {\bibfnamefont {A.~H.}\ \bibnamefont
  {Romero}}, \bibinfo {author} {\bibfnamefont {D.~W.}\ \bibnamefont {Brown}}, \
  and\ \bibinfo {author} {\bibfnamefont {K.}~\bibnamefont {Lindenberg}},\
  }\bibfield  {title} {\enquote {\bibinfo {title} {Converging toward a
  practical solution of the {H}olstein molecular crystal model},}\ }\href
  {\doibase 10.1063/1.477305} {\bibfield  {journal} {\bibinfo  {journal} {J.
  Chem. Phys.}\ }\textbf {\bibinfo {volume} {109}},\ \bibinfo {pages}
  {6540--6549} (\bibinfo {year} {1998})}\BibitemShut {NoStop}%
\bibitem [{\citenamefont {Wellein}\ and\ \citenamefont
  {Fehske}(1998)}]{PhysRevB.58.6208}%
  \BibitemOpen
  \bibfield  {author} {\bibinfo {author} {\bibfnamefont {G.}~\bibnamefont
  {Wellein}}\ and\ \bibinfo {author} {\bibfnamefont {H.}~\bibnamefont
  {Fehske}},\ }\bibfield  {title} {\enquote {\bibinfo {title} {Self-trapping
  problem of electrons or excitons in one dimension},}\ }\href {\doibase
  10.1103/PhysRevB.58.6208} {\bibfield  {journal} {\bibinfo  {journal} {Phys.
  Rev. B}\ }\textbf {\bibinfo {volume} {58}},\ \bibinfo {pages} {6208--6218}
  (\bibinfo {year} {1998})}\BibitemShut {NoStop}%
\bibitem [{\citenamefont {Bon\ifmmode~\check{c}\else \v{c}\fi{}a}\ \emph
  {et~al.}(1999)\citenamefont {Bon\ifmmode~\check{c}\else \v{c}\fi{}a},
  \citenamefont {Trugman},\ and\ \citenamefont {Batisti\ifmmode~\acute{c}\else
  \'{c}\fi{}}}]{PhysRevB.60.1633}%
  \BibitemOpen
  \bibfield  {author} {\bibinfo {author} {\bibfnamefont {J.}~\bibnamefont
  {Bon\ifmmode~\check{c}\else \v{c}\fi{}a}}, \bibinfo {author} {\bibfnamefont
  {S.~A.}\ \bibnamefont {Trugman}}, \ and\ \bibinfo {author} {\bibfnamefont
  {I.}~\bibnamefont {Batisti\ifmmode~\acute{c}\else \'{c}\fi{}}},\ }\bibfield
  {title} {\enquote {\bibinfo {title} {Holstein polaron},}\ }\href {\doibase
  10.1103/PhysRevB.60.1633} {\bibfield  {journal} {\bibinfo  {journal} {Phys.
  Rev. B}\ }\textbf {\bibinfo {volume} {60}},\ \bibinfo {pages} {1633--1642}
  (\bibinfo {year} {1999})}\BibitemShut {NoStop}%
\bibitem [{\citenamefont {Ku}\ \emph {et~al.}(2002)\citenamefont {Ku},
  \citenamefont {Trugman},\ and\ \citenamefont {Bon\ifmmode~\check{c}\else
  \v{c}\fi{}a}}]{PhysRevB.65.174306}%
  \BibitemOpen
  \bibfield  {author} {\bibinfo {author} {\bibfnamefont {L.-C.}\ \bibnamefont
  {Ku}}, \bibinfo {author} {\bibfnamefont {S.~A.}\ \bibnamefont {Trugman}}, \
  and\ \bibinfo {author} {\bibfnamefont {J.}~\bibnamefont
  {Bon\ifmmode~\check{c}\else \v{c}\fi{}a}},\ }\bibfield  {title} {\enquote
  {\bibinfo {title} {Dimensionality effects on the {H}olstein polaron},}\
  }\href {\doibase 10.1103/PhysRevB.65.174306} {\bibfield  {journal} {\bibinfo
  {journal} {Phys. Rev. B}\ }\textbf {\bibinfo {volume} {65}},\ \bibinfo
  {pages} {174306} (\bibinfo {year} {2002})}\BibitemShut {NoStop}%
\bibitem [{\citenamefont {Berciu}(2006)}]{PhysRevLett.97.036402}%
  \BibitemOpen
  \bibfield  {author} {\bibinfo {author} {\bibfnamefont {M.}~\bibnamefont
  {Berciu}},\ }\bibfield  {title} {\enquote {\bibinfo {title} {Green's function
  of a dressed particle},}\ }\href {\doibase 10.1103/PhysRevLett.97.036402}
  {\bibfield  {journal} {\bibinfo  {journal} {Phys. Rev. Lett.}\ }\textbf
  {\bibinfo {volume} {97}},\ \bibinfo {pages} {036402} (\bibinfo {year}
  {2006})}\BibitemShut {NoStop}%
\bibitem [{\citenamefont {Goodvin}\ \emph {et~al.}(2006)\citenamefont
  {Goodvin}, \citenamefont {Berciu},\ and\ \citenamefont
  {Sawatzky}}]{PhysRevB.74.245104}%
  \BibitemOpen
  \bibfield  {author} {\bibinfo {author} {\bibfnamefont {G.~L.}\ \bibnamefont
  {Goodvin}}, \bibinfo {author} {\bibfnamefont {M.}~\bibnamefont {Berciu}}, \
  and\ \bibinfo {author} {\bibfnamefont {G.~A.}\ \bibnamefont {Sawatzky}},\
  }\bibfield  {title} {\enquote {\bibinfo {title} {Green's function of the
  {H}olstein polaron},}\ }\href {\doibase 10.1103/PhysRevB.74.245104}
  {\bibfield  {journal} {\bibinfo  {journal} {Phys. Rev. B}\ }\textbf {\bibinfo
  {volume} {74}},\ \bibinfo {pages} {245104} (\bibinfo {year}
  {2006})}\BibitemShut {NoStop}%
\bibitem [{\citenamefont {Paganelli}\ and\ \citenamefont
  {Ciuchi}(2006)}]{JPhysCondensMatter.18.7669}%
  \BibitemOpen
  \bibfield  {author} {\bibinfo {author} {\bibfnamefont {S.}~\bibnamefont
  {Paganelli}}\ and\ \bibinfo {author} {\bibfnamefont {S.}~\bibnamefont
  {Ciuchi}},\ }\bibfield  {title} {\enquote {\bibinfo {title} {Tunnelling
  system coupled to a harmonic oscillator: {A}n analytical treatment},}\ }\href
  {\doibase 10.1088/0953-8984/18/32/015} {\bibfield  {journal} {\bibinfo
  {journal} {J. Phys.: Condens. Matter}\ }\textbf {\bibinfo {volume} {18}},\
  \bibinfo {pages} {7669--7685} (\bibinfo {year} {2006})}\BibitemShut {NoStop}%
\bibitem [{\citenamefont {de~Mello}\ and\ \citenamefont
  {Ranninger}(1997)}]{PhysRevB.55.14872}%
  \BibitemOpen
  \bibfield  {author} {\bibinfo {author} {\bibfnamefont {E.~V.~L.}\
  \bibnamefont {de~Mello}}\ and\ \bibinfo {author} {\bibfnamefont
  {J.}~\bibnamefont {Ranninger}},\ }\bibfield  {title} {\enquote {\bibinfo
  {title} {Dynamical properties of small polarons},}\ }\href {\doibase
  10.1103/PhysRevB.55.14872} {\bibfield  {journal} {\bibinfo  {journal} {Phys.
  Rev. B}\ }\textbf {\bibinfo {volume} {55}},\ \bibinfo {pages} {14872--14885}
  (\bibinfo {year} {1997})}\BibitemShut {NoStop}%
\bibitem [{\citenamefont {Ciuchi}\ \emph {et~al.}(1997)\citenamefont {Ciuchi},
  \citenamefont {de~Pasquale}, \citenamefont {Fratini},\ and\ \citenamefont
  {Feinberg}}]{PhysRevB.56.4494}%
  \BibitemOpen
  \bibfield  {author} {\bibinfo {author} {\bibfnamefont {S.}~\bibnamefont
  {Ciuchi}}, \bibinfo {author} {\bibfnamefont {F.}~\bibnamefont {de~Pasquale}},
  \bibinfo {author} {\bibfnamefont {S.}~\bibnamefont {Fratini}}, \ and\
  \bibinfo {author} {\bibfnamefont {D.}~\bibnamefont {Feinberg}},\ }\bibfield
  {title} {\enquote {\bibinfo {title} {Dynamical mean-field theory of the small
  polaron},}\ }\href {\doibase 10.1103/PhysRevB.56.4494} {\bibfield  {journal}
  {\bibinfo  {journal} {Phys. Rev. B}\ }\textbf {\bibinfo {volume} {56}},\
  \bibinfo {pages} {4494--4512} (\bibinfo {year} {1997})}\BibitemShut {NoStop}%
\bibitem [{\citenamefont {Bon\ifmmode~\check{c}\else \v{c}\fi{}a}\ \emph
  {et~al.}(2019)\citenamefont {Bon\ifmmode~\check{c}\else \v{c}\fi{}a},
  \citenamefont {Trugman},\ and\ \citenamefont {Berciu}}]{PhysRevB.100.094307}%
  \BibitemOpen
  \bibfield  {author} {\bibinfo {author} {\bibfnamefont {J.}~\bibnamefont
  {Bon\ifmmode~\check{c}\else \v{c}\fi{}a}}, \bibinfo {author} {\bibfnamefont
  {S.~A.}\ \bibnamefont {Trugman}}, \ and\ \bibinfo {author} {\bibfnamefont
  {M.}~\bibnamefont {Berciu}},\ }\bibfield  {title} {\enquote {\bibinfo {title}
  {Spectral function of the {H}olstein polaron at finite temperature},}\ }\href
  {\doibase 10.1103/PhysRevB.100.094307} {\bibfield  {journal} {\bibinfo
  {journal} {Phys. Rev. B}\ }\textbf {\bibinfo {volume} {100}},\ \bibinfo
  {pages} {094307} (\bibinfo {year} {2019})}\BibitemShut {NoStop}%
\bibitem [{\citenamefont {Bon\ifmmode~\check{c}\else \v{c}\fi{}a}\ and\
  \citenamefont {Trugman}(2021)}]{PhysRevB.103.054304}%
  \BibitemOpen
  \bibfield  {author} {\bibinfo {author} {\bibfnamefont {J.}~\bibnamefont
  {Bon\ifmmode~\check{c}\else \v{c}\fi{}a}}\ and\ \bibinfo {author}
  {\bibfnamefont {S.~A.}\ \bibnamefont {Trugman}},\ }\bibfield  {title}
  {\enquote {\bibinfo {title} {Dynamic properties of a polaron coupled to
  dispersive optical phonons},}\ }\href {\doibase 10.1103/PhysRevB.103.054304}
  {\bibfield  {journal} {\bibinfo  {journal} {Phys. Rev. B}\ }\textbf {\bibinfo
  {volume} {103}},\ \bibinfo {pages} {054304} (\bibinfo {year}
  {2021})}\BibitemShut {NoStop}%
\bibitem [{\citenamefont {Jansen}\ \emph {et~al.}(2020)\citenamefont {Jansen},
  \citenamefont {Bon\ifmmode~\check{c}\else \v{c}\fi{}a},\ and\ \citenamefont
  {Heidrich-Meisner}}]{PhysRevB.102.165155}%
  \BibitemOpen
  \bibfield  {author} {\bibinfo {author} {\bibfnamefont {D.}~\bibnamefont
  {Jansen}}, \bibinfo {author} {\bibfnamefont {J.}~\bibnamefont
  {Bon\ifmmode~\check{c}\else \v{c}\fi{}a}}, \ and\ \bibinfo {author}
  {\bibfnamefont {F.}~\bibnamefont {Heidrich-Meisner}},\ }\bibfield  {title}
  {\enquote {\bibinfo {title} {Finite-temperature density-matrix
  renormalization group method for electron-phonon systems: Thermodynamics and
  {H}olstein-polaron spectral functions},}\ }\href {\doibase
  10.1103/PhysRevB.102.165155} {\bibfield  {journal} {\bibinfo  {journal}
  {Phys. Rev. B}\ }\textbf {\bibinfo {volume} {102}},\ \bibinfo {pages}
  {165155} (\bibinfo {year} {2020})}\BibitemShut {NoStop}%
\bibitem [{\citenamefont {Grover}\ and\ \citenamefont
  {Silbey}(1971)}]{JChemPhys.54.4843}%
  \BibitemOpen
  \bibfield  {author} {\bibinfo {author} {\bibfnamefont {M.}~\bibnamefont
  {Grover}}\ and\ \bibinfo {author} {\bibfnamefont {R.}~\bibnamefont
  {Silbey}},\ }\bibfield  {title} {\enquote {\bibinfo {title} {Exciton
  migration in molecular crystals},}\ }\href {\doibase 10.1063/1.1674761}
  {\bibfield  {journal} {\bibinfo  {journal} {J. Chem. Phys.}\ }\textbf
  {\bibinfo {volume} {54}},\ \bibinfo {pages} {4843--4851} (\bibinfo {year}
  {1971})}\BibitemShut {NoStop}%
\bibitem [{\citenamefont {Yarkony}\ and\ \citenamefont
  {Silbey}(1977)}]{JChemPhys.67.5818}%
  \BibitemOpen
  \bibfield  {author} {\bibinfo {author} {\bibfnamefont {D.~R.}\ \bibnamefont
  {Yarkony}}\ and\ \bibinfo {author} {\bibfnamefont {R.}~\bibnamefont
  {Silbey}},\ }\bibfield  {title} {\enquote {\bibinfo {title} {Variational
  approach to exciton transport in molecular crystals},}\ }\href {\doibase
  10.1063/1.434789} {\bibfield  {journal} {\bibinfo  {journal} {J. Chem.
  Phys.}\ }\textbf {\bibinfo {volume} {67}},\ \bibinfo {pages} {5818--5827}
  (\bibinfo {year} {1977})}\BibitemShut {NoStop}%
\bibitem [{\citenamefont {Silbey}\ and\ \citenamefont
  {Munn}(1980)}]{JChemPhys.72.2763}%
  \BibitemOpen
  \bibfield  {author} {\bibinfo {author} {\bibfnamefont {R.}~\bibnamefont
  {Silbey}}\ and\ \bibinfo {author} {\bibfnamefont {R.~W.}\ \bibnamefont
  {Munn}},\ }\bibfield  {title} {\enquote {\bibinfo {title} {General theory of
  electronic transport in molecular crystals. {I.} {L}ocal linear
  electron–phonon coupling},}\ }\href {\doibase 10.1063/1.439425} {\bibfield
  {journal} {\bibinfo  {journal} {J. Chem. Phys.}\ }\textbf {\bibinfo {volume}
  {72}},\ \bibinfo {pages} {2763--2773} (\bibinfo {year} {1980})}\BibitemShut
  {NoStop}%
\bibitem [{\citenamefont {Hannewald}\ and\ \citenamefont
  {Bobbert}(2004)}]{PhysRevB.69.075212}%
  \BibitemOpen
  \bibfield  {author} {\bibinfo {author} {\bibfnamefont {K.}~\bibnamefont
  {Hannewald}}\ and\ \bibinfo {author} {\bibfnamefont {P.~A.}\ \bibnamefont
  {Bobbert}},\ }\bibfield  {title} {\enquote {\bibinfo {title} {Anisotropy
  effects in phonon-assisted charge-carrier transport in organic molecular
  crystals},}\ }\href {\doibase 10.1103/PhysRevB.69.075212} {\bibfield
  {journal} {\bibinfo  {journal} {Phys. Rev. B}\ }\textbf {\bibinfo {volume}
  {69}},\ \bibinfo {pages} {075212} (\bibinfo {year} {2004})}\BibitemShut
  {NoStop}%
\bibitem [{\citenamefont {Cheng}\ and\ \citenamefont
  {Silbey}(2008)}]{jcp.128.114713}%
  \BibitemOpen
  \bibfield  {author} {\bibinfo {author} {\bibfnamefont {Y.-C.}\ \bibnamefont
  {Cheng}}\ and\ \bibinfo {author} {\bibfnamefont {R.~J.}\ \bibnamefont
  {Silbey}},\ }\bibfield  {title} {\enquote {\bibinfo {title} {A unified theory
  for charge-carrier transport in organic crystals},}\ }\href {\doibase
  http://dx.doi.org/10.1063/1.2894840} {\bibfield  {journal} {\bibinfo
  {journal} {J. Chem. Phys.}\ }\textbf {\bibinfo {volume} {128}},\ \bibinfo
  {eid} {114713} (\bibinfo {year} {2008})}\BibitemShut {NoStop}%
\bibitem [{\citenamefont {Ortmann}\ \emph {et~al.}(2009)\citenamefont
  {Ortmann}, \citenamefont {Bechstedt},\ and\ \citenamefont
  {Hannewald}}]{PhysRevB.79.235206}%
  \BibitemOpen
  \bibfield  {author} {\bibinfo {author} {\bibfnamefont {F.}~\bibnamefont
  {Ortmann}}, \bibinfo {author} {\bibfnamefont {F.}~\bibnamefont {Bechstedt}},
  \ and\ \bibinfo {author} {\bibfnamefont {K.}~\bibnamefont {Hannewald}},\
  }\bibfield  {title} {\enquote {\bibinfo {title} {Theory of charge transport
  in organic crystals: Beyond {H}olstein's small-polaron model},}\ }\href
  {\doibase 10.1103/PhysRevB.79.235206} {\bibfield  {journal} {\bibinfo
  {journal} {Phys. Rev. B}\ }\textbf {\bibinfo {volume} {79}},\ \bibinfo
  {pages} {235206} (\bibinfo {year} {2009})}\BibitemShut {NoStop}%
\bibitem [{\citenamefont {Prodanovi\ifmmode~\acute{c}\else \'{c}\fi{}}\ and\
  \citenamefont {Vukmirovi\ifmmode~\acute{c}\else
  \'{c}\fi{}}(2019)}]{PhysRevB.99.104304}%
  \BibitemOpen
  \bibfield  {author} {\bibinfo {author} {\bibfnamefont {N.}~\bibnamefont
  {Prodanovi\ifmmode~\acute{c}\else \'{c}\fi{}}}\ and\ \bibinfo {author}
  {\bibfnamefont {N.}~\bibnamefont {Vukmirovi\ifmmode~\acute{c}\else
  \'{c}\fi{}}},\ }\bibfield  {title} {\enquote {\bibinfo {title} {Charge
  carrier mobility in systems with local electron-phonon interaction},}\ }\href
  {\doibase 10.1103/PhysRevB.99.104304} {\bibfield  {journal} {\bibinfo
  {journal} {Phys. Rev. B}\ }\textbf {\bibinfo {volume} {99}},\ \bibinfo
  {pages} {104304} (\bibinfo {year} {2019})}\BibitemShut {NoStop}%
\bibitem [{\citenamefont {Fetherolf}\ \emph {et~al.}(2020)\citenamefont
  {Fetherolf}, \citenamefont {Gole\ifmmode~\check{z}\else \v{z}\fi{}},\ and\
  \citenamefont {Berkelbach}}]{PhysRevX.10.021062}%
  \BibitemOpen
  \bibfield  {author} {\bibinfo {author} {\bibfnamefont {J.~H.}\ \bibnamefont
  {Fetherolf}}, \bibinfo {author} {\bibfnamefont {D.}~\bibnamefont
  {Gole\ifmmode~\check{z}\else \v{z}\fi{}}}, \ and\ \bibinfo {author}
  {\bibfnamefont {T.~C.}\ \bibnamefont {Berkelbach}},\ }\bibfield  {title}
  {\enquote {\bibinfo {title} {A unification of the {H}olstein polaron and
  dynamic disorder pictures of charge transport in organic crystals},}\ }\href
  {\doibase 10.1103/PhysRevX.10.021062} {\bibfield  {journal} {\bibinfo
  {journal} {Phys. Rev. X}\ }\textbf {\bibinfo {volume} {10}},\ \bibinfo
  {pages} {021062} (\bibinfo {year} {2020})}\BibitemShut {NoStop}%
\bibitem [{\citenamefont {Fratini}\ and\ \citenamefont
  {Ciuchi}(2003)}]{PhysRevLett.91.256403}%
  \BibitemOpen
  \bibfield  {author} {\bibinfo {author} {\bibfnamefont {S.}~\bibnamefont
  {Fratini}}\ and\ \bibinfo {author} {\bibfnamefont {S.}~\bibnamefont
  {Ciuchi}},\ }\bibfield  {title} {\enquote {\bibinfo {title} {Dynamical
  mean-field theory of transport of small polarons},}\ }\href {\doibase
  10.1103/PhysRevLett.91.256403} {\bibfield  {journal} {\bibinfo  {journal}
  {Phys. Rev. Lett.}\ }\textbf {\bibinfo {volume} {91}},\ \bibinfo {pages}
  {256403} (\bibinfo {year} {2003})}\BibitemShut {NoStop}%
\bibitem [{\citenamefont {Fratini}\ and\ \citenamefont
  {Ciuchi}(2006)}]{PhysRevB.74.075101}%
  \BibitemOpen
  \bibfield  {author} {\bibinfo {author} {\bibfnamefont {S.}~\bibnamefont
  {Fratini}}\ and\ \bibinfo {author} {\bibfnamefont {S.}~\bibnamefont
  {Ciuchi}},\ }\bibfield  {title} {\enquote {\bibinfo {title} {Optical
  properties of small polarons from dynamical mean-field theory},}\ }\href
  {\doibase 10.1103/PhysRevB.74.075101} {\bibfield  {journal} {\bibinfo
  {journal} {Phys. Rev. B}\ }\textbf {\bibinfo {volume} {74}},\ \bibinfo
  {pages} {075101} (\bibinfo {year} {2006})}\BibitemShut {NoStop}%
\bibitem [{\citenamefont {Goodvin}\ \emph {et~al.}(2011)\citenamefont
  {Goodvin}, \citenamefont {Mishchenko},\ and\ \citenamefont
  {Berciu}}]{PhysRevLett.107.076403}%
  \BibitemOpen
  \bibfield  {author} {\bibinfo {author} {\bibfnamefont {G.~L.}\ \bibnamefont
  {Goodvin}}, \bibinfo {author} {\bibfnamefont {A.~S.}\ \bibnamefont
  {Mishchenko}}, \ and\ \bibinfo {author} {\bibfnamefont {M.}~\bibnamefont
  {Berciu}},\ }\bibfield  {title} {\enquote {\bibinfo {title} {Optical
  conductivity of the {H}olstein polaron},}\ }\href {\doibase
  10.1103/PhysRevLett.107.076403} {\bibfield  {journal} {\bibinfo  {journal}
  {Phys. Rev. Lett.}\ }\textbf {\bibinfo {volume} {107}},\ \bibinfo {pages}
  {076403} (\bibinfo {year} {2011})}\BibitemShut {NoStop}%
\bibitem [{\citenamefont {Mishchenko}\ \emph {et~al.}(2015)\citenamefont
  {Mishchenko}, \citenamefont {Nagaosa}, \citenamefont {De~Filippis},
  \citenamefont {de~Candia},\ and\ \citenamefont
  {Cataudella}}]{PhysRevLett.114.146401}%
  \BibitemOpen
  \bibfield  {author} {\bibinfo {author} {\bibfnamefont {A.~S.}\ \bibnamefont
  {Mishchenko}}, \bibinfo {author} {\bibfnamefont {N.}~\bibnamefont {Nagaosa}},
  \bibinfo {author} {\bibfnamefont {G.}~\bibnamefont {De~Filippis}}, \bibinfo
  {author} {\bibfnamefont {A.}~\bibnamefont {de~Candia}}, \ and\ \bibinfo
  {author} {\bibfnamefont {V.}~\bibnamefont {Cataudella}},\ }\bibfield  {title}
  {\enquote {\bibinfo {title} {Mobility of {H}olstein polaron at finite
  temperature: An unbiased approach},}\ }\href {\doibase
  10.1103/PhysRevLett.114.146401} {\bibfield  {journal} {\bibinfo  {journal}
  {Phys. Rev. Lett.}\ }\textbf {\bibinfo {volume} {114}},\ \bibinfo {pages}
  {146401} (\bibinfo {year} {2015})}\BibitemShut {NoStop}%
\bibitem [{\citenamefont {Li}\ \emph {et~al.}(2020)\citenamefont {Li},
  \citenamefont {Ren},\ and\ \citenamefont {Shuai}}]{JPhysChemLett.11.4930}%
  \BibitemOpen
  \bibfield  {author} {\bibinfo {author} {\bibfnamefont {W.}~\bibnamefont
  {Li}}, \bibinfo {author} {\bibfnamefont {J.}~\bibnamefont {Ren}}, \ and\
  \bibinfo {author} {\bibfnamefont {Z.}~\bibnamefont {Shuai}},\ }\bibfield
  {title} {\enquote {\bibinfo {title} {Finite-temperature {T}{D}-{D}{M}{R}{G}
  for the carrier mobility of organic semiconductors},}\ }\href {\doibase
  10.1021/acs.jpclett.0c01072} {\bibfield  {journal} {\bibinfo  {journal} {J.
  Phys. Chem. Lett.}\ }\textbf {\bibinfo {volume} {11}},\ \bibinfo {pages}
  {4930--4936} (\bibinfo {year} {2020})}\BibitemShut {NoStop}%
\bibitem [{\citenamefont {Tanimura}(2006)}]{JPhysSocJpn.75.082001}%
  \BibitemOpen
  \bibfield  {author} {\bibinfo {author} {\bibfnamefont {Y.}~\bibnamefont
  {Tanimura}},\ }\bibfield  {title} {\enquote {\bibinfo {title} {Stochastic
  {L}iouville, {L}angevin, {F}okker--{P}lanck, and master equation approaches
  to quantum dissipative systems},}\ }\href {\doibase 10.1143/JPSJ.75.082001}
  {\bibfield  {journal} {\bibinfo  {journal} {J. Phys. Soc. Jpn.}\ }\textbf
  {\bibinfo {volume} {75}},\ \bibinfo {pages} {082001} (\bibinfo {year}
  {2006})}\BibitemShut {NoStop}%
\bibitem [{\citenamefont {Xu}\ and\ \citenamefont
  {Yan}(2007)}]{PhysRevE.75.031107}%
  \BibitemOpen
  \bibfield  {author} {\bibinfo {author} {\bibfnamefont {R.-X.}\ \bibnamefont
  {Xu}}\ and\ \bibinfo {author} {\bibfnamefont {Y.~J.}\ \bibnamefont {Yan}},\
  }\bibfield  {title} {\enquote {\bibinfo {title} {Dynamics of quantum
  dissipation systems interacting with bosonic canonical bath: Hierarchical
  equations of motion approach},}\ }\href {\doibase 10.1103/PhysRevE.75.031107}
  {\bibfield  {journal} {\bibinfo  {journal} {Phys. Rev. E}\ }\textbf {\bibinfo
  {volume} {75}},\ \bibinfo {pages} {031107} (\bibinfo {year}
  {2007})}\BibitemShut {NoStop}%
\bibitem [{\citenamefont {Ishizaki}\ and\ \citenamefont
  {Fleming}(2009{\natexlab{b}})}]{JChemPhys.130.234111}%
  \BibitemOpen
  \bibfield  {author} {\bibinfo {author} {\bibfnamefont {A.}~\bibnamefont
  {Ishizaki}}\ and\ \bibinfo {author} {\bibfnamefont {G.~R.}\ \bibnamefont
  {Fleming}},\ }\bibfield  {title} {\enquote {\bibinfo {title} {Unified
  treatment of quantum coherent and incoherent hopping dynamics in electronic
  energy transfer: Reduced hierarchy equation approach},}\ }\href {\doibase
  10.1063/1.3155372} {\bibfield  {journal} {\bibinfo  {journal} {J. Chem.
  Phys.}\ }\textbf {\bibinfo {volume} {130}},\ \bibinfo {pages} {234111}
  (\bibinfo {year} {2009}{\natexlab{b}})}\BibitemShut {NoStop}%
\bibitem [{\citenamefont {Tanimura}(2020)}]{JChemPhys.153.020901}%
  \BibitemOpen
  \bibfield  {author} {\bibinfo {author} {\bibfnamefont {Y.}~\bibnamefont
  {Tanimura}},\ }\bibfield  {title} {\enquote {\bibinfo {title} {Numerically
  “exact” approach to open quantum dynamics: The hierarchical equations of
  motion ({H}{E}{O}{M})},}\ }\href {\doibase 10.1063/5.0011599} {\bibfield
  {journal} {\bibinfo  {journal} {J. Chem. Phys.}\ }\textbf {\bibinfo {volume}
  {153}},\ \bibinfo {pages} {020901} (\bibinfo {year} {2020})}\BibitemShut
  {NoStop}%
\bibitem [{\citenamefont {Feynman}\ and\ \citenamefont
  {Vernon}(1963)}]{AnnPhys.24.118}%
  \BibitemOpen
  \bibfield  {author} {\bibinfo {author} {\bibfnamefont {R.P}\ \bibnamefont
  {Feynman}}\ and\ \bibinfo {author} {\bibfnamefont {F.L}\ \bibnamefont
  {Vernon}},\ }\bibfield  {title} {\enquote {\bibinfo {title} {The theory of a
  general quantum system interacting with a linear dissipative system},}\
  }\href {\doibase https://doi.org/10.1016/0003-4916(63)90068-X} {\bibfield
  {journal} {\bibinfo  {journal} {Ann. Phys.}\ }\textbf {\bibinfo {volume}
  {24}},\ \bibinfo {pages} {118--173} (\bibinfo {year} {1963})}\BibitemShut
  {NoStop}%
\bibitem [{\citenamefont {Shi}\ \emph {et~al.}(2009)\citenamefont {Shi},
  \citenamefont {Chen}, \citenamefont {Nan}, \citenamefont {Xu},\ and\
  \citenamefont {Yan}}]{JChemPhys.130.084105}%
  \BibitemOpen
  \bibfield  {author} {\bibinfo {author} {\bibfnamefont {Q.}~\bibnamefont
  {Shi}}, \bibinfo {author} {\bibfnamefont {L.}~\bibnamefont {Chen}}, \bibinfo
  {author} {\bibfnamefont {G.}~\bibnamefont {Nan}}, \bibinfo {author}
  {\bibfnamefont {R.-X.}\ \bibnamefont {Xu}}, \ and\ \bibinfo {author}
  {\bibfnamefont {Y.}~\bibnamefont {Yan}},\ }\bibfield  {title} {\enquote
  {\bibinfo {title} {Efficient hierarchical {L}iouville space propagator to
  quantum dissipative dynamics},}\ }\href {\doibase 10.1063/1.3077918}
  {\bibfield  {journal} {\bibinfo  {journal} {J. Chem. Phys.}\ }\textbf
  {\bibinfo {volume} {130}},\ \bibinfo {pages} {084105} (\bibinfo {year}
  {2009})}\BibitemShut {NoStop}%
\bibitem [{\citenamefont {Zhang}\ \emph {et~al.}(2018)\citenamefont {Zhang},
  \citenamefont {Xu}, \citenamefont {Zheng},\ and\ \citenamefont
  {Yan}}]{MolPhys.116.780}%
  \BibitemOpen
  \bibfield  {author} {\bibinfo {author} {\bibfnamefont {H.-D.}\ \bibnamefont
  {Zhang}}, \bibinfo {author} {\bibfnamefont {R.-X.}\ \bibnamefont {Xu}},
  \bibinfo {author} {\bibfnamefont {X.}~\bibnamefont {Zheng}}, \ and\ \bibinfo
  {author} {\bibfnamefont {Y.}~\bibnamefont {Yan}},\ }\bibfield  {title}
  {\enquote {\bibinfo {title} {Statistical quasi-particle theory for open
  quantum systems},}\ }\href {\doibase 10.1080/00268976.2018.1431407}
  {\bibfield  {journal} {\bibinfo  {journal} {Mol. Phys.}\ }\textbf {\bibinfo
  {volume} {116}},\ \bibinfo {pages} {780--812} (\bibinfo {year}
  {2018})}\BibitemShut {NoStop}%
\bibitem [{\citenamefont {Ishizaki}\ and\ \citenamefont
  {Fleming}(2009{\natexlab{c}})}]{ProcNatlAcadSci.106.17255}%
  \BibitemOpen
  \bibfield  {author} {\bibinfo {author} {\bibfnamefont {A.}~\bibnamefont
  {Ishizaki}}\ and\ \bibinfo {author} {\bibfnamefont {G.~R.}\ \bibnamefont
  {Fleming}},\ }\bibfield  {title} {\enquote {\bibinfo {title} {Theoretical
  examination of quantum coherence in a photosynthetic system at physiological
  temperature},}\ }\href {\doibase 10.1073/pnas.0908989106} {\bibfield
  {journal} {\bibinfo  {journal} {Proc. Natl. Acad. Sci.}\ }\textbf {\bibinfo
  {volume} {106}},\ \bibinfo {pages} {17255--17260} (\bibinfo {year}
  {2009}{\natexlab{c}})}\BibitemShut {NoStop}%
\bibitem [{\citenamefont {Str{\"{u}}mpfer}\ and\ \citenamefont
  {Schulten}(2012)}]{JChemTheoryComput.8.2808}%
  \BibitemOpen
  \bibfield  {author} {\bibinfo {author} {\bibfnamefont {J.}~\bibnamefont
  {Str{\"{u}}mpfer}}\ and\ \bibinfo {author} {\bibfnamefont {K.}~\bibnamefont
  {Schulten}},\ }\bibfield  {title} {\enquote {\bibinfo {title} {Open quantum
  dynamics calculations with the hierarchy equations of motion on parallel
  computers},}\ }\href {\doibase 10.1021/ct3003833} {\bibfield  {journal}
  {\bibinfo  {journal} {J. Chem. Theory Comput.}\ }\textbf {\bibinfo {volume}
  {8}},\ \bibinfo {pages} {2808--2816} (\bibinfo {year} {2012})}\BibitemShut
  {NoStop}%
\bibitem [{\citenamefont {Wilkins}\ and\ \citenamefont
  {Dattani}(2015)}]{JChemTheoryComput.11.3411}%
  \BibitemOpen
  \bibfield  {author} {\bibinfo {author} {\bibfnamefont {D.~M.}\ \bibnamefont
  {Wilkins}}\ and\ \bibinfo {author} {\bibfnamefont {N.~S.}\ \bibnamefont
  {Dattani}},\ }\bibfield  {title} {\enquote {\bibinfo {title} {Why quantum
  coherence is not important in the {F}enna--{M}atthews--{O}lsen complex},}\
  }\href {\doibase 10.1021/ct501066k} {\bibfield  {journal} {\bibinfo
  {journal} {J. Chem. Theory Comput.}\ }\textbf {\bibinfo {volume} {11}},\
  \bibinfo {pages} {3411--3419} (\bibinfo {year} {2015})}\BibitemShut {NoStop}%
\bibitem [{\citenamefont {Kramer}\ \emph {et~al.}(2018)\citenamefont {Kramer},
  \citenamefont {Noack}, \citenamefont {Reinefeld}, \citenamefont
  {Rodr{\'{i}}guez},\ and\ \citenamefont {Zelinskyy}}]{JComputChem.39.1779}%
  \BibitemOpen
  \bibfield  {author} {\bibinfo {author} {\bibfnamefont {T.}~\bibnamefont
  {Kramer}}, \bibinfo {author} {\bibfnamefont {M.}~\bibnamefont {Noack}},
  \bibinfo {author} {\bibfnamefont {A.}~\bibnamefont {Reinefeld}}, \bibinfo
  {author} {\bibfnamefont {M.}~\bibnamefont {Rodr{\'{i}}guez}}, \ and\ \bibinfo
  {author} {\bibfnamefont {Y.}~\bibnamefont {Zelinskyy}},\ }\bibfield  {title}
  {\enquote {\bibinfo {title} {Efficient calculation of open quantum system
  dynamics and time-resolved spectroscopy with distributed memory {H}{E}{O}{M}
  ({D}{M}-{H}{E}{O}{M})},}\ }\href {\doibase https://doi.org/10.1002/jcc.25354}
  {\bibfield  {journal} {\bibinfo  {journal} {J. Comput. Chem.}\ }\textbf
  {\bibinfo {volume} {39}},\ \bibinfo {pages} {1779--1794} (\bibinfo {year}
  {2018})}\BibitemShut {NoStop}%
\bibitem [{\citenamefont {Jankovi{\'c}}\ and\ \citenamefont {Man{\v
  c}al}(2020{\natexlab{a}})}]{JChemPhys.153.244122}%
  \BibitemOpen
  \bibfield  {author} {\bibinfo {author} {\bibfnamefont {V.}~\bibnamefont
  {Jankovi{\'c}}}\ and\ \bibinfo {author} {\bibfnamefont {T.}~\bibnamefont
  {Man{\v c}al}},\ }\bibfield  {title} {\enquote {\bibinfo {title} {Exact
  description of excitonic dynamics in molecular aggregates weakly driven by
  light},}\ }\href {\doibase 10.1063/5.0029914} {\bibfield  {journal} {\bibinfo
   {journal} {J. Chem. Phys.}\ }\textbf {\bibinfo {volume} {153}},\ \bibinfo
  {pages} {244122} (\bibinfo {year} {2020}{\natexlab{a}})}\BibitemShut
  {NoStop}%
\bibitem [{\citenamefont {Jankovi{\'c}}\ and\ \citenamefont {Man{\v
  c}al}(2020{\natexlab{b}})}]{JChemPhys.153.244110}%
  \BibitemOpen
  \bibfield  {author} {\bibinfo {author} {\bibfnamefont {V.}~\bibnamefont
  {Jankovi{\'c}}}\ and\ \bibinfo {author} {\bibfnamefont {T.}~\bibnamefont
  {Man{\v c}al}},\ }\bibfield  {title} {\enquote {\bibinfo {title}
  {Nonequilibrium steady-state picture of incoherent light-induced excitation
  harvesting},}\ }\href {\doibase 10.1063/5.0029918} {\bibfield  {journal}
  {\bibinfo  {journal} {J. Chem. Phys.}\ }\textbf {\bibinfo {volume} {153}},\
  \bibinfo {pages} {244110} (\bibinfo {year} {2020}{\natexlab{b}})}\BibitemShut
  {NoStop}%
\bibitem [{\citenamefont {Wang}\ \emph {et~al.}(2010)\citenamefont {Wang},
  \citenamefont {Chen}, \citenamefont {Zheng}, \citenamefont {Wang},\ and\
  \citenamefont {Shi}}]{JChemPhys.132.081101}%
  \BibitemOpen
  \bibfield  {author} {\bibinfo {author} {\bibfnamefont {D.}~\bibnamefont
  {Wang}}, \bibinfo {author} {\bibfnamefont {L.}~\bibnamefont {Chen}}, \bibinfo
  {author} {\bibfnamefont {R.}~\bibnamefont {Zheng}}, \bibinfo {author}
  {\bibfnamefont {L.}~\bibnamefont {Wang}}, \ and\ \bibinfo {author}
  {\bibfnamefont {Q.}~\bibnamefont {Shi}},\ }\bibfield  {title} {\enquote
  {\bibinfo {title} {Communications: A nonperturbative quantum master equation
  approach to charge carrier transport in organic molecular crystals},}\ }\href
  {\doibase 10.1063/1.3328107} {\bibfield  {journal} {\bibinfo  {journal} {J.
  Chem. Phys.}\ }\textbf {\bibinfo {volume} {132}},\ \bibinfo {pages} {081101}
  (\bibinfo {year} {2010})}\BibitemShut {NoStop}%
\bibitem [{\citenamefont {Song}\ and\ \citenamefont
  {Shi}(2015{\natexlab{a}})}]{JChemPhys.142.174103}%
  \BibitemOpen
  \bibfield  {author} {\bibinfo {author} {\bibfnamefont {L.}~\bibnamefont
  {Song}}\ and\ \bibinfo {author} {\bibfnamefont {Q.}~\bibnamefont {Shi}},\
  }\bibfield  {title} {\enquote {\bibinfo {title} {A new approach to calculate
  charge carrier transport mobility in organic molecular crystals from
  imaginary time path integral simulations},}\ }\href {\doibase
  10.1063/1.4919061} {\bibfield  {journal} {\bibinfo  {journal} {J. Chem.
  Phys.}\ }\textbf {\bibinfo {volume} {142}},\ \bibinfo {pages} {174103}
  (\bibinfo {year} {2015}{\natexlab{a}})}\BibitemShut {NoStop}%
\bibitem [{\citenamefont {Chen}\ \emph {et~al.}(2015)\citenamefont {Chen},
  \citenamefont {Zhao},\ and\ \citenamefont {Tanimura}}]{JPhysChemLett.6.3110}%
  \BibitemOpen
  \bibfield  {author} {\bibinfo {author} {\bibfnamefont {L.}~\bibnamefont
  {Chen}}, \bibinfo {author} {\bibfnamefont {Y.}~\bibnamefont {Zhao}}, \ and\
  \bibinfo {author} {\bibfnamefont {Y.}~\bibnamefont {Tanimura}},\ }\bibfield
  {title} {\enquote {\bibinfo {title} {Dynamics of a one-dimensional {H}olstein
  polaron with the hierarchical equations of motion approach},}\ }\href
  {\doibase 10.1021/acs.jpclett.5b01368} {\bibfield  {journal} {\bibinfo
  {journal} {J. Phys. Chem. Lett.}\ }\textbf {\bibinfo {volume} {6}},\ \bibinfo
  {pages} {3110--3115} (\bibinfo {year} {2015})}\BibitemShut {NoStop}%
\bibitem [{\citenamefont {Song}\ and\ \citenamefont
  {Shi}(2015{\natexlab{b}})}]{JChemPhys.143.194106}%
  \BibitemOpen
  \bibfield  {author} {\bibinfo {author} {\bibfnamefont {L.}~\bibnamefont
  {Song}}\ and\ \bibinfo {author} {\bibfnamefont {Q.}~\bibnamefont {Shi}},\
  }\bibfield  {title} {\enquote {\bibinfo {title} {Calculation of correlated
  initial state in the hierarchical equations of motion method using an
  imaginary time path integral approach},}\ }\href {\doibase 10.1063/1.4935799}
  {\bibfield  {journal} {\bibinfo  {journal} {J. Chem. Phys.}\ }\textbf
  {\bibinfo {volume} {143}},\ \bibinfo {pages} {194106} (\bibinfo {year}
  {2015}{\natexlab{b}})}\BibitemShut {NoStop}%
\bibitem [{\citenamefont {Kuhn}(1998)}]{kuhncontribution}%
  \BibitemOpen
  \bibfield  {author} {\bibinfo {author} {\bibfnamefont {T.}~\bibnamefont
  {Kuhn}},\ }\bibfield  {title} {\enquote {\bibinfo {title} {Density matrix
  theory of coherent ultrafast dynamics},}\ }in\ \href@noop {} {\emph {\bibinfo
  {booktitle} {Theory of Transport Properties of Semiconductor
  Nanostructures}}},\ \bibinfo {editor} {edited by\ \bibinfo {editor}
  {\bibfnamefont {E.}~\bibnamefont {Sch{\"{o}}l}}}\ (\bibinfo  {publisher}
  {Springer Science+Business Media},\ \bibinfo {address} {Dordrecht},\ \bibinfo
  {year} {1998})\BibitemShut {NoStop}%
\bibitem [{\citenamefont {Kira}\ and\ \citenamefont {Koch}(2012)}]{Kirabook}%
  \BibitemOpen
  \bibfield  {author} {\bibinfo {author} {\bibfnamefont {M.}~\bibnamefont
  {Kira}}\ and\ \bibinfo {author} {\bibfnamefont {S.W.}\ \bibnamefont {Koch}},\
  }\href@noop {} {\emph {\bibinfo {title} {Semiconductor Quantum Optics}}}\
  (\bibinfo  {publisher} {Cambridge University Press, New York},\ \bibinfo
  {year} {2012})\BibitemShut {NoStop}%
\bibitem [{\citenamefont {Dunn}\ \emph {et~al.}(2019)\citenamefont {Dunn},
  \citenamefont {Tempelaar},\ and\ \citenamefont
  {Reichman}}]{JChemPhys.150.184109}%
  \BibitemOpen
  \bibfield  {author} {\bibinfo {author} {\bibfnamefont {I.~S.}\ \bibnamefont
  {Dunn}}, \bibinfo {author} {\bibfnamefont {R.}~\bibnamefont {Tempelaar}}, \
  and\ \bibinfo {author} {\bibfnamefont {D.~R.}\ \bibnamefont {Reichman}},\
  }\bibfield  {title} {\enquote {\bibinfo {title} {Removing instabilities in
  the hierarchical equations of motion: Exact and approximate projection
  approaches},}\ }\href {\doibase 10.1063/1.5092616} {\bibfield  {journal}
  {\bibinfo  {journal} {J. Chem. Phys.}\ }\textbf {\bibinfo {volume} {150}},\
  \bibinfo {pages} {184109} (\bibinfo {year} {2019})}\BibitemShut {NoStop}%
\bibitem [{\citenamefont {Liu}\ \emph {et~al.}(2014)\citenamefont {Liu},
  \citenamefont {Zhu}, \citenamefont {Bai},\ and\ \citenamefont
  {Shi}}]{JChemPhys.140.134106}%
  \BibitemOpen
  \bibfield  {author} {\bibinfo {author} {\bibfnamefont {H.}~\bibnamefont
  {Liu}}, \bibinfo {author} {\bibfnamefont {L.}~\bibnamefont {Zhu}}, \bibinfo
  {author} {\bibfnamefont {S.}~\bibnamefont {Bai}}, \ and\ \bibinfo {author}
  {\bibfnamefont {Q.}~\bibnamefont {Shi}},\ }\bibfield  {title} {\enquote
  {\bibinfo {title} {Reduced quantum dynamics with arbitrary bath spectral
  densities: Hierarchical equations of motion based on several different bath
  decomposition schemes},}\ }\href {\doibase 10.1063/1.4870035} {\bibfield
  {journal} {\bibinfo  {journal} {J. Chem. Phys.}\ }\textbf {\bibinfo {volume}
  {140}},\ \bibinfo {pages} {134106} (\bibinfo {year} {2014})}\BibitemShut
  {NoStop}%
\bibitem [{\citenamefont {Seibt}\ and\ \citenamefont
  {Man{\v{c}}al}(2018)}]{ChemPhys.515.129}%
  \BibitemOpen
  \bibfield  {author} {\bibinfo {author} {\bibfnamefont {J.}~\bibnamefont
  {Seibt}}\ and\ \bibinfo {author} {\bibfnamefont {T.}~\bibnamefont
  {Man{\v{c}}al}},\ }\bibfield  {title} {\enquote {\bibinfo {title} {Treatment
  of {H}erzberg-{T}eller and non-{C}ondon effects in optical spectra with
  hierarchical equations of motion},}\ }\href {\doibase
  10.1016/j.chemphys.2018.08.026} {\bibfield  {journal} {\bibinfo  {journal}
  {Chem. Phys.}\ }\textbf {\bibinfo {volume} {515}},\ \bibinfo {pages}
  {129--140} (\bibinfo {year} {2018})}\BibitemShut {NoStop}%
\bibitem [{\citenamefont {Yan}\ \emph {et~al.}(2020)\citenamefont {Yan},
  \citenamefont {Xing},\ and\ \citenamefont {Shi}}]{JChemPhys.153.204109}%
  \BibitemOpen
  \bibfield  {author} {\bibinfo {author} {\bibfnamefont {Y.}~\bibnamefont
  {Yan}}, \bibinfo {author} {\bibfnamefont {T.}~\bibnamefont {Xing}}, \ and\
  \bibinfo {author} {\bibfnamefont {Q.}~\bibnamefont {Shi}},\ }\bibfield
  {title} {\enquote {\bibinfo {title} {A new method to improve the numerical
  stability of the hierarchical equations of motion for discrete harmonic
  oscillator modes},}\ }\href {\doibase 10.1063/5.0027962} {\bibfield
  {journal} {\bibinfo  {journal} {J. Chem. Phys.}\ }\textbf {\bibinfo {volume}
  {153}},\ \bibinfo {pages} {204109} (\bibinfo {year} {2020})}\BibitemShut
  {NoStop}%
\bibitem [{\citenamefont {Mahan}(2000)}]{Mahanbook}%
  \BibitemOpen
  \bibfield  {author} {\bibinfo {author} {\bibfnamefont {G.}~\bibnamefont
  {Mahan}},\ }\href@noop {} {\emph {\bibinfo {title} {Many-Particle Physics}}}\
  (\bibinfo  {publisher} {Kluwer Academic, New York},\ \bibinfo {year}
  {2000})\BibitemShut {NoStop}%
\bibitem [{com()}]{comment071021}%
  \BibitemOpen
  \href@noop {} {}\bibinfo {howpublished} {{See the Supplemental Material for
  more detailed discussions on (i) the definition of and interrelations between
  the greater and lesser Green's functions and the spectral function; (ii) the
  QMC method for thermodynamic expectation values; (iii) the weak-coupling
  limit of the HEOM method; (iv) the strong-coupling limit of the HEOM method;
  (v) the interpretation of the artificial broadening of spectral lines within
  the HEOM method; (vi) time- and frequency-domain HEOM-method results for
  different maximum hierarchy depths; (vii) HEOM and QMC data for the
  electronic momentum distribution; (viii) HEOM-method results in the adiabatic
  regime; (ix) the reduction of the statistical noise in the QMC data with
  increasing the sample size; (x) the HEOM and QMC imaginary-time correlation
  functions in the antiadiabatic and adiabatic regime.}}\BibitemShut {Stop}%
\bibitem [{\citenamefont {Tanimura}(2014)}]{JChemPhys.141.044114}%
  \BibitemOpen
  \bibfield  {author} {\bibinfo {author} {\bibfnamefont {Y.}~\bibnamefont
  {Tanimura}},\ }\bibfield  {title} {\enquote {\bibinfo {title} {Reduced
  hierarchical equations of motion in real and imaginary time: Correlated
  initial states and thermodynamic quantities},}\ }\href {\doibase
  10.1063/1.4890441} {\bibfield  {journal} {\bibinfo  {journal} {J. Chem.
  Phys.}\ }\textbf {\bibinfo {volume} {141}},\ \bibinfo {pages} {044114}
  (\bibinfo {year} {2014})}\BibitemShut {NoStop}%
\bibitem [{\citenamefont {Moix}\ \emph {et~al.}(2012)\citenamefont {Moix},
  \citenamefont {Zhao},\ and\ \citenamefont {Cao}}]{PhysRevB.85.115412}%
  \BibitemOpen
  \bibfield  {author} {\bibinfo {author} {\bibfnamefont {J.~M.}\ \bibnamefont
  {Moix}}, \bibinfo {author} {\bibfnamefont {Y.}~\bibnamefont {Zhao}}, \ and\
  \bibinfo {author} {\bibfnamefont {J.}~\bibnamefont {Cao}},\ }\bibfield
  {title} {\enquote {\bibinfo {title} {Equilibrium-reduced density matrix
  formulation: Influence of noise, disorder, and temperature on localization in
  excitonic systems},}\ }\href {\doibase 10.1103/PhysRevB.85.115412} {\bibfield
   {journal} {\bibinfo  {journal} {Phys. Rev. B}\ }\textbf {\bibinfo {volume}
  {85}},\ \bibinfo {pages} {115412} (\bibinfo {year} {2012})}\BibitemShut
  {NoStop}%
\bibitem [{\citenamefont {Zhang}\ \emph {et~al.}(2017)\citenamefont {Zhang},
  \citenamefont {Qiao}, \citenamefont {Xu}, \citenamefont {Zheng},\ and\
  \citenamefont {Yan}}]{JChemPhys.147.044105}%
  \BibitemOpen
  \bibfield  {author} {\bibinfo {author} {\bibfnamefont {H.-D.}\ \bibnamefont
  {Zhang}}, \bibinfo {author} {\bibfnamefont {Q.}~\bibnamefont {Qiao}},
  \bibinfo {author} {\bibfnamefont {R.-X.}\ \bibnamefont {Xu}}, \bibinfo
  {author} {\bibfnamefont {X.}~\bibnamefont {Zheng}}, \ and\ \bibinfo {author}
  {\bibfnamefont {Y.}~\bibnamefont {Yan}},\ }\bibfield  {title} {\enquote
  {\bibinfo {title} {Efficient steady-state solver for hierarchical quantum
  master equations},}\ }\href {\doibase 10.1063/1.4995424} {\bibfield
  {journal} {\bibinfo  {journal} {J. Chem. Phys.}\ }\textbf {\bibinfo {volume}
  {147}},\ \bibinfo {pages} {044105} (\bibinfo {year} {2017})}\BibitemShut
  {NoStop}%
\bibitem [{\citenamefont {Zhu}\ \emph {et~al.}(2012)\citenamefont {Zhu},
  \citenamefont {Liu}, \citenamefont {Xie},\ and\ \citenamefont
  {Shi}}]{JChemPhys.137.194106}%
  \BibitemOpen
  \bibfield  {author} {\bibinfo {author} {\bibfnamefont {L.}~\bibnamefont
  {Zhu}}, \bibinfo {author} {\bibfnamefont {H.}~\bibnamefont {Liu}}, \bibinfo
  {author} {\bibfnamefont {W.}~\bibnamefont {Xie}}, \ and\ \bibinfo {author}
  {\bibfnamefont {Q.}~\bibnamefont {Shi}},\ }\bibfield  {title} {\enquote
  {\bibinfo {title} {Explicit system-bath correlation calculated using the
  hierarchical equations of motion method},}\ }\href {\doibase
  10.1063/1.4766358} {\bibfield  {journal} {\bibinfo  {journal} {J. Chem.
  Phys.}\ }\textbf {\bibinfo {volume} {137}},\ \bibinfo {pages} {194106}
  (\bibinfo {year} {2012})}\BibitemShut {NoStop}%
\bibitem [{\citenamefont {Xu}\ \emph {et~al.}(2005)\citenamefont {Xu},
  \citenamefont {Cui}, \citenamefont {Li}, \citenamefont {Mo},\ and\
  \citenamefont {Yan}}]{JChemPhys.122.041103}%
  \BibitemOpen
  \bibfield  {author} {\bibinfo {author} {\bibfnamefont {R.-X.}\ \bibnamefont
  {Xu}}, \bibinfo {author} {\bibfnamefont {P.}~\bibnamefont {Cui}}, \bibinfo
  {author} {\bibfnamefont {X.-Q.}\ \bibnamefont {Li}}, \bibinfo {author}
  {\bibfnamefont {Y.}~\bibnamefont {Mo}}, \ and\ \bibinfo {author}
  {\bibfnamefont {Y.~J.}\ \bibnamefont {Yan}},\ }\bibfield  {title} {\enquote
  {\bibinfo {title} {Exact quantum master equation via the calculus on path
  integrals},}\ }\href {\doibase 10.1063/1.1850899} {\bibfield  {journal}
  {\bibinfo  {journal} {J. Chem. Phys.}\ }\textbf {\bibinfo {volume} {122}},\
  \bibinfo {pages} {041103} (\bibinfo {year} {2005})}\BibitemShut {NoStop}%
\bibitem [{\citenamefont {Lang}\ and\ \citenamefont
  {Firsov}(1962)}]{SovPhysJETP.16.1301}%
  \BibitemOpen
  \bibfield  {author} {\bibinfo {author} {\bibfnamefont {I.G.}\ \bibnamefont
  {Lang}}\ and\ \bibinfo {author} {\bibfnamefont {Yu.~A.}\ \bibnamefont
  {Firsov}},\ }\bibfield  {title} {\enquote {\bibinfo {title} {Kinetic theory
  of semiconductors with low mobility},}\ }\href
  {http://jetp.ras.ru/cgi-bin/e/index/e/16/5/p1301?a=list} {\bibfield
  {journal} {\bibinfo  {journal} {Sov. Phys. JETP}\ }\textbf {\bibinfo {volume}
  {16}},\ \bibinfo {pages} {1301--1312} (\bibinfo {year} {1962})}\BibitemShut
  {NoStop}%
\bibitem [{\citenamefont {May}\ and\ \citenamefont
  {K{\"{u}}hn}(2011)}]{May-Kuhn-book}%
  \BibitemOpen
  \bibfield  {author} {\bibinfo {author} {\bibfnamefont {V.}~\bibnamefont
  {May}}\ and\ \bibinfo {author} {\bibfnamefont {O.}~\bibnamefont
  {K{\"{u}}hn}},\ }\href@noop {} {\emph {\bibinfo {title} {Charge and Energy
  Transfer Dynamics in Molecular Systems}}},\ \bibinfo {edition} {3rd}\ ed.\
  (\bibinfo  {publisher} {WILEY-VCH Verlag GmbH \& Co. KGaA},\ \bibinfo
  {address} {Weinheim},\ \bibinfo {year} {2011})\BibitemShut {NoStop}%
\bibitem [{\citenamefont {Mukamel}(1995)}]{Mukamel-book}%
  \BibitemOpen
  \bibfield  {author} {\bibinfo {author} {\bibfnamefont {S.}~\bibnamefont
  {Mukamel}},\ }\href@noop {} {\emph {\bibinfo {title} {Principles of Nonlinear
  Optical Spectroscopy}}}\ (\bibinfo  {publisher} {Oxford University Press,
  Inc., New York},\ \bibinfo {year} {1995})\BibitemShut {NoStop}%
\bibitem [{\citenamefont {Lin}\ \emph {et~al.}(2001)\citenamefont {Lin},
  \citenamefont {Zong},\ and\ \citenamefont {Ceperley}}]{PhysRevE.64.016702}%
  \BibitemOpen
  \bibfield  {author} {\bibinfo {author} {\bibfnamefont {C.}~\bibnamefont
  {Lin}}, \bibinfo {author} {\bibfnamefont {F.~H.}\ \bibnamefont {Zong}}, \
  and\ \bibinfo {author} {\bibfnamefont {D.~M.}\ \bibnamefont {Ceperley}},\
  }\bibfield  {title} {\enquote {\bibinfo {title} {Twist-averaged boundary
  conditions in continuum quantum {M}onte {C}arlo algorithms},}\ }\href
  {\doibase 10.1103/PhysRevE.64.016702} {\bibfield  {journal} {\bibinfo
  {journal} {Phys. Rev. E}\ }\textbf {\bibinfo {volume} {64}},\ \bibinfo
  {pages} {016702} (\bibinfo {year} {2001})}\BibitemShut {NoStop}%
\bibitem [{\citenamefont {LeBlanc}\ \emph {et~al.}(2015)\citenamefont
  {LeBlanc}, \citenamefont {Antipov}, \citenamefont {Becca}, \citenamefont
  {Bulik}, \citenamefont {Chan}, \citenamefont {Chung}, \citenamefont {Deng},
  \citenamefont {Ferrero}, \citenamefont {Henderson}, \citenamefont
  {Jim\'enez-Hoyos}, \citenamefont {Kozik}, \citenamefont {Liu}, \citenamefont
  {Millis}, \citenamefont {Prokof'ev}, \citenamefont {Qin}, \citenamefont
  {Scuseria}, \citenamefont {Shi}, \citenamefont {Svistunov}, \citenamefont
  {Tocchio}, \citenamefont {Tupitsyn}, \citenamefont {White}, \citenamefont
  {Zhang}, \citenamefont {Zheng}, \citenamefont {Zhu},\ and\ \citenamefont
  {Gull}}]{PhysRevX.5.041041}%
  \BibitemOpen
  \bibfield  {author} {\bibinfo {author} {\bibfnamefont {J.~P.~F.}\
  \bibnamefont {LeBlanc}}, \bibinfo {author} {\bibfnamefont {A.~E.}\
  \bibnamefont {Antipov}}, \bibinfo {author} {\bibfnamefont {F.}~\bibnamefont
  {Becca}}, \bibinfo {author} {\bibfnamefont {I.~W.}\ \bibnamefont {Bulik}},
  \bibinfo {author} {\bibfnamefont {G.~K.-L.}\ \bibnamefont {Chan}}, \bibinfo
  {author} {\bibfnamefont {C.-M.}\ \bibnamefont {Chung}}, \bibinfo {author}
  {\bibfnamefont {Y.}~\bibnamefont {Deng}}, \bibinfo {author} {\bibfnamefont
  {M.}~\bibnamefont {Ferrero}}, \bibinfo {author} {\bibfnamefont {T.~M.}\
  \bibnamefont {Henderson}}, \bibinfo {author} {\bibfnamefont {C.~A.}\
  \bibnamefont {Jim\'enez-Hoyos}}, \bibinfo {author} {\bibfnamefont
  {E.}~\bibnamefont {Kozik}}, \bibinfo {author} {\bibfnamefont {X.-W.}\
  \bibnamefont {Liu}}, \bibinfo {author} {\bibfnamefont {A.~J.}\ \bibnamefont
  {Millis}}, \bibinfo {author} {\bibfnamefont {N.~V.}\ \bibnamefont
  {Prokof'ev}}, \bibinfo {author} {\bibfnamefont {M.}~\bibnamefont {Qin}},
  \bibinfo {author} {\bibfnamefont {G.~E.}\ \bibnamefont {Scuseria}}, \bibinfo
  {author} {\bibfnamefont {H.}~\bibnamefont {Shi}}, \bibinfo {author}
  {\bibfnamefont {B.~V.}\ \bibnamefont {Svistunov}}, \bibinfo {author}
  {\bibfnamefont {L.~F.}\ \bibnamefont {Tocchio}}, \bibinfo {author}
  {\bibfnamefont {I.~S.}\ \bibnamefont {Tupitsyn}}, \bibinfo {author}
  {\bibfnamefont {S.~R.}\ \bibnamefont {White}}, \bibinfo {author}
  {\bibfnamefont {S.}~\bibnamefont {Zhang}}, \bibinfo {author} {\bibfnamefont
  {B.-X.}\ \bibnamefont {Zheng}}, \bibinfo {author} {\bibfnamefont
  {Z.}~\bibnamefont {Zhu}}, \ and\ \bibinfo {author} {\bibfnamefont
  {E.}~\bibnamefont {Gull}} (\bibinfo {collaboration} {Simons Collaboration on
  the Many-Electron Problem}),\ }\bibfield  {title} {\enquote {\bibinfo {title}
  {Solutions of the two-dimensional {H}ubbard model: Benchmarks and results
  from a wide range of numerical algorithms},}\ }\href {\doibase
  10.1103/PhysRevX.5.041041} {\bibfield  {journal} {\bibinfo  {journal} {Phys.
  Rev. X}\ }\textbf {\bibinfo {volume} {5}},\ \bibinfo {pages} {041041}
  (\bibinfo {year} {2015})}\BibitemShut {NoStop}%
\bibitem [{\citenamefont {Karakuzu}\ \emph {et~al.}(2017)\citenamefont
  {Karakuzu}, \citenamefont {Tocchio}, \citenamefont {Sorella},\ and\
  \citenamefont {Becca}}]{PhysRevB.96.205145}%
  \BibitemOpen
  \bibfield  {author} {\bibinfo {author} {\bibfnamefont {S.}~\bibnamefont
  {Karakuzu}}, \bibinfo {author} {\bibfnamefont {L.~F.}\ \bibnamefont
  {Tocchio}}, \bibinfo {author} {\bibfnamefont {S.}~\bibnamefont {Sorella}}, \
  and\ \bibinfo {author} {\bibfnamefont {F.}~\bibnamefont {Becca}},\ }\bibfield
   {title} {\enquote {\bibinfo {title} {Superconductivity, charge-density
  waves, antiferromagnetism, and phase separation in the {H}ubbard-{H}olstein
  model},}\ }\href {\doibase 10.1103/PhysRevB.96.205145} {\bibfield  {journal}
  {\bibinfo  {journal} {Phys. Rev. B}\ }\textbf {\bibinfo {volume} {96}},\
  \bibinfo {pages} {205145} (\bibinfo {year} {2017})}\BibitemShut {NoStop}%
\bibitem [{\citenamefont {Hohenadler}\ \emph {et~al.}(2003)\citenamefont
  {Hohenadler}, \citenamefont {Aichhorn},\ and\ \citenamefont {von~der
  Linden}}]{PhysRevB.68.184304}%
  \BibitemOpen
  \bibfield  {author} {\bibinfo {author} {\bibfnamefont {M.}~\bibnamefont
  {Hohenadler}}, \bibinfo {author} {\bibfnamefont {M.}~\bibnamefont
  {Aichhorn}}, \ and\ \bibinfo {author} {\bibfnamefont {W.}~\bibnamefont
  {von~der Linden}},\ }\bibfield  {title} {\enquote {\bibinfo {title} {Spectral
  function of electron-phonon models by cluster perturbation theory},}\ }\href
  {\doibase 10.1103/PhysRevB.68.184304} {\bibfield  {journal} {\bibinfo
  {journal} {Phys. Rev. B}\ }\textbf {\bibinfo {volume} {68}},\ \bibinfo
  {pages} {184304} (\bibinfo {year} {2003})}\BibitemShut {NoStop}%
\bibitem [{\citenamefont {Mitrić}\ \emph {et~al.}(2021)\citenamefont
  {Mitrić}, \citenamefont {Janković}, \citenamefont {Vukmirović},\ and\
  \citenamefont {Tanasković}}]{2112.15542}%
  \BibitemOpen
  \bibfield  {author} {\bibinfo {author} {\bibfnamefont {P.}~\bibnamefont
  {Mitrić}}, \bibinfo {author} {\bibfnamefont {V.}~\bibnamefont {Janković}},
  \bibinfo {author} {\bibfnamefont {N.}~\bibnamefont {Vukmirović}}, \ and\
  \bibinfo {author} {\bibfnamefont {D.}~\bibnamefont {Tanasković}},\
  }\href@noop {} {\enquote {\bibinfo {title} {Spectral functions of the
  {Holstein} polaron: exact and approximate solutions},}\ } (\bibinfo {year}
  {2021}),\ \Eprint {http://arxiv.org/abs/2112.15542} {arXiv:2112.15542
  [cond-mat.str-el]} \BibitemShut {NoStop}%
\bibitem [{\citenamefont {Kessing}\ \emph {et~al.}(2021)\citenamefont
  {Kessing}, \citenamefont {Yang}, \citenamefont {Manmana},\ and\ \citenamefont
  {Cao}}]{2111.06137}%
  \BibitemOpen
  \bibfield  {author} {\bibinfo {author} {\bibfnamefont {R.~Kevin}\
  \bibnamefont {Kessing}}, \bibinfo {author} {\bibfnamefont {P.-Y.}\
  \bibnamefont {Yang}}, \bibinfo {author} {\bibfnamefont {S.~R.}\ \bibnamefont
  {Manmana}}, \ and\ \bibinfo {author} {\bibfnamefont {J.}~\bibnamefont
  {Cao}},\ }\href@noop {} {\enquote {\bibinfo {title} {Long-range
  non-equilibrium coherent tunneling induced by fractional vibronic
  resonances},}\ } (\bibinfo {year} {2021}),\ \Eprint
  {http://arxiv.org/abs/2111.06137} {arXiv:2111.06137 [quant-ph]} \BibitemShut
  {NoStop}%
\bibitem [{\citenamefont {Vu\ifmmode \check{c}\else \v{c}\fi{}i\ifmmode
  \check{c}\else \v{c}\fi{}evi\ifmmode~\acute{c}\else \'{c}\fi{}}\ \emph
  {et~al.}(2019)\citenamefont {Vu\ifmmode \check{c}\else \v{c}\fi{}i\ifmmode
  \check{c}\else \v{c}\fi{}evi\ifmmode~\acute{c}\else \'{c}\fi{}},
  \citenamefont {Kokalj}, \citenamefont {\ifmmode~\check{Z}\else
  \v{Z}\fi{}itko}, \citenamefont {Wentzell}, \citenamefont
  {Tanaskovi\ifmmode~\acute{c}\else \'{c}\fi{}},\ and\ \citenamefont
  {Mravlje}}]{PhysRevLett.123.036601}%
  \BibitemOpen
  \bibfield  {author} {\bibinfo {author} {\bibfnamefont {J.}~\bibnamefont
  {Vu\ifmmode \check{c}\else \v{c}\fi{}i\ifmmode \check{c}\else
  \v{c}\fi{}evi\ifmmode~\acute{c}\else \'{c}\fi{}}}, \bibinfo {author}
  {\bibfnamefont {J.}~\bibnamefont {Kokalj}}, \bibinfo {author} {\bibfnamefont
  {R.}~\bibnamefont {\ifmmode~\check{Z}\else \v{Z}\fi{}itko}}, \bibinfo
  {author} {\bibfnamefont {N.}~\bibnamefont {Wentzell}}, \bibinfo {author}
  {\bibfnamefont {D.}~\bibnamefont {Tanaskovi\ifmmode~\acute{c}\else
  \'{c}\fi{}}}, \ and\ \bibinfo {author} {\bibfnamefont {J.}~\bibnamefont
  {Mravlje}},\ }\bibfield  {title} {\enquote {\bibinfo {title} {Conductivity in
  the square lattice {H}ubbard model at high temperatures: {I}mportance of
  vertex corrections},}\ }\href {\doibase 10.1103/PhysRevLett.123.036601}
  {\bibfield  {journal} {\bibinfo  {journal} {Phys. Rev. Lett.}\ }\textbf
  {\bibinfo {volume} {123}},\ \bibinfo {pages} {036601} (\bibinfo {year}
  {2019})}\BibitemShut {NoStop}%
\bibitem [{\citenamefont {Ponc{\'{e}}}\ \emph {et~al.}(2020)\citenamefont
  {Ponc{\'{e}}}, \citenamefont {Li}, \citenamefont {Reichardt},\ and\
  \citenamefont {Giustino}}]{RepProgPhys.83.036501}%
  \BibitemOpen
  \bibfield  {author} {\bibinfo {author} {\bibfnamefont {S.}~\bibnamefont
  {Ponc{\'{e}}}}, \bibinfo {author} {\bibfnamefont {W.}~\bibnamefont {Li}},
  \bibinfo {author} {\bibfnamefont {S.}~\bibnamefont {Reichardt}}, \ and\
  \bibinfo {author} {\bibfnamefont {F.}~\bibnamefont {Giustino}},\ }\bibfield
  {title} {\enquote {\bibinfo {title} {First-principles calculations of charge
  carrier mobility and conductivity in bulk semiconductors and two-dimensional
  materials},}\ }\href {\doibase 10.1088/1361-6633/ab6a43} {\bibfield
  {journal} {\bibinfo  {journal} {Rep. Prog. Phys.}\ }\textbf {\bibinfo
  {volume} {83}},\ \bibinfo {pages} {036501} (\bibinfo {year}
  {2020})}\BibitemShut {NoStop}%
\bibitem [{\citenamefont {Hsieh}\ and\ \citenamefont
  {Cao}(2018{\natexlab{a}})}]{JChemPhys.148.014103}%
  \BibitemOpen
  \bibfield  {author} {\bibinfo {author} {\bibfnamefont {C.-Y.}\ \bibnamefont
  {Hsieh}}\ and\ \bibinfo {author} {\bibfnamefont {J.}~\bibnamefont {Cao}},\
  }\bibfield  {title} {\enquote {\bibinfo {title} {A unified stochastic
  formulation of dissipative quantum dynamics. {I}. {G}eneralized hierarchical
  equations},}\ }\href {\doibase 10.1063/1.5018725} {\bibfield  {journal}
  {\bibinfo  {journal} {J. Chem. Phys.}\ }\textbf {\bibinfo {volume} {148}},\
  \bibinfo {pages} {014103} (\bibinfo {year} {2018}{\natexlab{a}})}\BibitemShut
  {NoStop}%
\bibitem [{\citenamefont {Hsieh}\ and\ \citenamefont
  {Cao}(2018{\natexlab{b}})}]{JChemPhys.148.014104}%
  \BibitemOpen
  \bibfield  {author} {\bibinfo {author} {\bibfnamefont {C.-Y.}\ \bibnamefont
  {Hsieh}}\ and\ \bibinfo {author} {\bibfnamefont {J.}~\bibnamefont {Cao}},\
  }\bibfield  {title} {\enquote {\bibinfo {title} {A unified stochastic
  formulation of dissipative quantum dynamics. {II}. {B}eyond linear response
  of spin baths},}\ }\href {\doibase 10.1063/1.5018726} {\bibfield  {journal}
  {\bibinfo  {journal} {J. Chem. Phys.}\ }\textbf {\bibinfo {volume} {148}},\
  \bibinfo {pages} {014104} (\bibinfo {year} {2018}{\natexlab{b}})}\BibitemShut
  {NoStop}%
\bibitem [{\citenamefont {Loos}\ \emph {et~al.}(2006)\citenamefont {Loos},
  \citenamefont {Hohenadler}, \citenamefont {Alvermann},\ and\ \citenamefont
  {Fehske}}]{JPhysCondensMatter.18.7299}%
  \BibitemOpen
  \bibfield  {author} {\bibinfo {author} {\bibfnamefont {J.}~\bibnamefont
  {Loos}}, \bibinfo {author} {\bibfnamefont {M.}~\bibnamefont {Hohenadler}},
  \bibinfo {author} {\bibfnamefont {A.}~\bibnamefont {Alvermann}}, \ and\
  \bibinfo {author} {\bibfnamefont {H.}~\bibnamefont {Fehske}},\ }\bibfield
  {title} {\enquote {\bibinfo {title} {Phonon spectral function of the
  {H}olstein polaron},}\ }\href {\doibase 10.1088/0953-8984/18/31/023}
  {\bibfield  {journal} {\bibinfo  {journal} {J. Phys.: Condens. Matter}\
  }\textbf {\bibinfo {volume} {18}},\ \bibinfo {pages} {7299--7312} (\bibinfo
  {year} {2006})}\BibitemShut {NoStop}%
\bibitem [{\citenamefont {Vidmar}\ \emph {et~al.}(2010)\citenamefont {Vidmar},
  \citenamefont {Bon\ifmmode~\check{c}\else \v{c}\fi{}a},\ and\ \citenamefont
  {Trugman}}]{PhysRevB.82.104304}%
  \BibitemOpen
  \bibfield  {author} {\bibinfo {author} {\bibfnamefont {L.}~\bibnamefont
  {Vidmar}}, \bibinfo {author} {\bibfnamefont {J.}~\bibnamefont
  {Bon\ifmmode~\check{c}\else \v{c}\fi{}a}}, \ and\ \bibinfo {author}
  {\bibfnamefont {S.~A.}\ \bibnamefont {Trugman}},\ }\bibfield  {title}
  {\enquote {\bibinfo {title} {Emergence of states in the phonon spectral
  function of the {H}olstein polaron below and above the one-phonon
  continuum},}\ }\href {\doibase 10.1103/PhysRevB.82.104304} {\bibfield
  {journal} {\bibinfo  {journal} {Phys. Rev. B}\ }\textbf {\bibinfo {volume}
  {82}},\ \bibinfo {pages} {104304} (\bibinfo {year} {2010})}\BibitemShut
  {NoStop}%
\bibitem [{\citenamefont {Choro{\v{s}}ajev}\ \emph {et~al.}(2016)\citenamefont
  {Choro{\v{s}}ajev}, \citenamefont {Rancova},\ and\ \citenamefont
  {Abramavicius}}]{PhysChemChemPhys.18.7966}%
  \BibitemOpen
  \bibfield  {author} {\bibinfo {author} {\bibfnamefont {V.}~\bibnamefont
  {Choro{\v{s}}ajev}}, \bibinfo {author} {\bibfnamefont {O.}~\bibnamefont
  {Rancova}}, \ and\ \bibinfo {author} {\bibfnamefont {D.}~\bibnamefont
  {Abramavicius}},\ }\bibfield  {title} {\enquote {\bibinfo {title} {Polaronic
  effects at finite temperatures in the {B}850 ring of the {LH}2 complex},}\
  }\href {\doibase 10.1039/C5CP06871A} {\bibfield  {journal} {\bibinfo
  {journal} {Phys. Chem. Chem. Phys.}\ }\textbf {\bibinfo {volume} {18}},\
  \bibinfo {pages} {7966--7977} (\bibinfo {year} {2016})}\BibitemShut {NoStop}%
\bibitem [{\citenamefont {Axt}\ and\ \citenamefont
  {Mukamel}(1998)}]{RevModPhys.70.145}%
  \BibitemOpen
  \bibfield  {author} {\bibinfo {author} {\bibfnamefont {V.~M.}\ \bibnamefont
  {Axt}}\ and\ \bibinfo {author} {\bibfnamefont {S.}~\bibnamefont {Mukamel}},\
  }\bibfield  {title} {\enquote {\bibinfo {title} {Nonlinear optics of
  semiconductor and molecular nanostructures; a common perspective},}\ }\href
  {\doibase 10.1103/RevModPhys.70.145} {\bibfield  {journal} {\bibinfo
  {journal} {Rev. Mod. Phys.}\ }\textbf {\bibinfo {volume} {70}},\ \bibinfo
  {pages} {145--174} (\bibinfo {year} {1998})}\BibitemShut {NoStop}%
\bibitem [{\citenamefont {Berciu}(2007)}]{PhysRevB.75.081101}%
  \BibitemOpen
  \bibfield  {author} {\bibinfo {author} {\bibfnamefont {M.}~\bibnamefont
  {Berciu}},\ }\bibfield  {title} {\enquote {\bibinfo {title} {Exact {G}reen's
  functions for the two-site {H}ubbard-{H}olstein {H}amiltonian},}\ }\href
  {\doibase 10.1103/PhysRevB.75.081101} {\bibfield  {journal} {\bibinfo
  {journal} {Phys. Rev. B}\ }\textbf {\bibinfo {volume} {75}},\ \bibinfo
  {pages} {081101} (\bibinfo {year} {2007})}\BibitemShut {NoStop}%
\bibitem [{\citenamefont {Nakamura}\ and\ \citenamefont
  {Tanimura}(2021)}]{JChemPhys.155.064106}%
  \BibitemOpen
  \bibfield  {author} {\bibinfo {author} {\bibfnamefont {K.}~\bibnamefont
  {Nakamura}}\ and\ \bibinfo {author} {\bibfnamefont {Y.}~\bibnamefont
  {Tanimura}},\ }\bibfield  {title} {\enquote {\bibinfo {title} {Optical
  response of laser-driven charge-transfer complex described by
  {H}olstein–{H}ubbard model coupled to heat baths: {H}ierarchical equations
  of motion approach},}\ }\href {\doibase 10.1063/5.0060208} {\bibfield
  {journal} {\bibinfo  {journal} {J. Chem. Phys.}\ }\textbf {\bibinfo {volume}
  {155}},\ \bibinfo {pages} {064106} (\bibinfo {year} {2021})}\BibitemShut
  {NoStop}%
\bibitem [{\citenamefont {Gelzinis}\ \emph {et~al.}(2011)\citenamefont
  {Gelzinis}, \citenamefont {Abramavicius},\ and\ \citenamefont
  {Valkunas}}]{PhysRevB.84.245430}%
  \BibitemOpen
  \bibfield  {author} {\bibinfo {author} {\bibfnamefont {A.}~\bibnamefont
  {Gelzinis}}, \bibinfo {author} {\bibfnamefont {D.}~\bibnamefont
  {Abramavicius}}, \ and\ \bibinfo {author} {\bibfnamefont {L.}~\bibnamefont
  {Valkunas}},\ }\bibfield  {title} {\enquote {\bibinfo {title}
  {Non-{M}arkovian effects in time-resolved fluorescence spectrum of molecular
  aggregates: {T}racing polaron formation},}\ }\href {\doibase
  10.1103/PhysRevB.84.245430} {\bibfield  {journal} {\bibinfo  {journal} {Phys.
  Rev. B}\ }\textbf {\bibinfo {volume} {84}},\ \bibinfo {pages} {245430}
  (\bibinfo {year} {2011})}\BibitemShut {NoStop}%
\bibitem [{\citenamefont {Zurek}(2003)}]{RevModPhys.75.715}%
  \BibitemOpen
  \bibfield  {author} {\bibinfo {author} {\bibfnamefont {W.~H.}\ \bibnamefont
  {Zurek}},\ }\bibfield  {title} {\enquote {\bibinfo {title} {Decoherence,
  einselection, and the quantum origins of the classical},}\ }\href {\doibase
  10.1103/RevModPhys.75.715} {\bibfield  {journal} {\bibinfo  {journal} {Rev.
  Mod. Phys.}\ }\textbf {\bibinfo {volume} {75}},\ \bibinfo {pages} {715--775}
  (\bibinfo {year} {2003})}\BibitemShut {NoStop}%
\bibitem [{\citenamefont {Schlosshauer}(2007)}]{Schlosshauer-book}%
  \BibitemOpen
  \bibfield  {author} {\bibinfo {author} {\bibfnamefont {M.}~\bibnamefont
  {Schlosshauer}},\ }\href@noop {} {\emph {\bibinfo {title} {Decoherence and
  the Quantum-To-Classical Transition}}}\ (\bibinfo  {publisher}
  {Springer-Verlag Berlin Heidelberg},\ \bibinfo {year} {2007})\BibitemShut
  {NoStop}%
\bibitem [{\citenamefont {Yan}\ \emph {et~al.}(2021)\citenamefont {Yan},
  \citenamefont {Xu}, \citenamefont {Li},\ and\ \citenamefont
  {Shi}}]{JChemPhys.154.194104}%
  \BibitemOpen
  \bibfield  {author} {\bibinfo {author} {\bibfnamefont {Y.}~\bibnamefont
  {Yan}}, \bibinfo {author} {\bibfnamefont {M.}~\bibnamefont {Xu}}, \bibinfo
  {author} {\bibfnamefont {T.}~\bibnamefont {Li}}, \ and\ \bibinfo {author}
  {\bibfnamefont {Q.}~\bibnamefont {Shi}},\ }\bibfield  {title} {\enquote
  {\bibinfo {title} {Efficient propagation of the hierarchical equations of
  motion using the {T}ucker and hierarchical {T}ucker tensors},}\ }\href
  {\doibase 10.1063/5.0050720} {\bibfield  {journal} {\bibinfo  {journal} {J.
  Chem. Phys.}\ }\textbf {\bibinfo {volume} {154}},\ \bibinfo {pages} {194104}
  (\bibinfo {year} {2021})}\BibitemShut {NoStop}%
\bibitem [{\citenamefont {Krotz}\ and\ \citenamefont
  {Tempelaar}(2022)}]{JChemPhys.156.024105}%
  \BibitemOpen
  \bibfield  {author} {\bibinfo {author} {\bibfnamefont {A.}~\bibnamefont
  {Krotz}}\ and\ \bibinfo {author} {\bibfnamefont {R.}~\bibnamefont
  {Tempelaar}},\ }\bibfield  {title} {\enquote {\bibinfo {title} {A
  reciprocal-space formulation of surface hopping},}\ }\href {\doibase
  10.1063/5.0076070} {\bibfield  {journal} {\bibinfo  {journal} {J. Chem.
  Phys.}\ }\textbf {\bibinfo {volume} {156}},\ \bibinfo {pages} {024105}
  (\bibinfo {year} {2022})}\BibitemShut {NoStop}%
\bibitem [{\citenamefont {Collini}\ and\ \citenamefont
  {Scholes}(2009)}]{Science.323.369}%
  \BibitemOpen
  \bibfield  {author} {\bibinfo {author} {\bibfnamefont {E.}~\bibnamefont
  {Collini}}\ and\ \bibinfo {author} {\bibfnamefont {G.~D.}\ \bibnamefont
  {Scholes}},\ }\bibfield  {title} {\enquote {\bibinfo {title} {Coherent
  intrachain energy migration in a conjugated polymer at room temperature},}\
  }\href {\doibase 10.1126/science.1164016} {\bibfield  {journal} {\bibinfo
  {journal} {Science}\ }\textbf {\bibinfo {volume} {323}},\ \bibinfo {pages}
  {369--373} (\bibinfo {year} {2009})}\BibitemShut {NoStop}%
\end{thebibliography}%


\begin{thebibliography}{3}%
\makeatletter
\providecommand \@ifxundefined [1]{%
 \@ifx{#1\undefined}
}%
\providecommand \@ifnum [1]{%
 \ifnum #1\expandafter \@firstoftwo
 \else \expandafter \@secondoftwo
 \fi
}%
\providecommand \@ifx [1]{%
 \ifx #1\expandafter \@firstoftwo
 \else \expandafter \@secondoftwo
 \fi
}%
\providecommand \natexlab [1]{#1}%
\providecommand \enquote  [1]{``#1''}%
\providecommand \bibnamefont  [1]{#1}%
\providecommand \bibfnamefont [1]{#1}%
\providecommand \citenamefont [1]{#1}%
\providecommand \href@noop [0]{\@secondoftwo}%
\providecommand \href [0]{\begingroup \@sanitize@url \@href}%
\providecommand \@href[1]{\@@startlink{#1}\@@href}%
\providecommand \@@href[1]{\endgroup#1\@@endlink}%
\providecommand \@sanitize@url [0]{\catcode `\\12\catcode `\$12\catcode
  `\&12\catcode `\#12\catcode `\^12\catcode `\_12\catcode `\%12\relax}%
\providecommand \@@startlink[1]{}%
\providecommand \@@endlink[0]{}%
\providecommand \url  [0]{\begingroup\@sanitize@url \@url }%
\providecommand \@url [1]{\endgroup\@href {#1}{\urlprefix }}%
\providecommand \urlprefix  [0]{URL }%
\providecommand \Eprint [0]{\href }%
\providecommand \doibase [0]{http://dx.doi.org/}%
\providecommand \selectlanguage [0]{\@gobble}%
\providecommand \bibinfo  [0]{\@secondoftwo}%
\providecommand \bibfield  [0]{\@secondoftwo}%
\providecommand \translation [1]{[#1]}%
\providecommand \BibitemOpen [0]{}%
\providecommand \bibitemStop [0]{}%
\providecommand \bibitemNoStop [0]{.\EOS\space}%
\providecommand \EOS [0]{\spacefactor3000\relax}%
\providecommand \BibitemShut  [1]{\csname bibitem#1\endcsname}%
\let\auto@bib@innerbib\@empty
\bibitem [{\citenamefont {Mahan}(2000)}]{Mahanbook}%
  \BibitemOpen
  \bibfield  {author} {\bibinfo {author} {\bibfnamefont {G.}~\bibnamefont
  {Mahan}},\ }\href@noop {} {\emph {\bibinfo {title} {Many-Particle Physics}}}\
  (\bibinfo  {publisher} {Kluwer Academic, New York},\ \bibinfo {year}
  {2000})\BibitemShut {NoStop}%
\bibitem [{\citenamefont {Gradshteyn}\ and\ \citenamefont
  {Ryzhik}(2007)}]{Gradshteyn_Ryzhik}%
  \BibitemOpen
  \bibfield  {author} {\bibinfo {author} {\bibfnamefont {I.~S.}\ \bibnamefont
  {Gradshteyn}}\ and\ \bibinfo {author} {\bibfnamefont {I.~M.}\ \bibnamefont
  {Ryzhik}},\ }\href@noop {} {\emph {\bibinfo {title} {Table of Integrals,
  Series, and Products}}}\ (\bibinfo  {publisher} {Elsevier Academic Press},\
  \bibinfo {year} {2007})\BibitemShut {NoStop}%
\bibitem [{\citenamefont {Liu}\ \emph {et~al.}(2014)\citenamefont {Liu},
  \citenamefont {Zhu}, \citenamefont {Bai},\ and\ \citenamefont
  {Shi}}]{JChemPhys.140.134106}%
  \BibitemOpen
  \bibfield  {author} {\bibinfo {author} {\bibfnamefont {H.}~\bibnamefont
  {Liu}}, \bibinfo {author} {\bibfnamefont {L.}~\bibnamefont {Zhu}}, \bibinfo
  {author} {\bibfnamefont {S.}~\bibnamefont {Bai}}, \ and\ \bibinfo {author}
  {\bibfnamefont {Q.}~\bibnamefont {Shi}},\ }\bibfield  {title} {\enquote
  {\bibinfo {title} {Reduced quantum dynamics with arbitrary bath spectral
  densities: Hierarchical equations of motion based on several different bath
  decomposition schemes},}\ }\href {\doibase 10.1063/1.4870035} {\bibfield
  {journal} {\bibinfo  {journal} {J. Chem. Phys.}\ }\textbf {\bibinfo {volume}
  {140}},\ \bibinfo {pages} {134106} (\bibinfo {year} {2014})}\BibitemShut
  {NoStop}%
\end{thebibliography}%

\end{document}


\title{Supplemental Material for\\Spectral and thermodynamic properties of the Holstein polaron: Hierarchical equations of motion approach}
\author{Veljko Jankovi\'c} \email{veljko.jankovic@ipb.ac.rs}
\affiliation{Institute of Physics Belgrade, University of Belgrade, 
Pregrevica 118, 11080 Belgrade, Serbia}
\author{Nenad Vukmirovi\'c} \email{nenad.vukmirovic@ipb.ac.rs}
\affiliation{Institute of Physics Belgrade, University of Belgrade, 
Pregrevica 118, 11080 Belgrade, Serbia}

\maketitle{}

\section{Green's functions and spectral function}
The fluctuation--dissipation theorem for single-particle Green's functions states that the spectral function, which is defined as
\begin{equation}
 A(k,\omega)=-\frac{1}{\pi}\mathrm{Im}\:G^R(k,\omega),
\end{equation}
is related to the greater and lesser Green's functions as follows~\cite{Mahanbook}
\begin{equation}
 A(k,\omega)=-\frac{1}{2\pi}\mathrm{Im}\:G^>(k,\omega)\left(1+e^{-\beta\hbar\omega}\right),
\end{equation}
\begin{equation}
 A(k,\omega)=\frac{1}{2\pi}\mathrm{Im}\:G^<(k,\omega)\left(1+e^{\beta\hbar\omega}\right).
\end{equation}
The Lehmann representation of $G^>(k,\omega)$ reads as
\begin{equation}
 G^>(k,\omega)=-\frac{2\pi i}{Z}\sum_{nm}e^{-\beta E_n}\:\delta(\omega+\omega_n-\omega_m)\left|\left\langle n\left|c_k\right|m\right\rangle\right|^2,
\end{equation}
where state $|m\rangle$ belongs to the single-excitation subspace, while state $|n\rangle$ belongs to the zero-excitation subspace. It is thus advantageous to separate out the vertical excitation energy $\varepsilon_e$ from $\omega_n$ by $\omega_n=\varepsilon_e/\hbar+\Delta\omega_n$ and to shift the zero of the frequency axis to $\varepsilon_e/\hbar$ by defining $G_\mathrm{shifted}^>(k,\omega)=G^>(\omega+\varepsilon_e/\hbar)$. The corresponding spectral function $A_\mathrm{shifted}(k,\omega)=A(k,\omega+\varepsilon_e/\hbar)$ then reads as
\begin{equation}
 {A}_\mathrm{shifted}(k,\omega)=-\frac{1}{2\pi}\mathrm{Im}\:{G}_\mathrm{shifted}^>(k,\omega)\left(1+e^{-\beta\hbar\omega}e^{-\beta\varepsilon_e}\right)\approx-\frac{1}{2\pi}\mathrm{Im}\:{G}_\mathrm{shifted}^>(k,\omega)
\end{equation}
under the physically plausible assumption that $\beta\varepsilon_e\gg 1$.

Using similar ideas, the Lehmann representation of $G^<(k,\omega)$ is transformed as follows
\begin{equation}
 G^<(k,\omega)=\frac{2\pi i}{Z}\:e^{-\beta\varepsilon_e}\sum_{nm}e^{-\beta\hbar\Delta\omega_m}\:\delta(\omega-\varepsilon_e/\hbar+\omega_n-\Delta\omega_m)\left|\left\langle n\left|c_k\right|m\right\rangle\right|^2.
\end{equation}
Defining
\begin{equation}
 G_\mathrm{shifted}^<(k,\omega)=e^{\beta\varepsilon_e}\:G^<(k,\omega+\varepsilon_e/\hbar),
\end{equation}
and employing $\beta\varepsilon_e\gg 1$, we finally obtain that
\begin{equation}
 A_\mathrm{shifted}(k,\omega)=\frac{1}{2\pi}\mathrm{Im}\:{G}_\mathrm{shifted}^<(k,\omega)\cdot e^{\beta\hbar\omega}.
\end{equation}
In the main body of the manuscript, we will not insist on notation $G_\mathrm{shifted}^{>/<}(k,\omega),A_\mathrm{shifted}(k,\omega)$, which simply shifts the origin of the energy scale from the energy of the unexcited state to the vertical excitation energy of the singly-excited state. The latter choice is far more common in the literature.

\newpage

\section{QMC method for thermodynamic expectation values}

\subsection{Operators that depend on electronic coordinates only}
We now describe the calculation of the expectation value of the operators that depend on electronic coordinates only, such as the free electronic operator $H_{\mathrm{e}}$ and the operator $n_k=\ket{k}\bra{k}$ of the number of electrons of momentum $k$.
The expectation value of such an operator $O$ is given as
\begin{equation}\label{eq:O9321}
 \expval{O}=
 \frac{
 \sum_{ab}\int\dd{\qty{x}}
 \mel{a\qty{x}}{e^{-{\beta}H}}{b\qty{x}}
 \mel{b}{O}{a} 
 }
 {
 \sum_{ab}\int\dd{\qty{x}}
 \mel{a\qty{x}}{e^{-{\beta}H}}{b\qty{x}}
 \delta_{ab}
 }
\end{equation}
To evaluate the integrals, we express $\mel{a\qty{x}}{e^{-{\beta}H}}{b\qty{x}}$ using Eq.~(\textcolor{red}{63}) in the main part of the paper and we analytically calculate the Gaussian integrals over the phononic coordinates using Eq.~(\textcolor{red}{68}) in the main part of the paper.

Equation~\eqref{eq:O9321} then takes the form
\begin{equation}
 \expval{O}=\frac{\mathcal{N}_1}{\mathcal{D}_1}   
\end{equation}
with
\begin{equation}
 \mathcal{N}_1=
 \sum_{ab} \sum_{j_1\ldots j_{K-1}}
f\qty(j_1)
f\qty(j_2)
\ldots
f\qty(j_{K-1})
  f\qty(a-b-\sum_{i=1}^{K-1}j_i)
e^{\frac{1}{2}\vb{b}\cdot \vb{A}^{-1} \cdot \vb{b}}
 \mel{b}{O}{a} 
\end{equation}
and
\begin{equation}
 \mathcal{D}_1=\sum_{ab}\sum_{j_1\ldots j_{K-1}}
f\qty(j_1)
f\qty(j_2)
\ldots
f\qty(j_{K-1})
  f\qty(a-b-\sum_{i=1}^{K-1}j_i)
e^{\frac{1}{2}\vb{b}\cdot \vb{A}^{-1} \cdot \vb{b}}
 \delta_{ab}. 
\end{equation}
We then perform Monte Carlo summation in the same way as for the correlation function.

The results presented in Fig.~\textcolor{red}{2(a)} and \textcolor{red}{2(c)} of the paper were obtained with $10^7$ samples and $K=80$.

The results presented in Figs.~\ref{Fig:Fig2a_SI_T_1_4e-1} and \ref{Fig:Fig6a_SI_gamma_1_2} in this Supplemental Material were obtained with $10^5$ samples and $K=80$, except for Fig.~\ref{Fig:Fig6a_SI_gamma_1_2}(b) where $K=200$.

\subsection{Interaction energy and fermion-boson correlation function}
Next, we describe Monte Carlo calculation of averages of the operator of electron-phonon interaction $H_{\mathrm{e-ph}}$ and the fermion-boson correlation function $\mathcal{C}\qty(l)=\sum_i \ket{i}\bra{i}\qty(b_{i+l}^\dagger+b_{i+l})$.
Expectation values of these operators read
\begin{equation}
 \expval{H_{\mathrm{e-ph}}}= 
 \frac{
 \sum_{a}\int\dd{\qty{x}}
 \mel{a\qty{x}}{e^{-{\beta}H}}{a\qty{x}}
 x_a \gamma\sqrt{\frac{2m\omega_0}{\hbar}}
 }
 {
 \sum_{a}\int\dd{\qty{x}}
 \mel{a\qty{x}}{e^{-{\beta}H}}{a\qty{x}}
 }
\end{equation}
and
\begin{equation}
 \expval{\mathcal{C}\qty(l)}= 
 \frac{
 \sum_{a}\int\dd{\qty{x}}
 \mel{a\qty{x}}{e^{-{\beta}H}}{a\qty{x}}
 x_{a+l} \sqrt{\frac{2m\omega_0}{\hbar}}
 }
 {
 \sum_{a}\int\dd{\qty{x}}
 \mel{a\qty{x}}{e^{-{\beta}H}}{a\qty{x}}
 } .
\end{equation}
 To evaluate the integrals in the numerator, we exploit the identity
\begin{equation}\label{eq:gaus3691}
 \int \dd[n]{\vb{z}} z_a e^{-\frac{1}{2}\vb{z}\cdot\vb{A}\cdot\vb{z}}
 e^{\vb{b}\cdot \vb{z}}= \qty(2\pi)^{n/2} \qty(\det\vb{A})^{-1/2} 
  e^{\frac{1}{2}\vb{b}\cdot \vb{A}^{-1} \cdot \vb{b}} \qty(\vb{A}^{-1}\cdot\vb{b})_a
 \end{equation}
 and obtain
 \begin{equation}
     \expval{H_{\mathrm{e-ph}}}=\frac{\mathcal{N}_2}{\mathcal{D}_2}
 \end{equation}
 with
 \begin{equation}
\mathcal{N}_2= \sum_{a}
 \sum_{j_1\ldots j_{K-1}}
f\qty(j_1)
f\qty(j_2)
\ldots
f\qty(j_{K-1})  
 f\qty(-\sum_{i=1}^{K-1}j_i)
e^{\frac{1}{2}\vb{b}\cdot \vb{A}^{-1} \vb{b}}
 \qty(\vb{A}^{-1} \vb{b})_a
  \gamma\sqrt{\frac{2m\omega_0}{\hbar}}
 \end{equation}
and 
\begin{equation}
    \mathcal{D}_2=\sum_{a}
 \sum_{j_1\ldots j_{K-1}}
f\qty(j_1)
f\qty(j_2)
\ldots
f\qty(j_{K-1})  
f\qty(-\sum_{i=1}^{K=1}j_i)
e^{\frac{1}{2}\vb{b}\cdot \vb{A}^{-1} \cdot \vb{b}}.
\end{equation} 
We find as well
\begin{equation}
 \expval{\mathcal{C}\qty(l)}=\frac{\mathcal{N}_3}{\mathcal{D}_2}   
\end{equation}
with
\begin{equation}
        \mathcal{N}_3=\sum_{a}
 \sum_{j_1\ldots j_{K-1}}
f\qty(j_1)
f\qty(j_2)
\ldots
f\qty(j_{K-1})  
 f\qty(-\sum_{i=1}^{K-1}j_i)
e^{\frac{1}{2}\vb{b}\cdot \vb{A}^{-1} \vb{b}}
 \qty(\vb{A}^{-1} \vb{b})_{a+l}
 \sqrt{\frac{2m\omega_0}{\hbar}}.
\end{equation}
The sums are then evaluated using the Monte Carlo method in the same way as in previous cases.

The results presented in Fig.~\textcolor{red}{2(b)} of the paper were obtained with $10^6$ samples and $K=80$, while the results in Fig.~\textcolor{red}{2(d)} of the paper were obtained with $10^7$ samples and $K=80$.

\newpage

\section{Weak-coupling limit: Rayleigh--Schr\"odinger perturbation theory}
Truncating the HEOM [Eq.~(30) of the main text] at depth 1, we remain with the following equations
\begin{equation}
\label{Eq:wc_0}
 \partial_t G^>(k,t)=-i\Omega_k G^>(k,t)+i\sum_{qm}G^{(>,1)}_{\mathbf{0}_{qm}^+}(k-q,t),
\end{equation}
\begin{equation}
\label{Eq:wc_1}
 \partial_t G^{(>,1)}_{\mathbf{0}_{qm}^+}(k-q,t)=-i\left[\Omega_{k-q}+\omega_q(\delta_{m0}-\delta_{m1})\right]G^{(>,1)}_{\mathbf{0}_{qm}^+}(k-q,t)+i\omega_0^2c_{qm}G^>(k,t).
\end{equation}
Separating out the free-rotation contribution from $G^>$ by defining $G^>(k,t)=e^{-i\Omega_kt}\mathcal{G}^>(k,t)$, the integration of Eq.~\eqref{Eq:wc_1} leads to
\begin{equation}
 G^{(>,1)}_{\mathbf{0}_{qm}^+}(k-q,t)=i\omega_0^2c_{qm}\int_0^t ds\:\exp\left\{i\left[\Omega_k-\Omega_{k-q}-\omega_q(\delta_{m0}-\delta_{m1})\right]s\right\}e^{-i\Omega_k t}\mathcal{G}^>(k,t-s).
\end{equation}
The Markov approximation $\mathcal{G}^>(k,t-s)\approx\mathcal{G}^>(t)$ assumes that $\mathcal{G}^>$ changes very slowly in time, so that its value at time $t-s$ is approximately equal to its value at time $t$. The remaining integration can be performed analytically. Nevertheless, one may resort to the so-called adiabatic approximation, in which the upper limit of the integral $t$ is replaced by $+\infty$. Physically, this means that the last scattering on phonons that is relevant to the properties at time $t$ happened in the distant past. Moreover, to ensure the integral convergence, one adds a small and positive imaginary part to the frequency difference. Finally,
\begin{equation}
\label{Eq:1st_order_final}
\begin{split}
 G^{(>,1)}_{\mathbf{0}_{qm}^+}(k-q,t)&\approx G^>(k,t)\times i\omega_0^2c_{qm}\int_0^{+\infty} ds\:\exp\left\{i\left[\Omega_k-\Omega_{k-q}-\omega_q(\delta_{m0}-\delta_{m1})+i\eta\right]s\right\}\\
 &=\frac{-\omega_0^2 c_{qm}}{\Omega_k-\Omega_{k-q}-\omega_q(\delta_{m0}-\delta_{m1})+i\eta}G^>(k,t).
 \end{split}
\end{equation}
We now insert Eq.~\eqref{Eq:1st_order_final} into Eq.~\eqref{Eq:wc_0} and transfer to the frequency domain. Keeping in mind that Eq.~\eqref{Eq:wc_0} is solved for $t>0$ under the initial condition $G^>(k,t=0)=-i$, we finally obtain
\begin{equation}
 G^>(k,\omega)=\frac{1}{\omega-\Omega_k-\Sigma_k}
\end{equation}
where the one-phonon self-energy reads as
\begin{equation}
 \Sigma_k=\sum_{qm}\frac{\omega_0^2c_{qm}}{\Omega_k-\Omega_{k-q}-\omega_q(\delta_{m0}-\delta_{m1})+i\eta}.
\end{equation}
The real part of the self-energy brings about the energy renormalization
\begin{equation}
 \mathrm{Re}\:\Sigma_k=\frac{\gamma^2}{\hbar}\frac{1}{N}\sum_q\left(\frac{1+n_\mathrm{BE}(\omega_q,T)}{\hbar\Omega_k-\hbar\Omega_{k-q}-\hbar\omega_q}+\frac{n_\mathrm{BE}(\omega_q,T)}{\hbar\Omega_k-\hbar\Omega_{k-q}+\hbar\omega_q}\right)
\end{equation}
while its imaginary part
\begin{equation}
 \mathrm{Im}\:\Sigma_k=-\pi\frac{\gamma^2}{\hbar}\frac{1}{N}\sum_q\left\{[1+n_\mathrm{BE}(\omega_q,T)]\delta(\hbar\Omega_k-\hbar\Omega_{k-q}-\hbar\omega_q)+n_\mathrm{BE}(\omega_q,T)\delta(\hbar\Omega_k-\hbar\Omega_{k-q}+\hbar\omega_q)\right\}
\end{equation}
is related to the second-order result for carrier scattering time $\tau_k$ via $\tau_k^{-1}=-2\mathrm{Im}\:\Sigma_k$.

\newpage

\section{Single-site limit: Lang--Firsov results}

Here, we explicitly present the steps that are needed to reduce the exact Feynman--Vernon results presented in the main text to the Lang--Firsov results in the limit of vanishing electronic coupling. While these results are by no means new (see, e.g., Ch.~4 of Ref.~\onlinecite{Mahanbook}), we feel that it is important to demonstrate how our approach reduces to this important limiting case.

We start from the computation of $G^>$. The exact solution in the single-site limit reduces to
\begin{equation}
 G^>(t)=-ie^{-i\varepsilon_et/\hbar}\exp\left[-\omega_0^2\int_0^t ds_2\int_0^{s_2}ds_1\sum_m c_me^{-\mu_m(s_2-s_1)}\right].
\end{equation}
To obtain this result, we used the fact that, in the single-site limit, the operator $V_q$ reduces to $|j\rangle\langle j|$, which is time-independent, so that the time-ordering sign entering the exact solution is not effective. Solving the integral under the exponential we finally obtain
\begin{equation}
\label{Eq:greater-w-o-broadening}
 G^>(t)=-ie^{-i\varepsilon_et/\hbar}\:e^{-(\gamma/(\hbar\omega_0))^2(2N_\mathrm{ph}+1)}\exp\left[\left(\frac{\gamma}{\hbar\omega_0}\right)^2\left((1+N_\mathrm{ph})e^{-i\omega_0t}+N_\mathrm{ph}e^{i\omega_0t}+i\omega_0t\right)\right],
\end{equation}
where $N_\mathrm{ph}=\left(e^{\beta\hbar\omega_0}-1\right)^{-1}$.
The exact result for $G^>(\omega)$ can then be obtained following the procedure from Ref.~\onlinecite{Mahanbook} 
The crux of the derivation is to employ the following identity~\cite{Gradshteyn_Ryzhik}
\begin{equation}
 e^{z\cos\theta}=\sum_{l=-\infty}^{+\infty}I_l(z)\:e^{il\theta},
\end{equation}
where $I_l(z)$ are the Bessel functions of complex argument.

A similar procedure can be repeated to obtain the  single-site limit of the exact result for $G^<$:
\begin{equation}
\label{Eq:lesser-w-o-broadening}
 G^<(t)=ie^{-i\varepsilon_et/\hbar}e^{-\beta[\varepsilon_e-\gamma^2/(\hbar\omega_0)]}\:e^{-(\gamma/(\hbar\omega_0))^2(2N_\mathrm{ph}+1)}\exp\left[\left(\frac{\gamma}{\hbar\omega_0}\right)^2\left((1+N_\mathrm{ph})e^{i\omega_0t}+N_\mathrm{ph}e^{-i\omega_0t}+i\omega_0t\right)\right].
\end{equation}
Apart from the prefactors, the only difference between the single-site expressions for $G^>$ and $G^<$ is the place of phonon factors $1+N_\mathrm{ph}\pm 1$ in front of exponentials $e^{\pm i\omega_0t}$.

\begin{figure}[htbp!]
 \includegraphics[scale=0.65]{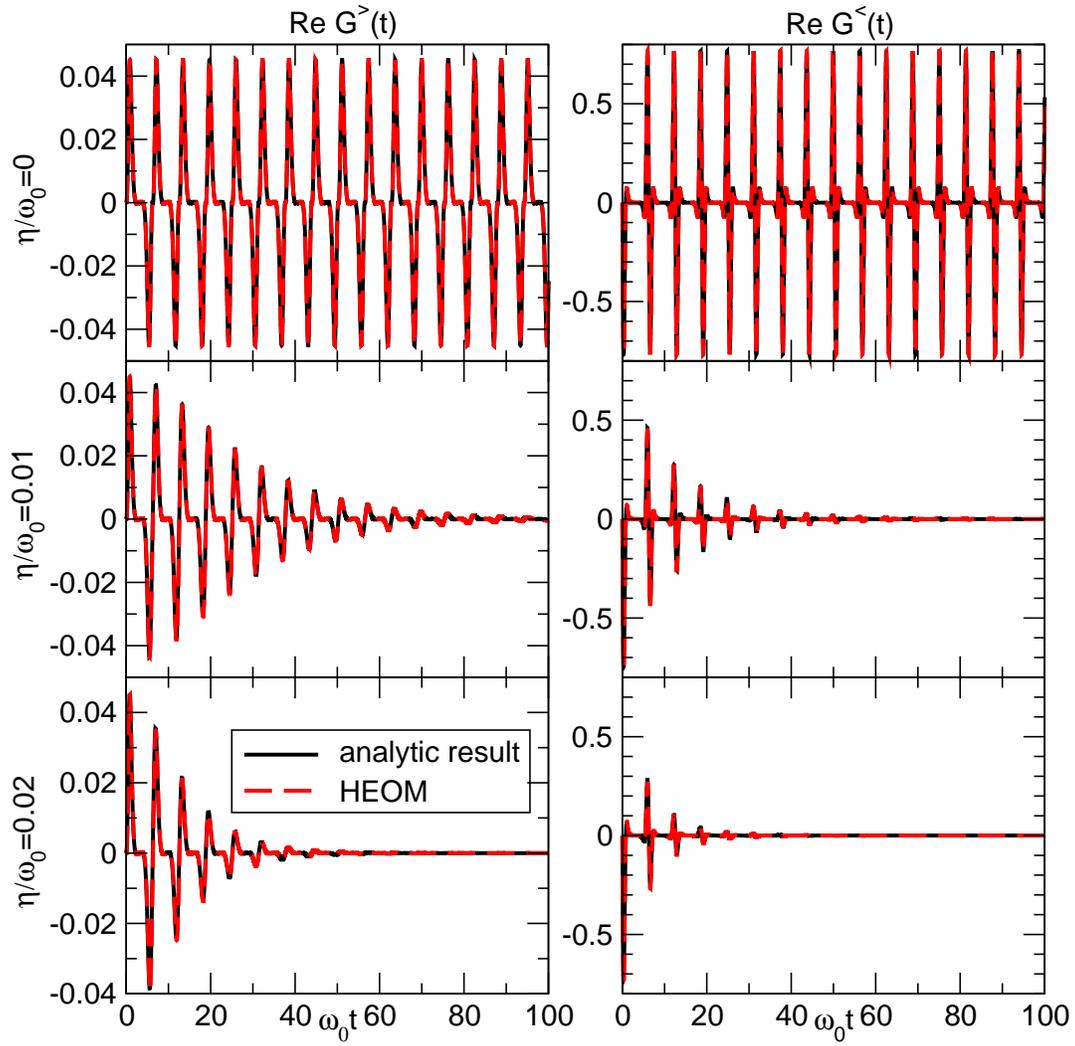}
 \caption{Results for $G^>(t)$ (left panels) and $G^<(t)$ (right panels) in the single-site limit for $\gamma/\omega_0=\sqrt{2}$, $k_BT/\omega_0=1$ and different values of the artificial broadening $\eta=0$ (upper panels), $\eta/\omega_0=0.01$ (middle panels), and $\eta/\omega_0=0.02$ (bottom panels). Without loss of generality, we assume that $\varepsilon_e=0$. Analytical results are obtained using Eqs.~\eqref{Eq:greater-w-broadening} and~\eqref{Eq:lesser-w-broadening}, while the kinetic terms in the HEOM assume the form of Eq.~\eqref{Eq:heom_kinetic_broadening}.}
 \label{Fig:fig_ss_t_280921}
\end{figure}

While $G^{>/<}(\omega)$ is an infinite series of equidistant $\delta$ peaks of varying intensity, one commonly introduces the artificial broadening $\eta$, i.e., replaces $\delta$ peaks by Lorentzians whose full width at half maximum is equal to $\eta$. To take the artificial broadening into account, it is enough to replace
\begin{equation}
\label{Eq:replacements}
 \mu_0=i\omega_0\to i\omega_0+\eta,\quad\mu_1=-i\omega_0\to -i\omega_0+\eta.
\end{equation}
Equation~\eqref{Eq:greater-w-o-broadening} then becomes
\begin{equation}
\begin{split}
 \label{Eq:greater-w-broadening}
 G^>(t)&=-ie^{-i\varepsilon_et/\hbar}\:e^{-(\gamma/(\hbar\omega_0))^2(2N_\mathrm{ph}+1)}\\&\times\exp\left[\left(\frac{\gamma}{\hbar\omega_0}\right)^2\left((1+N_\mathrm{ph})e^{-i(\omega_0-i\eta)t}+N_\mathrm{ph}e^{i(\omega_0+i\eta)t}+i(\omega_0+i(2N_\mathrm{ph}+1)\eta)t\right)\right],
 \end{split}
\end{equation}
while Eq.~\eqref{Eq:lesser-w-o-broadening} becomes
\begin{equation}
\label{Eq:lesser-w-broadening}
\begin{split}
 G^<(t)&=ie^{-i\varepsilon_et/\hbar}e^{-\beta[\varepsilon_e-\gamma^2/(\hbar\omega_0)]}\:e^{-(\gamma/(\hbar\omega_0))^2(2N_\mathrm{ph}+1)}\\&\times\exp\left[\left(\frac{\gamma}{\hbar\omega_0}\right)^2\left((1+N_\mathrm{ph})e^{i(\omega_0+i\eta)t}+N_\mathrm{ph}e^{-i(\omega_0-i\eta)t}+i(\omega_0+i(2N_\mathrm{ph}+1)\eta)t\right)\right].
 \end{split}
\end{equation}
On the other hand, the replacements of Eq.~\eqref{Eq:replacements} affect the hierarchy through its kinetic term only, while the hierarchical links remain unaffected. In more detail,
\begin{equation}
\label{Eq:heom_kinetic_broadening}
\begin{split}
 \left[\partial_t G^{(n)}_\mathbf{n}(k-k_\mathbf{n},t)\right]_\mathrm{kin}&=-i(\omega_{k-k_\mathbf{n}}+\mu_\mathbf{n})G^{(n)}_\mathbf{n}(k-k_\mathbf{n},t)\\
 &\to-i(\omega_{k-k_\mathbf{n}}+\mu_\mathbf{n}-in\eta)G^{(n)}_\mathbf{n}(k-k_\mathbf{n},t),
\end{split}
\end{equation}
so that the auxiliary Green's functions at depth $n$ are damped at rate $n\eta$.
Analytical and numerical result exhibit excellent agreement in the single-site limit, which is demonstrated in Fig.~\ref{Fig:fig_ss_t_280921} for $\gamma/\omega_0=\sqrt{2}$ and $k_BT/\omega_0=1$ and different values of the artificial broadening $\eta/\omega_0=0,0.01,0.02$.

\begin{figure}[htbp!]
    \centering
    \includegraphics[scale=0.6]{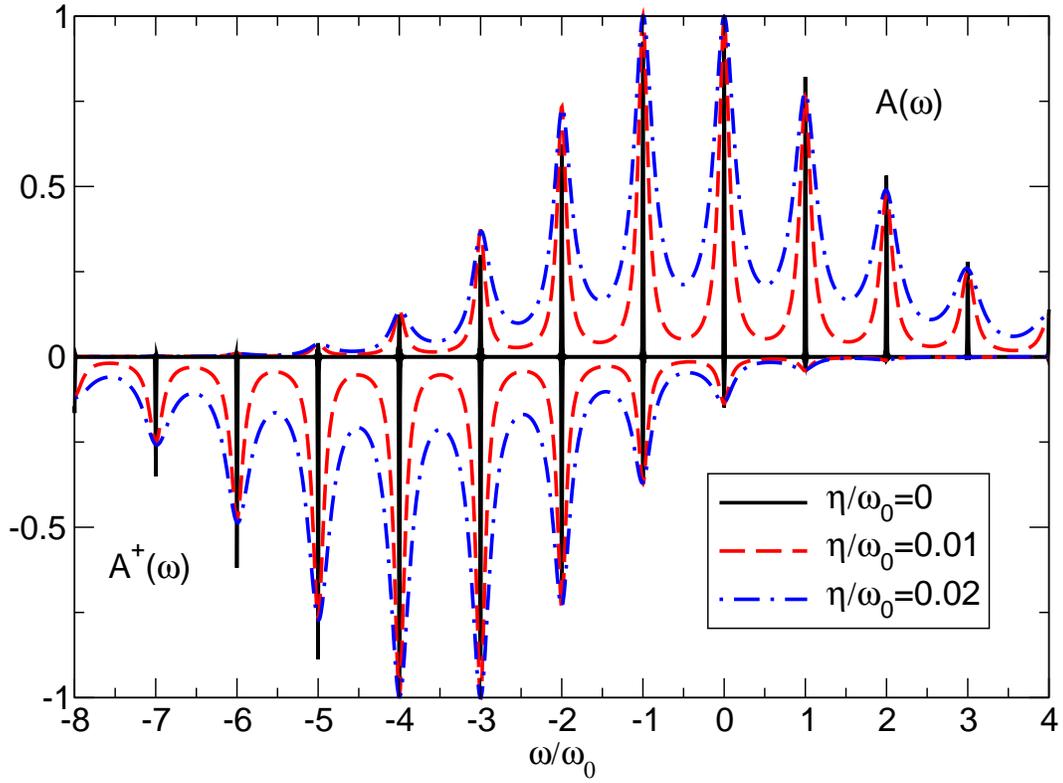}
    \caption{Electron-addition [$A(\omega)=-\frac{1}{2\pi}\mathrm{Im}\:G^>(\omega)$] and electron-removal [$A^+(\omega)=\frac{1}{2\pi}\mathrm{Im}\:G^<(\omega)$] spectral functions for a single site with $\gamma/\omega_0=\sqrt{2}$ and $k_BT/\omega_0=1$ and different levels of the artificial broadening $\eta/\omega_0=0,0.01,0.02$. Both $A(\omega)$ and $A^+(\omega)$ are normalized so that their maximum values are equal to 1. $A(\omega)$ is presented in the positive half-plane, while $A^+(\omega)$ is presented in the negative half-plane. This style of presentation emphasises the fact that $A(\omega)$ and $A^+(\omega)$ are mirror images of one another around $\omega_\mathrm{QP}/\omega_0=-(\gamma/\omega_0)^2=-2$.}
    \label{Fig:fig_ss_w_280921}
\end{figure}

The influence of $\eta$ on the spectral properties in the single-site limit are summarized in Fig.~\ref{Fig:fig_ss_w_280921}.

\pagebreak

\pagebreak

%
%

\pagebreak
\section{Artificial broadening of spectral lines}
When the spectral density of the electron--phonon coupling is that of an underdamped Brownian oscillator
\begin{equation}
 \mathcal{J}(\omega)=\Lambda\left[\frac{\omega\eta}{(\omega-\omega_0)^2+\eta^2}+\frac{\omega\eta}{(\omega+\omega_0)^2+\eta^2}\right]
\end{equation}
the bath correlation function assumes the following form~\cite{JChemPhys.140.134106}
\begin{equation}
 \mathcal{C}_{q_2q_1}(t)=\delta_{q_2,-q_1}(\hbar\omega_0)^2\sum_{m=0}^{+\infty}c_m\:e^{-\mu_m t}
\end{equation}
where
\begin{equation}
 c_0=\frac{\Lambda}{\hbar\omega_0}\left(1-i\frac{\eta}{\omega_0}\right)\left[1+n_\mathrm{BE}(\hbar\omega_0-i\hbar\eta,T)\right],\quad\mu_0=i\omega_0+\eta
\end{equation}
\begin{equation}
 c_1=\frac{\Lambda}{\hbar\omega_0}\left(1+i\frac{\eta}{\omega_0}\right)n_\mathrm{BE}(\hbar\omega_0+i\hbar\eta,T),\quad\mu_1=-i\omega_0+\eta
\end{equation}
\begin{equation}
\label{Eq:c_m_gtr_2}
 c_m=-4\frac{\eta}{\omega_0}\frac{\Lambda k_BT}{(\hbar\omega_0)^2}\rho_{m-1}\frac{\xi_{m-1}}{\beta\hbar\omega_0}\frac{1-\left(\frac{\xi_{m-1}}{\beta\hbar\omega_0}\right)^2+\left(\frac{\eta}{\omega_0}\right)^2}{\left[1-\left(\frac{\xi_{m-1}}{\beta\hbar\omega_0}\right)^2+\left(\frac{\eta}{\omega_0}\right)^2\right]^2+4\frac{\xi_{m-1}}{\beta\hbar\omega_0}},\quad\mu_m=\frac{\xi_{m-1}}{\beta\hbar},\quad m\geq 2
\end{equation}
In Eq.~\eqref{Eq:c_m_gtr_2}, $\xi_m$ and $\rho_m$ (for $m\geq 1$) are the poles and residues of the Bose--Einstein function $\left(e^z-1\right)^{-1}$ in the upper half of the complex plane. The corresponding decomposition of the Bose--Einstein function into simple poles reads as
\begin{equation}
 \frac{1}{e^z-1}=-\frac{1}{2}+\frac{1}{z}+\sum_{m=1}^{+\infty}\rho_m\left(\frac{1}{z-i\xi_m}+\frac{1}{z+i\xi_m}\right).
\end{equation}
In the Matsubara decomposition, $\xi_m=2\pi m$, $\rho_m=1$. There are other possible choices, e.g., Pad\'e decomposition.

Let us now assume that $\hbar\eta$ is the smallest energy scale in the problem, i.e., $\hbar\eta\ll k_BT$ and $\eta\ll\omega_0$. One may then neglect the imaginary parts of $c_0$ and $c_1$. For $m\geq 2$, $c_m$ is linear in the small quantity $\eta/\omega_0$, while the corresponding exponential factor $e^{-\mu_mt}$ decays much faster than $e^{-\mu_{0/1}t}$ because of $\xi_{m-1}/(\beta\hbar\eta)\gg 1$. We may thus completely discard the infinite-series part of the bath correlation function, after which it reduces to
\begin{equation}
 \mathcal{C}_{q_2q_1}(t)=\delta_{q_2,-q_1}\frac{\Lambda}{\hbar\omega_0}\left\{[1+n_\mathrm{BE}(\omega_0,T)]e^{-(i\omega_0+\eta)t}+n_\mathrm{BE}(\omega_0,T)e^{-(-i\omega_0+\eta)t}\right\}.
\end{equation}
By identifying $\Lambda=\gamma^2/(\hbar\omega_0)$, we obtain the expression that is identical to the bath correlation function presented in the main text, with the only difference that the oscillatory terms are replaced by damped oscillatory terms.

\pagebreak
\section{Comparison between time-domain and frequency-domain HEOM data for different maximum hierarchy depths}
\begin{figure}[htbp!]
    \centering
    \includegraphics[scale=1.2]{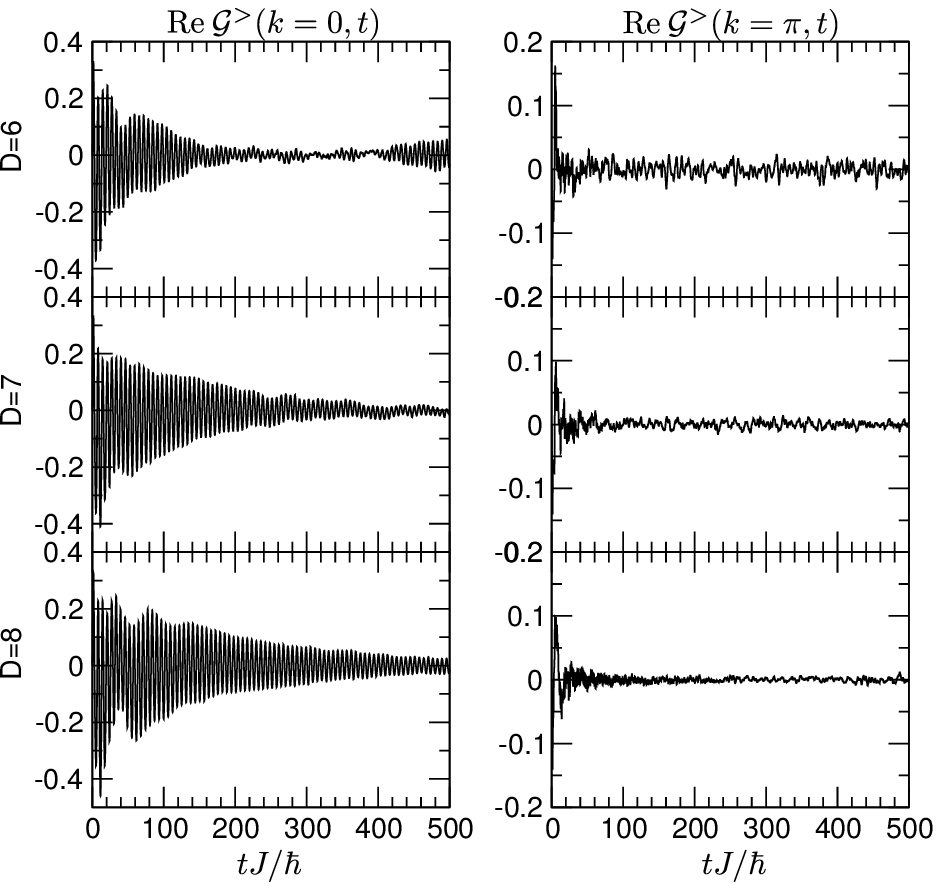}
    \caption{Time dependence of the real part of the envelope of the greater Green's function, $\mathrm{Re}\:\mathcal{G}^>(k,t)$, for three different values of the maximum hierarchy depth ($D=6$ in the upper panels, $D=7$ in the middle panels, and $D=8$ in the lower panels) and for two different values of the dimensionless wave number ($k=0$ in the left column, $k=\pi$ in the right column). The ranges on the vertical axes on all three left-column plots are the same ($[-0.5,0.4]$), as are their counterparts on all three right-column plots ($[-0.2,0.2]$).}
    \label{Fig:regime_5b_sm_real_time}
\end{figure}

\begin{figure}[htbp!]
    \centering
    \includegraphics[scale=0.75]{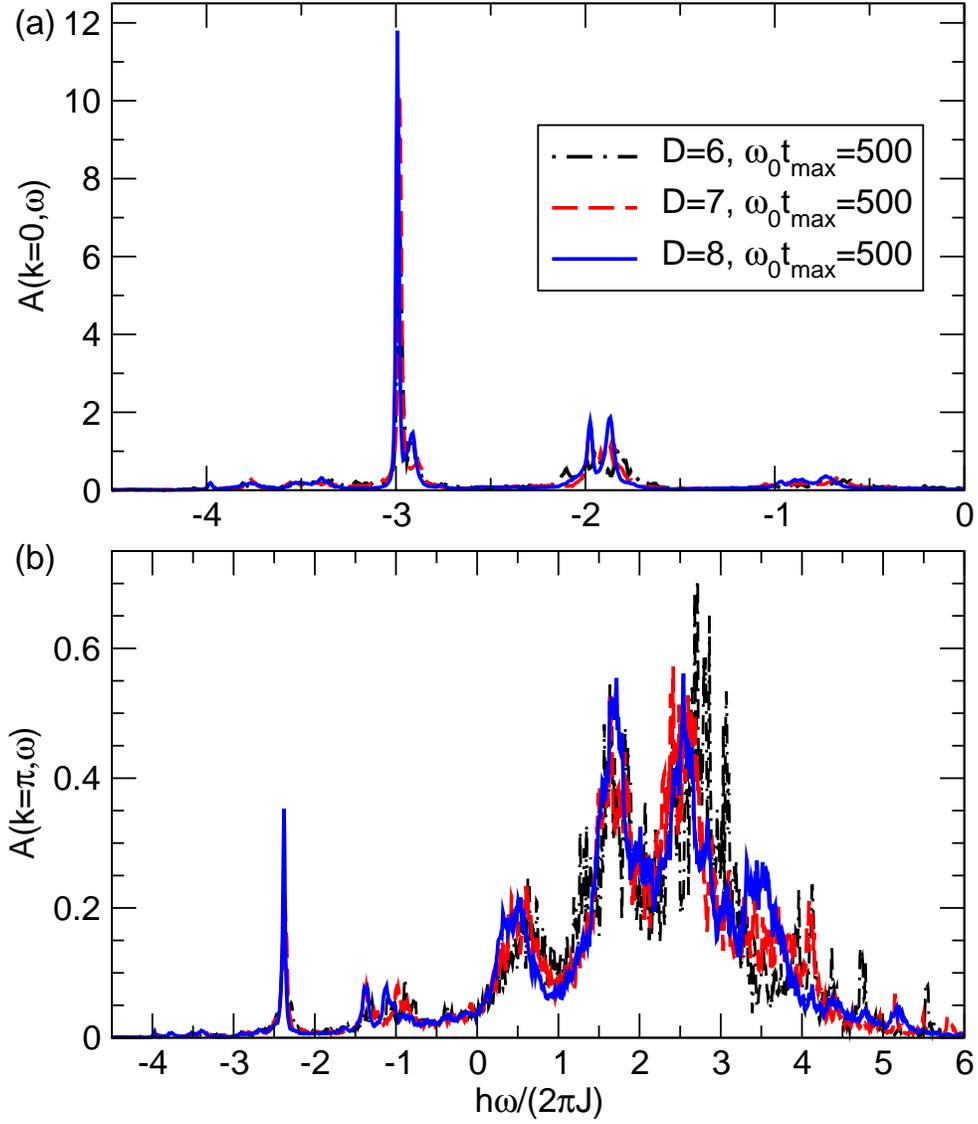}
    \caption{Spectral function $A(k,\omega)$ for three different values of the maximum hierarchy depth ($D=6$ is represented by dash-dotted black lines, $D=7$ is represented by dashed red lines, $D=8$ is represented by solid blue lines) and for two different values of the dimensionless wave number ($k=0$ in the upper panel and $k=\pi$ in the lower panel). All the time-domain data presented in Fig.~\ref{Fig:regime_5b_sm_real_time} are used to compute $A(k,\omega)$, which is signalized by the label $\omega_0t_\mathrm{max}=500$ in the legend.}
    \label{Fig:5b_diff_depths}
\end{figure}

In Figs.~\ref{Fig:regime_5b_sm_real_time} and~\ref{Fig:5b_diff_depths}, we compare the real part of the envelope $\mathcal{G}^>(k,t)$ of the greater Green's function and the spectral function $A(k,\omega)$ for three different values of the maximum hierarchy depth $D$ in the following regime of model parameters: $k_BT/J=0.4$, $\hbar\omega_0/J=1$, $\gamma/J=\sqrt{2}$, $N=8$.

\pagebreak

\pagebreak

\section{Comparison between QMC and HEOM results for the electronic momentum distribution}
\begin{figure}[htbp!]
 \centering
 \includegraphics[scale=0.6]{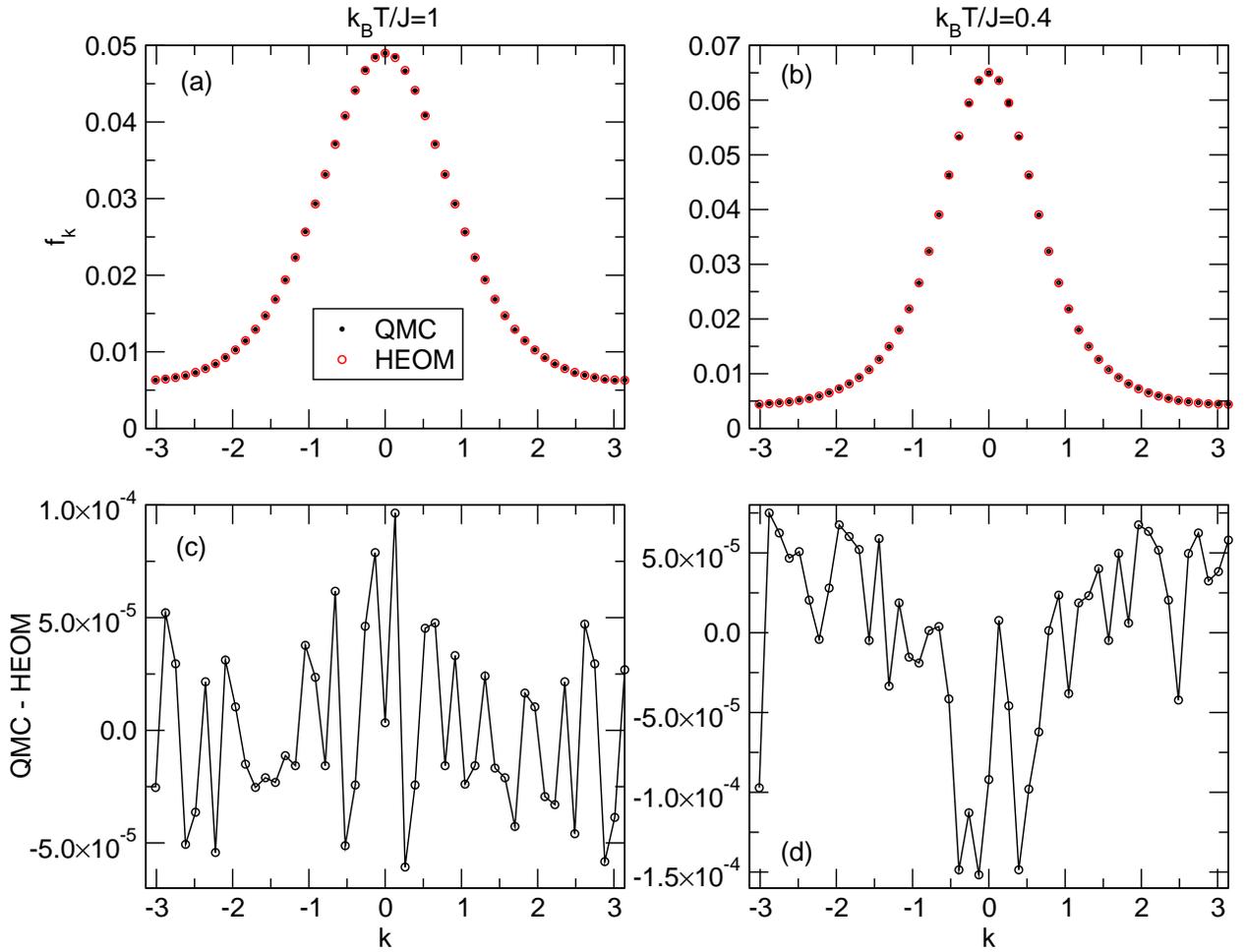}
 \caption{(a) and (b): Electronic momentum distribution $f_k$ computed using QMC (black dots) and HEOM (red empty circles) for (a) $k_BT/J=1$ and (b) $k_BT/J=0.4$. QMC error bars are smaller than the size of individual dots. (c) and (d): Difference between QMC and HEOM electronic momentum distributions for (c) $k_BT/J=1$ and (d) $k_BT/J=0.4$. The remaining parameters assume the following values: $\hbar\omega_0/J=1$, $\gamma/J=\sqrt{2}$.}
 \label{Fig:Fig2a_SI_T_1_4e-1}
\end{figure}

In Fig.~\ref{Fig:Fig2a_SI_T_1_4e-1} we compare QMC and HEOM predictions for the electronic momentum distribution for two different temperatures, $k_BT/J=1$ in Figs.~\ref{Fig:Fig2a_SI_T_1_4e-1}(a) and~\ref{Fig:Fig2a_SI_T_1_4e-1}(c), and $k_BT/J=0.4$ in Figs.~\ref{Fig:Fig2a_SI_T_1_4e-1}(b) and~\ref{Fig:Fig2a_SI_T_1_4e-1}~(d).

\begin{figure}[htbp!]
 \centering
 \includegraphics[scale=0.6]{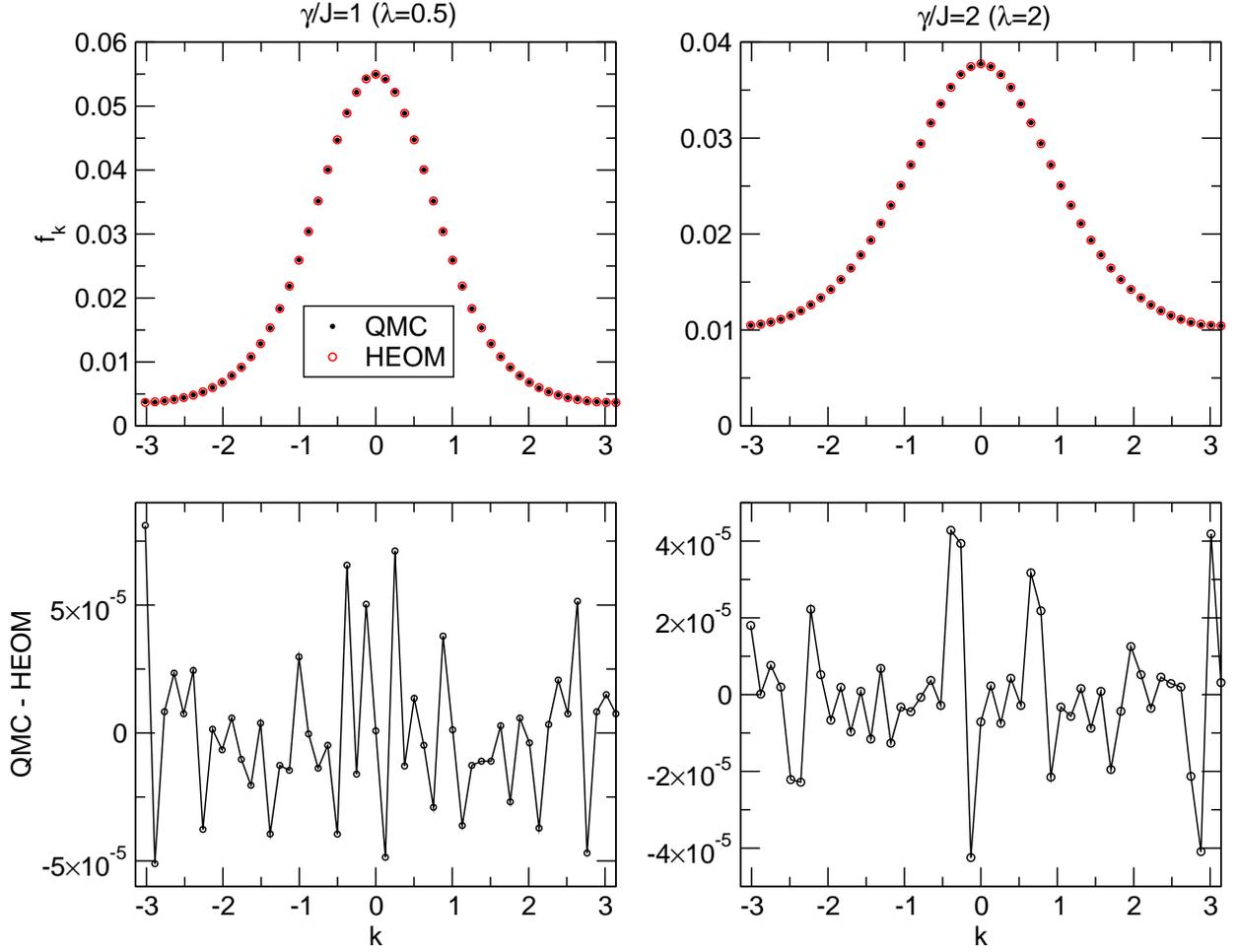}
 \caption{(a) and (b): Electronic momentum distribution $f_k$ computed using QMC (black dots) and HEOM (red empty circles) for (a) $\gamma/J=1$ and (b) $\gamma/J=2$. QMC error bars are smaller than the size of individual dots. (c) and (d): Difference between QMC and HEOM electronic momentum distributions for (c) $\gamma/J=1$ and (d) $\gamma/J=2$. The remaining parameters assume the following values: $\hbar\omega_0/J=1$, $k_BT/J=1$.}
 \label{Fig:Fig6a_SI_gamma_1_2}
\end{figure}

\pagebreak
In Fig.~\ref{Fig:Fig6a_SI_gamma_1_2} we compare QMC and HEOM predictions for the electronic momentum distribution for two different electron--phonon coupling strengths, $\gamma/J=1$ in Figs.~\ref{Fig:Fig6a_SI_gamma_1_2}(a) and~\ref{Fig:Fig6a_SI_gamma_1_2}(c), and $\gamma/J=2$ in Figs.~\ref{Fig:Fig6a_SI_gamma_1_2}(b) and~\ref{Fig:Fig6a_SI_gamma_1_2}~(d).

\pagebreak

\section{Discussion of the HEOM method results in the strong-coupling adiabatic regime}
Here, we discuss in greater detail the results of the HEOM method in the strong-coupling adiabatic regime, $\hbar\omega_0/J=0.2$, $\gamma/J=\sqrt{4/5}$, at 
temperature $k_BT/J=1$. 

\begin{figure}[htbp!]
    \centering
    \includegraphics[scale=1.2]{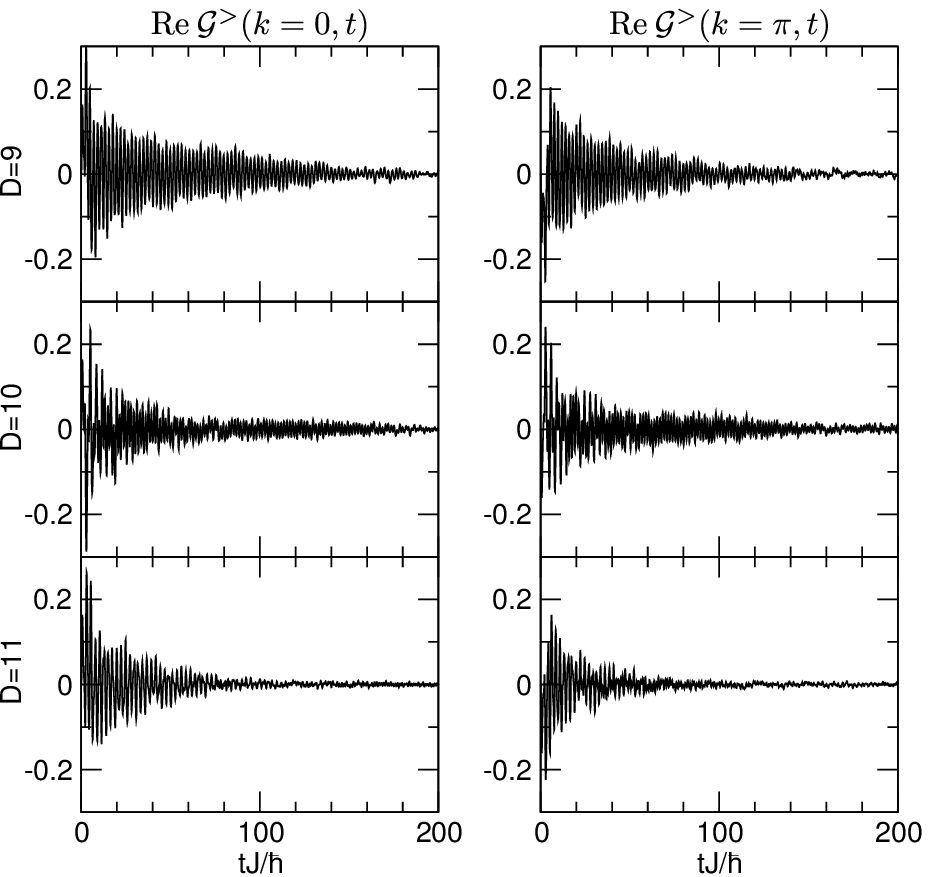}
    \caption{Time dependence of the real part of the envelope of the greater Green's function, $\mathrm{Re}\:\mathcal{G}^>(k,t)$, for three different values of the maximum hierarchy depth ($D=9$ in the upper panels, $D=10$ in the middle panels, and $D=11$ in the lower panels) and for two different values of the dimensionless wave number ($k=0$ in the left column and $k=\pi$ in the right column). The vertical-axes ranges on all six panels are the same, $[-0.3,0.3]$. The model parameters assume the following values: $k_BT/J=1$, $\hbar\omega_0/J=0.2$, $\gamma/J=\sqrt{4/5}$, $N=6$.}
    \label{Fig:time_domain_adiabatic_inkscape}
\end{figure}

\begin{figure}[htbp!]
    \centering
    \includegraphics[scale=0.65]{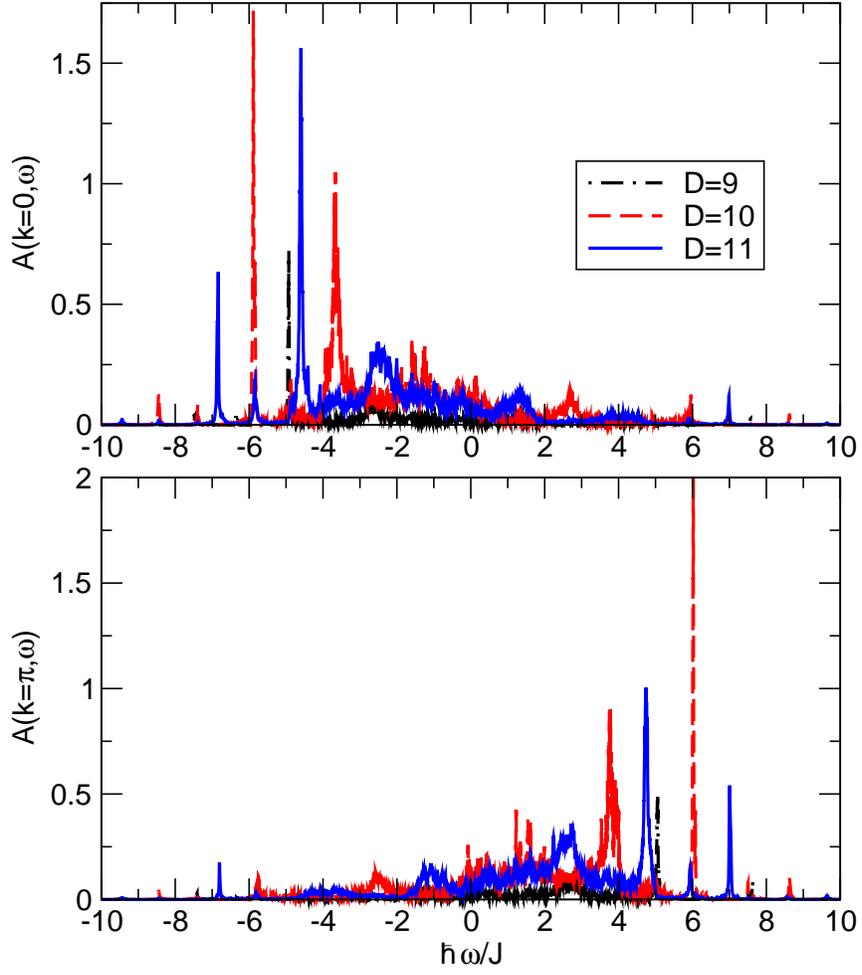}
    \caption{Spectral function $A(k,\omega)$ for three different values of the maximum hierarchy depth ($D=9$ is represented by dash-dotted black lines, $D=10$ is represented by dashed red lines, $D=11$ is represented by solid blue lines) and for two different values of the dimensionless wave number ($k=0$ in the upper panel and $k=\pi$ in the lower panel). The model parameters assume the following values: $k_BT/J=1$, $\hbar\omega_0/J=0.2$, $\gamma/J=\sqrt{4/5}$, $N=6$.}
    \label{Fig:frequency_domain_adiabatic}
\end{figure}

We start with Fig.~\ref{Fig:time_domain_adiabatic_inkscape}, in which we compare the real parts of the envelopes $\mathcal{G}^>(k,t)$ of the greater Green's functions at finite temperature for different maximum depths $D$ of the hierarchy. While the calculations are actually performed up to maximum time $\omega_0t_\mathrm{max}=400$, which translates to $Jt_\mathrm{max}/\hbar=2000$, Fig.~\ref{Fig:time_domain_adiabatic_inkscape} shows only the time window $0\leq Jt/\hbar\leq 200$. For $Jt/\hbar>200$, $\mathrm{Re}\:\mathcal{G}^>(k,t)$ exhibits small-amplitude oscillations around 0, the effect that is due to the finite size of the system studied. Although the differences between the results for different $D$ are difficult to appreciate in the real-time domain, they become more visible upon transformation in the real-frequency domain, which is demonstrated in Fig.~\ref{Fig:frequency_domain_adiabatic}. To obtain $A(k,\omega)$ shown in Fig.~\ref{Fig:frequency_domain_adiabatic}, we used $G^>(k,t)$ up to the maximum time $Jt_\mathrm{max}/\hbar=2000$ for which we performed the calculations. The positions and intensities of the peaks exhibit appreciable changes with $D$, which is very different from the situation presented in Fig.~\ref{Fig:5b_diff_depths}, where the positions of the peaks do not change with $D$, while the changes in their intensities are minor. The results presented in Figs.~\ref{Fig:time_domain_adiabatic_inkscape} and~\ref{Fig:frequency_domain_adiabatic} suggest that the HEOM method experiences problems in the adiabatic regime. 

\newpage

\begin{figure}[htbp!]
    \centering
    \includegraphics[scale=0.6]{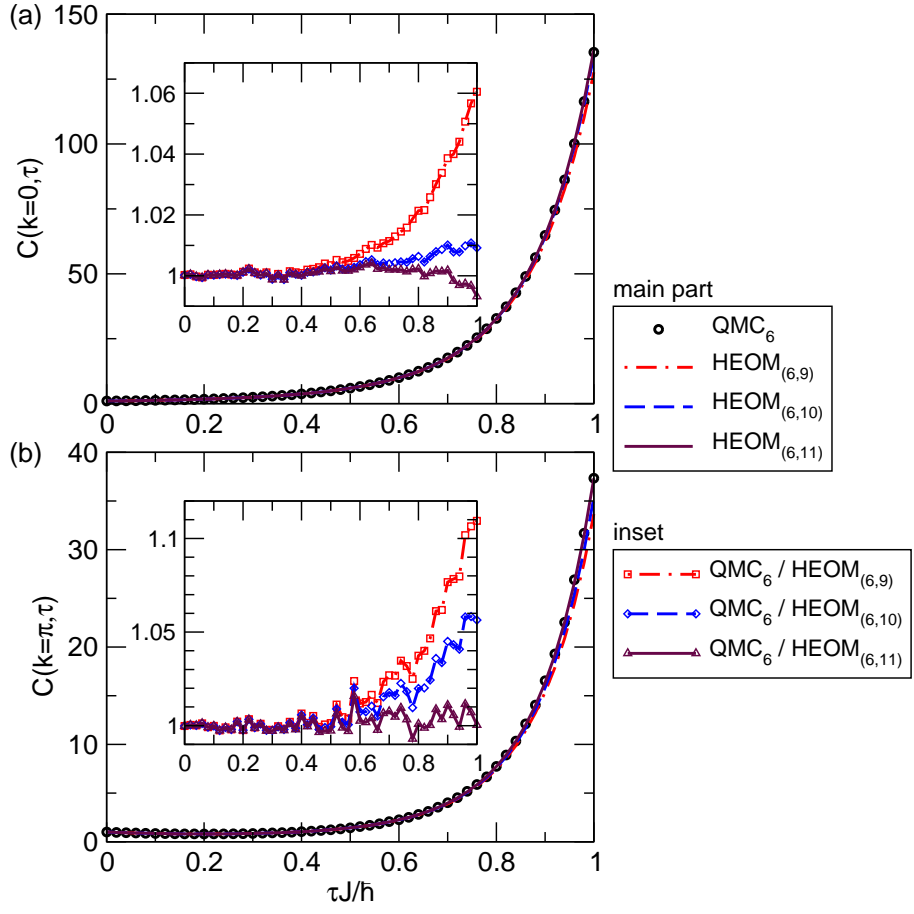}
    \caption{Imaginary-time correlation function $C(k,\tau)$ (a) in the zone center $k=0$ and (b) at the zone edge $k=\pi$ computed using QMC on a $6$-site chain (label QMC$_6$) and HEOM on a $6$-site chain with different maximum depths $D$ (label HEOM$_{(6,D)}$). Insets present the ratio QMC$_6$/HEOM$_{(6,D)}$ for different $D$. The model parameters assume the following values: $k_BT/J=1$, $\hbar\omega_0/J=0.2$, and $\gamma/J=\sqrt{4/5}$.}
    \label{Fig:figure_imaginary_time_adiabatic}
\end{figure}

In order to understand whether the HEOM-method results in the adiabatic limit are reasonable, in Figs.~\ref{Fig:figure_imaginary_time_adiabatic}(a) and~\ref{Fig:figure_imaginary_time_adiabatic}(b) we compare the imaginary-time correlation function $C(k,\tau)$ evaluated using the QMC method on a $N=6$-site chain and the HEOM method on a $N=6$-site chain with maximum depths $D=9,10,$ and $11$. The results show that, as $D$ is increased, the ratio QMC$_6$/HEOM$_{(6,D)}$ of the QMC and HEOM results (labels are precisely defined in the caption of Fig.~\ref{Fig:figure_imaginary_time_adiabatic}) becomes closer to unity on the whole imaginary-time interval $0\leq J\tau/\hbar\leq\beta J$. This suggests that the HEOM results for $D=11$ are reasonable in the parameter regime studied. These results are shown in Figs.\textcolor{red}{~9(a1) and~9(a2)} of the main body of the paper.

\newpage

\newpage
\section{Influence of the statistical sample size on QMC results in the intermediate-coupling low-temperature regime}
\begin{figure}[htbp!]
    \centering
    \includegraphics[scale=0.6]{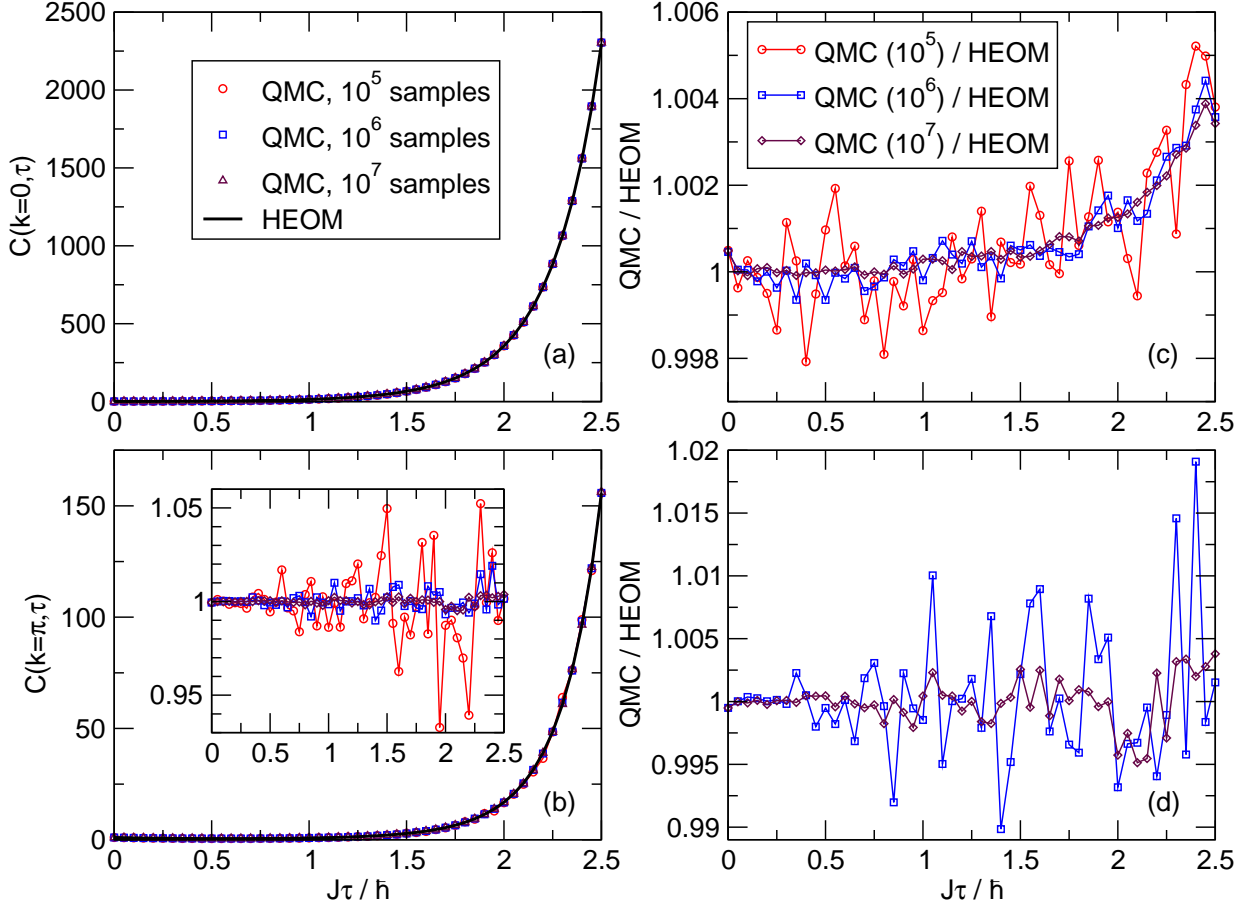}
    \caption{(a) and (b) Imaginary-time correlation function $C(k,\tau)$ in the zone center [(a)] and at the zone edge [(b)] computed using QMC with different statistical sample sizes (empty symbols) and HEOM (solid line). (c) and (d) Ratio of the QMC and HEOM results for different QMC statistical sample sizes. For the sake of clearer visibility, in (d) we show QMC/HEOM only for sample sizes $10^6$ and $10^7$, while the inset in (b) shows the ratio of the QMC and HEOM results for all sample sizes considered. The model parameters assume the following values: $k_BT/J=0.4$, $\hbar\omega_0/J=1$, $\gamma/J=\sqrt{2}$, $N=8$.}
    \label{Fig:r5b_011121}
\end{figure}

Here, we present the results for the QMC imaginary-time correlation function $C(k,\tau)$ obtained with different sizes of the statistical sample and compare them to the HEOM results for the same quantity. The calculations are performed on an $N=8$-site chain (the HEOM method is applied to the chain of the same length) with the following values of model parameters: $\hbar\omega_0/J=1$, $k_BT/J=0.4$, and $\gamma/J=\sqrt{2}$. In Figs.~\ref{Fig:r5b_011121}(a) and~\ref{Fig:r5b_011121}(b) we show HEOM and QMC data for $C(k=0,\tau)$ [Fig.~\ref{Fig:r5b_011121}(a)] and $C(k=\pi,\tau)$ [Fig.~\ref{Fig:r5b_011121}(b)], while in Figs.~\ref{Fig:r5b_011121}(c) and~\ref{Fig:r5b_011121}(d) we compare the ratios $C_\mathrm{QMC}(k,\tau)/C_\mathrm{HEOM}(k,\tau)$ for different sizes of the QMC sample.

The results presented in Figs.~\ref{Fig:r5b_011121}(a)--\ref{Fig:r5b_011121}(d) unambiguously show the reduction in the QMC statistical noise when the sample size is increased from $10^5$ to $10^6$ and $10^7$. While in the zone center decent results are obtained already with $10^5$ samples, at the zone edge we need as many as $10^7$ samples to reduce the level of the statistical noise.

\newpage

\section{Imaginary-time correlation function in the adiabatic and antiadiabatic regime}

\begin{figure}[htbp!]
    \centering
    \includegraphics[scale=0.5]{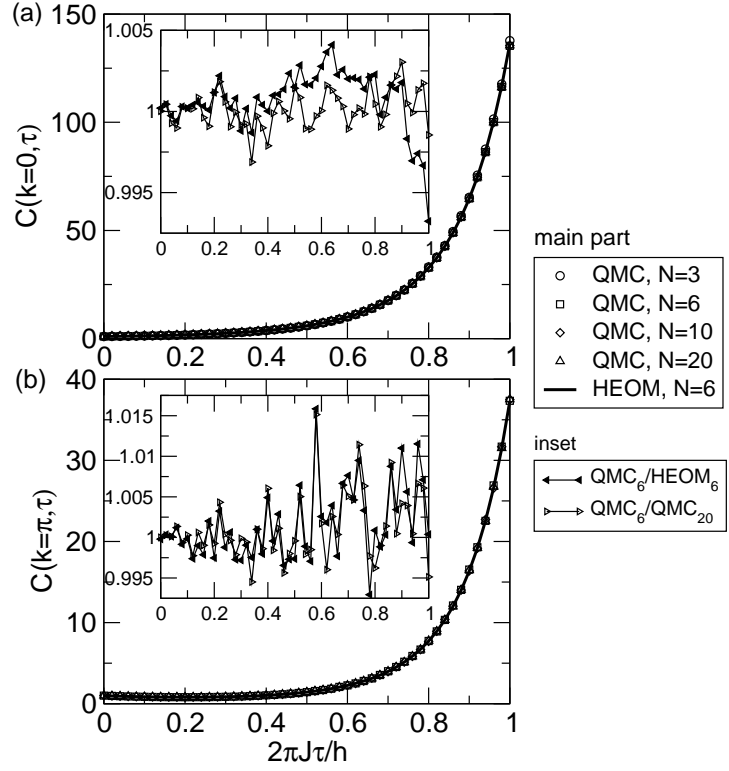}
    \caption{Imaginary time correlation function $C(k,\tau)$ (a) in the zone center $k=0$ and (b) at the zone edge $k=\pi$ computed using QMC with different chain lengths (empty symbols) and HEOM (solid line). Insets present the ratio of QMC and HEOM results for $N=6$ (full left-triangles) and the ratio of QMC results for $N=6$ and $N=20$ (empty right-triangles). The model parameters assume the following values: $k_BT/J=1$, $\hbar\omega_0/J=0.2$, $\gamma/J=\sqrt{4/5}$.}
    \label{Fig:fig_r50_240921}
\end{figure}

In Figs.~\ref{Fig:fig_r50_240921}(a) and~\ref{Fig:fig_r50_240921}(b) we present comparison between imaginary-time correlation functions in the zone center [Fig.~\ref{Fig:fig_r50_240921}(a)] and at the zone edge [Fig.~\ref{Fig:fig_r50_240921}(b)] obtained using the QMC and HEOM methods in the adiabatic regime, $k_BT/J=1$, $\hbar\omega_0/J=0.2$, $\gamma/J=\sqrt{4/5}$. The HEOM-method calculation is performed on a $N=6$-site chain. 

\begin{figure}[htbp!]
 \centering
 \includegraphics[scale=0.5]{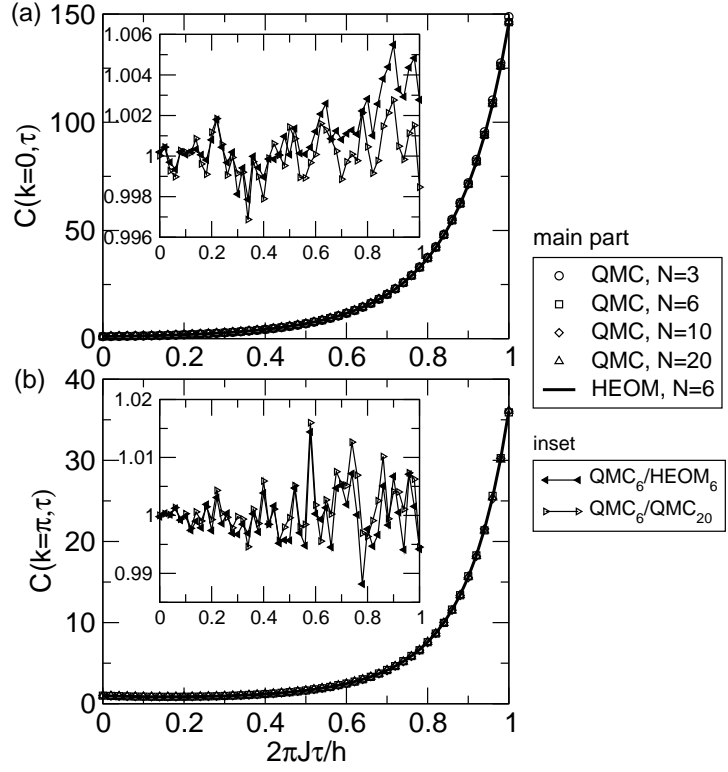}
 \caption{Imaginary time correlation function $C(k,\tau)$ (a) in the zone center $k=0$ and (b) at the zone edge $k=\pi$ computed using QMC with different chain lengths (empty symbols) and HEOM (solid line). Insets present the ratio of QMC and HEOM results for $N=6$ (full left-triangles) and the ratio of QMC results for $N=6$ and $N=20$ (empty right-triangles). The model parameters assume the following values: $k_BT/J=1$, $\hbar\omega_0/J=3$, $\gamma/J=\sqrt{12}$.}
 \label{Fig:Fig12}
\end{figure}

\newpage
In Figs.~\ref{Fig:Fig12}(a) and~\ref{Fig:Fig12}(b) we present comparison between imaginary-time correlation functions in the zone center [Fig.~\ref{Fig:Fig12}(a)] and at the zone edge [Fig.~\ref{Fig:Fig12}(b)] obtained using the QMC and HEOM methods in the antiadiabatic regime, $k_BT/J=1$, $\hbar\omega_0/J=3$, $\gamma/J=\sqrt{12}$. The HEOM-method calculation is performed on a $N=6$-site chain.

\newpage

\bibliography{refs}